\documentclass[fleqn,usenatbib,useAMS]{mnras}

\usepackage{graphicx}	
\usepackage{amsmath}	
\usepackage{amssymb}	
\usepackage{multicol}        
\usepackage{bm}		
\usepackage{pdflscape}	
\usepackage[normalem]{ulem}
\usepackage{tikz}
\usetikzlibrary{shapes.geometric, arrows}
\usepackage{caption}
\usepackage{subcaption}
\usepackage[export]{adjustbox}
\usepackage[utf8]{inputenc}
\usepackage[T1]{fontenc}
\usepackage{multirow}

\tikzstyle{input} = [rectangle, minimum width=1cm, minimum height=1cm, text centered, draw=black, fill=blue!10]

\tikzstyle{output} = [rectangle, minimum width=1cm, minimum height=1cm, text centered, draw=black, fill=white]

\tikzstyle{process} = [rectangle, minimum width=1cm, minimum height=1cm, text centered, draw=black, fill=orange!30]

\tikzstyle{decision} = [rectangle, minimum width=1cm, minimum height=1cm, text centered, draw=black, fill=green!30]

\tikzstyle{arrow} = [thick,->,>=stealth]

\usepackage[T1]{fontenc}
\usepackage{ae,aecompl}

\usepackage{newtxtext,newtxmath}

\title[eBOSS FS constraints on EDE]{Cosmological constraints on early dark energy from the full shape analysis of eBOSS DR16}

\author[R. Gsponer et al.]{
Rafaela Gsponer,$^{1}$\thanks{E-mail: rafaela.gsponer@port.ac.uk}
Ruiyang Zhao,$^{2,3,1}$
Jamie Donald-McCann,$^{1}$
David Bacon, $^{1}$ 
Kazuya Koyama,$^{1}$
\newauthor
Robert Crittenden,$^{1}$
Th\'eo Simon,$^{4}$
Eva-Maria Mueller$^{5}$
\vspace{5pt}\\
$^{1}$Institute of Cosmology \& Gravitation, University of Portsmouth, Dennis Sciama Building, Portsmouth, PO1 3FX, UK\\
$^{2}$ National Astronomy Observatories, Chinese Academy of Sciences, Beijing, 100101, P.R. China\\
$^{3}$University of Chinese Academy of Sciences, Beijing 100049, P.R.China \\
$^{4}$ Laboratoire Univers \& Particules de Montpellier (LUPM), CNRS \& Universit\'e de Montpellier (UMR-5299), F-34095 Montpellier Cedex 05, France\\
$^{5}$ Department of Physics and Astronomy, University of Sussex, Brighton BN1 9QH, UK\\
}

\date{Accepted XXX. Received YYY; in original form ZZZ}

\pubyear{2023}

\begin{document}
\label{firstpage}
\pagerange{\pageref{firstpage}--\pageref{lastpage}}
\maketitle

\begin{abstract}
We evaluate the effectiveness of Early Dark Energy (EDE) in addressing the Hubble tension using the luminous red galaxies (LRGs), quasars (QSOs), and emission line galaxies (ELGs) samples from the completed eBOSS survey. We perform cosmological parameter measurements based on full shape analysis of the power spectrum employing the effective field theory of large-scale structure (EFTofLSS). EDE is known to strongly suffer from volume projection effects, complicating cosmological constraints interpretation. To quantify the volume projection effects within an EDE full shape analysis, we explore the impact of different prior choices on the nuisance parameters of EFTofLSS through an extensive mock study. We compare classical Gaussian priors to the non-informative Jeffreys prior, known to mitigate volume projection effects in $\Lambda$CDM. Our full shape analysis combines eBOSS and BOSS data with Planck, external Baryon Acoustic Oscillation (BAO), PantheonPlus, and SH0ES supernova data. EDE reduces the tension from $5.2\sigma$ to $3\sigma$ compared to $\Lambda$CDM, yielding  $H_0=71.73_{-0.86}^{+0.82}$ km/s/Mpc with $f_\mathrm{EDE} = 0.1179_{-0.022}^{+0.025}$ (Gaussian priors) and $H_0=72.03_{-0.87}^{+0.82}$ km/s/Mpc with $f_\mathrm{EDE} = 0.1399_{-0.022}^{+0.023}$ (Jeffreys prior). Although the Hubble tension is mitigated compared to $\Lambda$CDM, the inclusion of eBOSS data amplifies the tension within EDE from $2\sigma$ to $3\sigma$, in contrast to the full shape analysis of BOSS data with Planck, external BAO, PantheonPlus, and SH0ES. This highlights the significance of incorporating additional large-scale structure data in discussions concerning models aiming to
resolve the Hubble tension. 
\end{abstract}





\section{Introduction}
Over the last century $\Lambda$ Cold Dark Matter ($\Lambda$CDM) has been established as the standard model in cosmology, but despite its successes, it faces both observational and theoretical challenges. One observational challenge which appears particularly persistent is the Hubble tension, which describes the difference in value of the Hubble constant $H_0$ as inferred from direct versus indirect measurements. The most significant discrepancy in its value comes from the cosmic microwave background (CMB) as measured by the Planck satellite, which implies $H_0 = 67.27 \pm 0.60$~km/s/Mpc~\citep{Planck:2018vyg} in $\Lambda$CDM, and the direct measurements of Cepheid-calibrated Type Ia supernovae by the SH0ES project, which reports $H_0 = 73.04 \pm 1.04$~km/s/Mpc~\citep{Riess:2021jrx, Riess:2022mme}. 
While each observational method presents unique systematic challenges, it is unlikely that a single source of systematic error can entirely account for the Hubble tension because the significance of the tension remains relatively stable when any individual measurement is excluded. Therefore, the tension may instead be indicating limitations in the current standard model. For comprehensive reviews on the Hubble tension and specific measurement systematics, readers are referred to~\citet{DiValentino:2021izs} and \citet{Abdalla:2022yfr}, and references therein. \\

Due to the robustness of the tension on the observational side, many proposals have been put forward to explain the Hubble tension by introducing new physics. Early-time solutions (models which alter the expansion history prior to recombination) as well as late-time solutions (models which alter the expansion history after recombination) have been extensively been studied; early-time solutions have demonstrated particular promise in alleviating the tension considering the current low redshift data, which tightly constrains the expansion history at $z \lesssim 2$~\citep{Bernal:2016gxb,Aylor:2018drw,Knox:2019rjx,Camarena:2021jlr,Schoneberg:2021qvd}.
A subclass of these early-time solutions that has gained much attention in recent years is early dark energy (EDE) models~\citep{Karwal:2016vyq,Poulin:2018cxd,Lin:2019qug,Smith:2019ihp,Sakstein:2019fmf,Niedermann:2020dwg,Niedermann:2019olb,Ye:2020btb,Karwal:2021vpk,Seto:2021xua,Rezazadeh:2022lsf,McDonough:2021pdg,Agrawal:2019lmo,Poulin:2023lkg}. EDE aims to alleviate the Hubble tension by introducing a light scalar degree of freedom just before recombination, which leads to a decreased sound horizon at
last scattering, so that the CMB is consistent with a larger Hubble constant $H_0$ at late times. \\

EDE has shown promise in reconciling the Hubble tension when analysing Planck CMB data, Baryon Acoustic Oscillation measurements, Pantheon supernova observations, and data from SH0ES~\citep{Poulin:2018cxd,Smith:2019ihp}. However, the inclusion of large-scale structure and weak lensing data imposes stringent upper bounds on the overall contribution of EDE~\citep{Hill:2020osr,PhysRevD.102.103502,DAmico:2020ods,Simon:2022adh, Reboucas:2023rjm,Goldstein:2023gnw,McDonough:2023qcu}. This limitation stems from the fact that EDE enhances clustering on small scales, exacerbating another prominent issue in $\Lambda$CDM: the $S_8$ tension. The $S_8$ tension quantifies the mismatch of the observed clustering in our universe when comparing Planck CMB data with weak lensing surveys.  The parameter $S_8 = \sigma_8 ( \Omega_m/ 0.3)^{0.5}$, is a combination of the matter fluctuations on a smoothing scale of $8$~Mpc/$h$ scale and the matter density today $\Omega_m$; its value inferred from the CMB differs by 2-3$\sigma$ from measurements obtained by weak lensing surveys such as CFHTLenS~\citep{Heymans:2012gg}, KiDS-1000~\citep{Heymans:2020gsg}, and DES~\citep{DES:2021wwk}. Discrepancies in the $S_8$ value have also been observed in large-scale structure studies, including the full shape analysis of BOSS data ~\citep{Zhang:2021yna,Philcox:2021kcw} - implying a moderate tension of below $2\sigma$. The cause of this tension is still under scrutiny, with recent reanalyses of BOSS data suggesting that lower values of $S_8$ might be attributed to projection effects \citep{DAmico:2022osl,Simon:2022lde,Carrilho:2022mon,Donald-McCann:2023kpx}. Furthermore, a recent combined analysis of cosmic shear of KiDS-1000 and DES suggests that the tension may only be 1.7$\sigma$~\citep{Kilo-DegreeSurvey:2023gfr}. \\

For EDE to effectively address the Hubble tension, a substantial energy injection of $\sim 12\%$ of the critical density around a redshift of 3500 is essential. The introduction of EDE leads to concurrent adjustments not only in the sound horizon $r_s$, but also in the spectral index $n_s$, and CDM energy density $\omega_\mathrm{CDM}$ to ensure a good fit to the CMB data~\citep{Poulin:2018cxd,Hill:2020osr,Vagnozzi:2021gjh}. 
These shifts lead to a worsening of the $S_8$ tension of around $0.5\sigma$~\citep{Hill:2020osr}. The enhancement in power of small-scale clustering seen in EDE can be directly probed by large-scale structure surveys such as BOSS~\citep{BOSS:2016psr} and eBOSS~\citep{eBOSS:2020yzd}. Large-scale structure analyses typically rely on a template fitting method, wherein the quasi-nonlinear galaxy power spectrum in redshift space is deduced from a given linear power spectrum, which is extended based on a perturbation theory model \citep[e.g.][]{Taruya2010} and galaxy bias scheme \citep[e.g.][]{McDonald:2009}. The resulting galaxy power spectrum is then fitted to data, enabling constraints on the Alcock-Paczynski (AP) and redshift-space distortion (RSD) parameters. While the template fitting method is computationally very efficient - the template needs to be evaluated only once -  the compression of the full power spectrum $P(k)$ to the AP and RSD parameters leads to some loss of information~\citep{Brieden:2021edu,Brieden:2021cfg,Chen:2021wdi,Maus:2023rtr}. Recently, there has been much effort to derive cosmological parameters from a direct fit to the power spectrum. Extending the modelling into modes which are mildly affected by non-linearities affords a higher $k_\mathrm{max}$, and therefore a higher number of modes ($N_{\mathrm{modes}} \propto k_\mathrm{max}^3$). The development of the effective field theory of large-scale structure \citep[EFTofLSS;][]{Baumann:2010tm,Carrasco:2012cv,senatore_bias_2015,Ivanov:2022review} was a crucial step towards this goal, and it is now possible to obtain constraints on cosmological parameters coming from direct fits, utilizing modes from the linear, as well as from the quasi-nonlinear regime~\citep{Ivanov:2019pdj,DAmico:2019fhj,Kobayashi:2021oud,Chen:2021wdi}.\\

In this work, we present constraints on EDE coming from the full shape analysis of the publicly available BOSS and eBOSS data. Previously, Bayesian analyses of the  EFTofLSS full shape fit suffered from projection effects~\citep{Simon:2022lde}, which originate from marginalization processes and lead to biases in the resulting posteriors. It was suggested in~\citet{Herold:2021ksg,Herold:2022iib,Holm:2023laa}, that by using a profile likelihood analysis, these projection effects are mitigated due to the prior independence and reparametrization invariance of the frequentist approach. 
In the context of Early Dark Energy (EDE), additional projection effects arise due to the introduction of new EDE parameters. \citet{Herold:2021ksg} and~\citet{Herold:2022iib} argue that, when considering large-scale structure data, a profile likelihood analysis favours a higher fractional amount of EDE compared to Bayesian analysis. In this paper, we address how the issue of projection effects in EDE can be tackled and mitigated within a Bayesian framework, and how this impacts the constraints on EDE. 
To do this, we extend the findings of~\citet{Donald-McCann:2023kpx} to a beyond $\Lambda$CDM context by utilizing a Jeffreys prior~\citep{jeffreys_theory_1998} on the marginalised nuisance parameters.\\

The paper is organised as follows. In
section~\ref{sec:model}, we briefly review the axion-like EDE model. In section~\ref{sec:dataandmethodology}, we further introduce the EFTofLSS and detail the data sets that we consider. We lay special focus on the BOSS and eBOSS measurements here, addressing the various systematics in the data sets. In section~\ref{sec:simulations}, we present a series of mock analyses for EDE, as well as $\Lambda$CDM, to test our inference pipeline and the effect of different prior choices. In section~\ref{sec:LCDM}, we briefly discuss results from the full shape analysis of eBOSS in $\Lambda$CDM, before moving on to section~\ref{sec:EDEresults}, where we present constraints on EDE coming from the EFTofLSS analysis of BOSS and eBOSS data and their combination with other measurements such as Planck, BAO, Pantheon+ and SH0ES. We conclude in section~\ref{sec:conclusion}.

\section{Early Dark Energy Model}
\label{sec:model}

The quantity measured by the CMB is the angular acoustic scale $\theta^*$ at recombination. It is constrained to a precision of $0.03 \% $ by Planck 2018~\citep{Planck:2018vyg}. This scale is a ratio of the sound horizon {at recombination $r_s^*$ and the associated angular diameter distance $D_A^*$, where $r_s^*$ is defined as:
\begin{equation}
    r_s^* = \int_{z_*}^\infty \frac{dz}{H(z)}c_s(z),
\end{equation}
and $D_A^*$:
\begin{equation}
    D_A^* = \int_0^{z_*} \frac{dz}{H(z)},
    \label{eq:diameter}
\end{equation}
where $z_* \approx 1100$ denotes the redshift at last scattering, $c_s(z)$ is the sound speed in the baryon-photon plasma and $H(z)$ is the Hubble parameter. Both integrals are dominated by the expansion history of their lower bounds. Assuming an underlying model for the expansion of the universe and determining the baryon and matter density from the CMB power spectra, it is possible to calculate $r_s^*$ and infer $H_0$ from $D_A^* = r_s^* / \theta^*$. EDE~\citep{KamiEDEstring,KarwalEDE} amends the assumptions of the underlying model with respect to $\Lambda \mathrm{CDM}$ by introducing an extra degree of freedom at early times. Most EDE models postulate the existence of a scalar field whose background dynamics is described by the Klein-Gordon equation and undergo the three following phases: \textit{i}) The field starts out initially frozen in its potential with an energy density constant in time, similarly to DE, \textit{ii}) a mechanism (i.e. the Hubble friction no longer holding the field in place or a spontaneous trip  change of the potential shape due to a phase transition) triggers the field to become dynamical, \textit{iii}) the field radiates away faster than matter in order to keep the expansion history unchanged at late times. The dilution ensures that the impact of EDE is localised in time. With the field becoming dynamical around recombination, EDE affects the sound horizon and decreases its value. In order to keep the value $\theta^*$ fixed, this decrease in $r_s^*$ needs to be compensated by a decrease in $D_A$, leading to a higher inferred $H_0$. To fully account for the Hubble tension, most models need to achieve a maximal fractional contribution of EDE of $\simeq 10 \%$ (roughly the percentage difference between the estimate of $H_0$ between Planck and SH0ES) around the matter-radiation equality scale.\\

EDE models come in various different flavours (see~\citet{Poulin:2023lkg} for a recent review). In this work, we adopt a canonical EDE model which introduces a pseudo scalar field with an axion-like potential of the form~\citep{PoulinEDE}:
\begin{equation}
    V (\phi) = V_0 (1 - \cos(\phi/f))^n, \quad V_0 \equiv m^2 f^2,
    \label{eq:axionpot}
\end{equation}
where $f$ is the decay constant of the axion-like field, $m$ denotes its mass and $n$ is an integer value, which controls the decay rate of the potential. For convenience, we define $\theta \equiv \phi / f$ as a renormalized field variable such that $- \pi \leq \theta \leq \pi$ (without loss of generality, we set $0 \leq \theta \leq \pi$ ). The parameter $\theta_i$ represents the renormalized initial field value of the frozen field and primarily controls the dynamics of field perturbations through an effective sound speed $c_s^2$. At early times, the field starts out in a slow-roll behaviour before the Hubble friction drops below a critical value ($H \sim m)$ and the field starts evolving to its potential minimum. Oscillating around its minimum, the field dilutes with a time averaged approximate equation of state $\omega_{\phi} = (n-1)/(n+1)$.
To make sure that EDE dilutes at least as fast as radiation, $n$ needs to be chosen $ \geq 2$. In~\citet{Smith:2019ihp}, it was found that the data are quite insensitive to this parameter in the range of $  2 < n \leq 6$, where the best fit was found to be $n =3$. For our analysis, we fix $n$ to this best fit value. In this configuration, the EDE final equation of state is $\omega_{\phi} = 1/2$, which is effectively higher than the radiation equation of state $\omega_{\rm r} = 1/3$. In principle, the background dynamics can be characterised by the three  theoretical parameters $m$, $f$ and $\theta_i$. For a more direct interpretation in terms of observables, we are mapping $m$ and $f$ to two phenomenological parameters $z_c$ and $f_\mathrm{EDE}$ following the shooting mechanism described in~\citep{Hill:2020osr}  where $z_c$ describes the critical redshift at which the field makes its maximal fractional contribution $f_\mathrm{EDE} (z_c) \equiv \rho_\mathrm{EDE}/(3 M_P^2 H^2)|_{z_c}$ to the energy budget of the universe. We therefore describe our cosmological model in terms of three additional EDE parameters - $f_\mathrm{EDE}$, $z_c$ and $\theta_i$ -  on top of the usual $\Lambda$CDM parameters.

\section{Data and Methodology}
\label{sec:dataandmethodology}
\subsection{Theory model: EFTofLSS}
\label{sec:theoryEFT}
In recent years, substantial efforts have been invested into the description of non linear modelling for large-scale structure data. Although modelling modes up to a fully non-linear regime remains a very challenging task, significant progress has been made in modelling modes which are just mildly affected by non-linearities. In this paper, we follow an effective field theory (EFT) approach, most often referred to as EFTofLSS~\citep{Baumann_2012,carrasco_effective_2012}\footnote{The initial formulation of the EFTofLSS was undertaken in Eulerian space in~\citet{Carrasco:2012cv,Baumann:2010tm}. \citet{Porto:2013qua} supplemented these efforts with a description of EFTofLSS in Lagrangian space. Following the establishment of these theoretical frameworks, various efforts were made to enhance their predictive capabilities. These efforts included the understanding of renormalization~\citep{Pajer:2013jj, Abolhasani:2015mra}, the IR-resummation of the long displacement fields~\citep{Senatore:2014vja, Baldauf:2015xfa, Senatore:2014via, Senatore:2017pbn, Lewandowski:2018ywf, Blas:2016sfa}, and calculating the two-loop matter power spectrum~\citep{Carrasco:2013sva, Carrasco:2013mua}. Subsequently, this theory was further developed within the context of biased tracers, such as galaxies and quasars, as described in the references~\citep{Senatore:2014eva, Mirbabayi:2014zca, Angulo:2015eqa, Fujita:2016dne, Perko:2016puo, Nadler:2017qto}.}. Within this framework, we take into account small non-linear corrections onto the long wavelength modes, while modelling the dark matter field as an imperfect fluid. EFTofLSS incorporates a cutoff scale, acting as an effective low pass filter and implying that the fluid equations are solved in terms of long-wavelength overdensity and velocity fields. 
Additionally, an effective stress energy tensor is introduced to account for the effects of small-scale physics onto the larger scales. At a given order $n$, the effect of these small scales and their backreactions can be captured by a finite number of so called "counterterms"
with coefficients $c_i$. These $c_i$ are free parameters and need to be either fitted to data or simulations. \\

\textsc{PyBird}~\citep{DAmico:2020kxu} generates predictions for the power spectrum of biased tracers up to 1-loop order on the basis of EFTofLSS. Employing a nonlinear bias scheme that correlates the underlying dark matter field with observed galaxy densities, the expression for the one-loop redshift-space galaxy power spectrum in Fourier space is derived as\footnote{To improve readability, we omit the redshift dependence in this and subsequent formulae if it is clear from context. In practice, all observables are computed at the effective redshift $z_{\text{eff}}$ of the respective samples.}:
\begin{align}
     P_g (k, \mu) &= Z_1 (\mu)^2 P_{11}(k) \nonumber \\
     &+ 2 \int \frac{d^3 q}{(2 \pi)^3} Z_2(\textbf{q},\textbf{k - q}, \mu)^2 P_{11}(|\textbf{k - q}|) P_{11}(q) \nonumber\\
     &+ 6 Z_1(\mu) P_{11}(k) \int \frac{d^3 q}{(2 \pi)^3} Z_3( \textbf{q},\textbf{-q},\textbf{k },\mu) P_{11}(q) \nonumber \\
     &+ 2 Z_1(\mu) P_{11}(k) \left( c_{ct} \frac{k^2}{k_M^2} + c_{r,1} \mu^2 \frac{k^2}{k_M^2} + c_{r,2} \mu^4 \frac{k^2}{k_M^2} \right) \nonumber \\
     &+ \frac{1}{\hat{n}_g} \left( c_{\epsilon, 1} + c_{\text{mono}} \frac{k^2}{k^2_M} + \frac{3}{2} c_{\text{quad}} \left( \mu^2 - \frac{1}{3}  \right)   \frac{k^2}{k^2_M} \right), 
     \label{eq:P_theory}
\end{align}
where $k$ is the norm of the wavenumber $\textbf{k}$ and $\mu \equiv \hat{k} \cdot \hat{z} $ is the cosine angle to the line of sight. The redshift-space galaxy kernels $Z_i$ encapsulate the velocity and density characteristics of galaxies (for their precise expressions, refer to~\citet{DAmico:2019fhj}). Additionally, $\hat{n}_g$ denotes the mean galaxy density, and $k_M^{-1}$ serves as the scale governing the expansion of spatial derivatives\footnote{\label{footnote:krkm}We note that EFTofLSS introduces in principle three different scales: $k_{NL}, k_M $ and $k_R$, where $k_R^{-1}$ controls the renormalization scale of the velocity products which appear in the redshift-space expansion~\citep{Senatore:2014vja}. In this analysis, we explicitly set $k_{NL}= k_M= k_R =0.7~h/ \mathrm{Mpc}$. Any deviation from this equivalence will then be represented in a deviation of $c_{r,1}$ and $c_{r,2}$ away from $\mathcal{O}(1)$ and by simple rescaling of $c_{r,1,2}$ the true value of $k_R$ can be inferred.}.\\

EFTofLSS introduces a total of 10 nuisance parameters up to first-loop order, denoted as: 
\begin{equation}
    \{ b_1, b_2, b_3, b_4, c_{ct}, c_{r,1}, c_{r,2}, c_{\epsilon,1}, c_\mathrm{mono}, c_\mathrm{quad} \}.
    \label{eq:nuis}
\end{equation}
Among these, the four galaxy bias parameters ($b_{1-4}$) emerge in the expansion of the galaxy density and velocity field with respect to the underlying dark matter field and are found in the galaxy kernels $Z_i$. In particular, $b_1$ corresponds to the linear galaxy bias parameter used in the Kaiser formula (i.e., the first term in eq.~\eqref{eq:P_theory}, which is the tree-level order of the EFTofLSS), while $b_{2-4}$ represent the non-linear galaxy bias parameters. Due to being highly degenerate~\citep{DAmico:2019fhj}, $b_2$ and $b_4$ are commonly reparameterized as follows:
\begin{align}
    c_2 = (b_2+b_4)\ / \ \sqrt{2}\ , \nonumber \\
    c_4 = (b_2-b_4)\ / \ \sqrt{2}\ .
\end{align}
The EFTofLSS model involves three stochastic parameters ($c_{\epsilon, 1}, c_{\text{mono}}, c_{\text{quad}}$) that are introduced to account for discrepancies between the actual observed galaxy field and its expected value. The first term describes a constant shot noise, while the other two terms correspond to the scale-dependant stochastic contributions of the monopole and the quadrupole. Additionally, there are three counter-terms used to incorporate the influence of UV physics: $c_\text{ct}$, which is a linear combination of the dark matter sound speed~\cite{Baumann:2010tm,Carrasco:2012cv} and a higher-derivative bias~\cite{Senatore:2014eva}, and $c_{r,1}$ and $c_{r,2}$, which are the redshift-space counterterms that govern the impact of small scales on redshift space distortions. The 2D power spectrum in Eq.~\eqref{eq:P_theory} can be decomposed into multipoles via
\begin{equation}
P_l(k) = \frac{2l+1}{2}\int_{-1}^1 P_g\left(k, \mu\right)\mathcal{L}_l\left(\mu \right)\mathrm{d}\mu\ ,
\label{eq:P_multi}
\end{equation}
where $\mathcal{L}_l (\mu)$ are the Legendre polynomial of order $l$.

\subsection{Data I: BOSS + eBOSS}
\label{sec:dataI}
The main results of this paper are based on the power spectrum analysis of the completed (extended) Baryon Oscillation Spectroscopic Survey (eBOSS) data. Here we give a brief overview of the data sets under consideration. For more details see~\cite{eBOSS:2020yzd} and references therein.

\subsubsection{eBOSS}
The eBOSS commenced full operations in July~2014. Its primary spectroscopic targets were luminous red galaxies (LRGs), emission line galaxies (ELGs) and quasars (QSOs). The survey concluded its operations on March 1, 2019~\footnote{All eBOSS data products are publicly available at: \url{https://svn.sdss.org/public/data/eboss/DR16cosmo/tags/v1_0_1/dataveccov/lrg_elg_qso/}}.  
\begin{itemize}
    \item LRGpCMASS: The eBOSS LRG sample consists of 174,816 redshift measurements over the redshift interval $0.6 < z < 1$, covering an effective volume of $V = 2.7~\mathrm{Gpc}^3$. In this work, we are supplementing the eBOSS LRG sample with BOSS DR12 LRGs which were observed in the high redshift tail  $z > 0.6$. The catalogue information can be found in~\cite{Ross:2020lqz}. The multipole power spectrum measurements for the monopole and quadrupole come from~\cite{Gil-Marin:2020bct}. The effective redshift $z_\mathrm{eff}$ is equal to $0.698$ for both galactic caps and the mean number density is set to $\hat{n}_g= 5 \cdot 10^{-5} (h/\mathrm{Mpc})^3$. In agreement with~\cite{Gil-Marin:2020bct}, we are setting the following scale cuts for our analysis: $k_\mathrm{min}= 0.02~h/\mathrm{Mpc}$ and $k_\mathrm{max} = 0.2~h/\mathrm{Mpc}$ with a binning of $0.01~h/\mathrm{Mpc}$.
    \item ELG:
    eBOSS obtained redshift measurements for 173,736 ELGs in the redshift range $0.6 < z < 1.1 $ with an effective volume of $V = 0.6~\mathrm{Gpc}^3$. The details of the ELG catalogs are described in~\cite{Raichoor:2020vio}. We use the full shape measurements of~\cite{deMattia:2020fkb} in a redshift range of $0.7 < z < 1.1 $ for our analysis of the multipoles, $l=0,2$, in the NGC and SGC sky patches. The effective redshift for NGC and SGC is $z_\mathrm{eff} = 0.86$ and $z_\mathrm{eff} = 0.853$, respectively. We set the mean number density $\hat{n}_g$ to $2.5 \cdot 10^{-4} (h/\mathrm{Mpc})^3$ for both caps. To be consistent with~\cite{deMattia:2020fkb}, we impose the following scale cuts: $k_\mathrm{min}= 0.03~h/\mathrm{Mpc}$ and $k_\mathrm{max} = 0.2~h/\mathrm{Mpc}$ with a binning of $0.01~h/\mathrm{Mpc}$.
    \item QSO: This sample includes 343,708 quasars in the redshift range $0.8 < z < 2.2 $. The description of the QSO catalogs can be found in~\cite{Ross:2020lqz}. We use the power spectrum multipoles, $l=0,2$, as measured from two different patches of the sky (NGC and SGC) in~\cite{Neveux:2020voa}. For our analysis, we use $z_\mathrm{eff} = 1.48$ as an effective redshift of the samples with an effective volume of $V = 0.6~\mathrm{Gpc}^3$ and a mean number density $\hat{n}_g= 1.5 \cdot 10^{-5} (h/\mathrm{Mpc})^3$. In line with~\cite{Neveux:2020voa}, we impose the following scale cuts: $k_\mathrm{min}= 0.02~h/\mathrm{Mpc}$ and $k_\mathrm{max} = 0.2~h/\mathrm{Mpc}$ with a binning of $0.01~h/\mathrm{Mpc}$.
    
\end{itemize}
\subsubsection{BOSS}
\label{sec:BOSS}
Between the years  2009 and 2014, the Baryon Oscillation Spectroscopic Survey (BOSS) conducted large scale structure spectroscopy within the redshift range of 0.2 to 0.75. The final galaxy catalogue used for clustering measurements counts redshift data from a total of 1,372,737 galaxies.
The sample was divided into three distinct redshift bins: $0.2 < z_1 < 0.5, 0.4 < z_2 < 0.6,$ and $0.5 < z_3 < 0.75$. The final galaxy catalogue is detailed in~\citet{Reid:2015gra}. We analyse the full shape of the BOSS power spectrum multipoles $l=0,2$ in the redshift bins $z_1$ and $z_3$ measured in~\citet{Beutler:2021eqq}\footnote{Publicly available at \url{https://fbeutler.github.io/hub/deconv_paper.html}}.  We deconvolve the data and the associated covariance matrices with the window functions provided in~\citet{Beutler:2021eqq}, such that one does not need to apply them to the theoretical predictions. The effective redshift for $z_1$ is set to $0.38$ with an effective volume $V = 3.6~\mathrm{Gpc}^3$ and mean number density $\hat{n}_g= 4 \cdot 10^{-4} (h/\mathrm{Mpc})^3$. For $z_3$, the effective redshift is $0.61$ with $V = 3.8~\mathrm{Gpc}^3$ and $\hat{n}_g= 4.5 \cdot 10^{-4} (h/\mathrm{Mpc})^3$. We set the following scale cuts for our analysis: $k_\mathrm{min}= 0.01~h/\mathrm{Mpc}$ and $k_\mathrm{max} = 0.2~h/\mathrm{Mpc}$ with a binning of $0.01~h/\mathrm{Mpc}$.

\subsubsection{BOSSz1+eBOSS}
In the context of combining data from the BOSS and eBOSS surveys, we adopt a simplified approach. Specifically, we include the low-z redshift bin ($z_1$) in BOSS, while substituting $z_3$ with LRGpCMASS. This simplification leads to the omission of BOSS galaxies falling within the redshift range of $0.5 < z < 0.6$. 

\subsection{Systematic effects}
\subsubsection{AP effect}
To convert measured celestial coordinates of galaxies and their redshifts into comoving distances, a reference cosmology must be assumed for this transformation. If the reference cosmology does not accurately represent the true cosmology, this transformation introduces artificial distortions along and parallel to the line of sight, which need to be accounted for. This distortion is known as the Alcock-Paczynski (AP) effect~\citep{Alcock:1979mp} and can be expressed using the parameters $q_{\perp}$ and $q_{\parallel}$: 
    \begin{align}
    q_\perp &= \frac{D_A(z)H(z=0)}{D_A^\mathrm{ref}(z)H^\mathrm{ref}(z=0)}\ , \nonumber \\
    q_\parallel &= \frac{H^\mathrm{ref}(z)H(z=0)}{H(z)H^\mathrm{ref}(z=0)}\ ,
\end{align}
    where $D_A(z)$ and $H(z)$ describe the angular-diameter distance and  Hubble parameter as a function of redshift, respectively. The true scales and angles are then related to the reference ones by:
    \begin{align}
        k = \frac{k^\mathrm{ref}}{q_\bot} \left[1+\left(\mu^\mathrm{ref}\right)^2\left(F^{-2}-1\right)\right]^{1/2}\ , \nonumber \\
        \mu =  \frac{\mu^\mathrm{ref}}{F}  \left[1+\left(\mu^\mathrm{ref}\right)^2\left(F^{-2}-1\right)\right]^{-1/2}\ 
    \end{align}
    with $F=q_\parallel\ / \ q_\perp$.
    \textsc{PyBird} gives predictions for the multipole of the galaxy power spectrum rather than for the 2D power spectrum (see Eq.~\eqref{eq:P_multi}). In order to account for the AP effect, the 2D power spectrum needs to be reconstructed from the theoretical multipoles via 
    \begin{equation}
    P_g(k, \mu) = \sum_{l=0} P_l(k)\mathcal{L}_l(\mu)\ .
    \label{eq:recon_2D}
    \end{equation}
    The multipole expansion of the power spectrum in the observed frame $k^\mathrm{ref}$, can then be related to the theoretical model of the line-of-sight power spectrum in terms of $k$ and $\mu$ by
    \begin{equation}
    P^{\mathrm{AP}}_l(k^\mathrm{ref}) = \frac{2l+1}{2q_\parallel q_\bot^2}\int_{-1}^1 P_g\left(k, \mu\right)\mathcal{L}_l\left(\mu^\mathrm{ref}\right)\mathrm{d}\mu^\mathrm{ref}\ .
    \label{eq:AP_multi}
\end{equation}

\begin{table*}
    \centering
    \caption{The relevant information about the fibre collision parameters for the different samples under consideration. $f_s$ describes the fraction of sky which is affected by fibre collision for each tracer and $D_\mathrm{FC}$ is the comoving distance of the 62'' fibre collision angular scale. Since the loss of targets due to fibre collision is minimal in the LRGpCMASS sample, we are not applying any fibre collision correction. }
    \begin{tabular}{l cc ccc}
    \hline
    \hline
    & $z_1/z_3$ BOSS~\citep{Hahn:2016kiy}   & ELG eBOSS~\citep{deMattia:2020fkb} & QSO eBOSS~\citep{Neveux:2020voa} & \\
    \hline\hline
    $f_s$ & 0.6 & 0.46 (NGC) / 0.38 (SGC) & 0.36 (NGC) / 0.45 (SGC) \\
    $D_\mathrm{FC}$ [Mpc/h]  & 0.43 & 0.62 & 0.9  \\
    \hline
    \end{tabular}
    \label{table:fibre}
\end{table*}

\subsubsection{Fibre collision}
To measure the spectroscopic redshift of a galaxy, a fibre needs to be placed on the location in the sky where the galaxy is located. However, if two galaxies are lying within the same angular scale that is covered by just one fibre, only one redshift measurement is obtained. This leads to a systematic bias in the clustering statistic because certain galaxies with close angular proximity are missing from the sample.
A common approach to correct for fibre collisions is by the nearest neighbour (NN) method~\citep{SDSS:2001fzq,SDSS:2004oes,BOSS:2013rlg}, where a statistical weight is assigned to the nearest angular neighbour of the fibre-collided galaxy with no redshift information. While this method provides a reasonable correction on scales much larger than the fibre collision scale, it fails elsewhere. In this paper, we therefore employ the effective window method as presented in~\citet{Hahn:2016kiy}, where the fibre collision effect is modelled at the prediction level. It was shown that the effective window method is able to mitigate fibre collision effects up to very small scales ($k  \gtrsim 0.3~h/$Mpc)~\citep{Hahn:2016kiy}, much smaller than those relevant to the scope of this paper.\\

As a first step, the correlation function $\xi_l$ is corrected using a spherical 2D top hat window function $\mathcal{W}_\mathrm{FC}$ and the fraction  of the sky $f_s$ affected by fibre collision:
    \begin{equation}
        \xi^\mathrm{FC}_l = \xi_l - \Delta \xi^\mathrm{FC}_l,
    \end{equation}
    where
    \begin{equation}
        \Delta \xi^\mathrm{FC} = \frac{2l+1}{2} \int_{-1}^1 f_s\mathcal{W}_\mathrm{FC}\left(s_\bot\right) (1 + \xi\left(s, \mu\right) )\mathcal{L}_l\left(\mu\right)\mathrm{d}\mu\ .
        \label{eq:C_fibre}
    \end{equation}
The step size of the top hat function corresponds to the comoving distance of the 62'' fibre collision angular scale, $D_\mathrm{FC}$ and  $s_\bot = s \sqrt{1 - \mu^2}$ is the transverse part of the separation vector $\boldsymbol{s} = |s|$. By Fourier transforming Eq.~\eqref{eq:C_fibre} we obtain the correction to the theoretical power spectrum. It consists of two terms: The first term corresponds to the Fourier transform of the top hat function, which accounts for uncorrelated pair collision. The second term accounts for correlated pair collision and is described by a convolution of the power spectrum with the top hat function:
    \begin{align}
        \Delta P^\mathrm{FC}_l &= \Delta P^\mathrm{uncorr}_l + \Delta P^\mathrm{corr}_l  \nonumber \\
        &= \Delta P^\mathrm{uncorr}_l + \left.\Delta P^\mathrm{corr}_l \right \vert_{q = 0}^{q = \mathrm{k_{trust}}}+ \left.\Delta P^\mathrm{corr}_l \right \vert^{q = \infty}_{q = \mathrm{k_{trust}}} \nonumber \\
        &= \Delta P^\mathrm{uncorr}_l + \left.\Delta P^\mathrm{corr}_l \right \vert_{q = 0}^{q = \mathrm{k_{trust}}}+ \sum_{l \leq n, n = 0,2,4...} C_{l,n} k^n . 
        \label{eq:P_fibre}
    \end{align}
The correlated part of the power spectrum correction involves an integration from scales 0 to $\infty$. However, in practice, it is not possible to compute the multipoles reliably up to an infinitely small distance. To address this, we split the integral in Eq.~\eqref{eq:P_fibre} into a low k integral (from 0 to $\mathrm{k_{trust}}$), which can be calculated numerically, while the high k part (from $\mathrm{k_{trust}}$ to $\infty$)  of the integral can be approximated by a polynomial in k. The form of the polynomial in Eq.~\eqref{eq:P_fibre} coincides with the stochastic contribution to the theoretical predictions of the monopole and quadrupole (compare with Eq.~\eqref{eq:P_theory}).  As a consequence, we incorporate the effect of the polynomial in the stochastic term governed by the parameters $c_\epsilon, c_\mathrm{mono}$ and $c_\mathrm{quad}$. By marginalising over the stochastic parameters, the impact of fibre collision is then fully accounted for~\footnote{In the case of the ELG, the fine-grained veto mask introduces window function kernels which are scale dependent even on small scales (see figure~1 in~\citet{deMattia:2020fkb}). The form of the stochastic contribution and the fibre collision polynomial are therefore no longer completely degenerate~\citep{Ivanov:2021zmi}. Nevertheless we expect the remaining difference to be small and safe to ignore due to the low constraining power of the sample.}. For our analysis, we set $k_\mathrm{trust}$ to be $0.25~h/\mathrm{Mpc}$.
Table~\ref{table:fibre} displays the corresponding values of $D_\mathrm{FC}$ and $f_s$ for the BOSS and eBOSS data sets. We are not applying any fibre collision to the LRGpCMASS sample since less than 4 percent of targets were not observed due to the fibre collision~\citep{Bautista:2020ahg}. 

\begin{figure}
	\includegraphics[width=1\columnwidth]{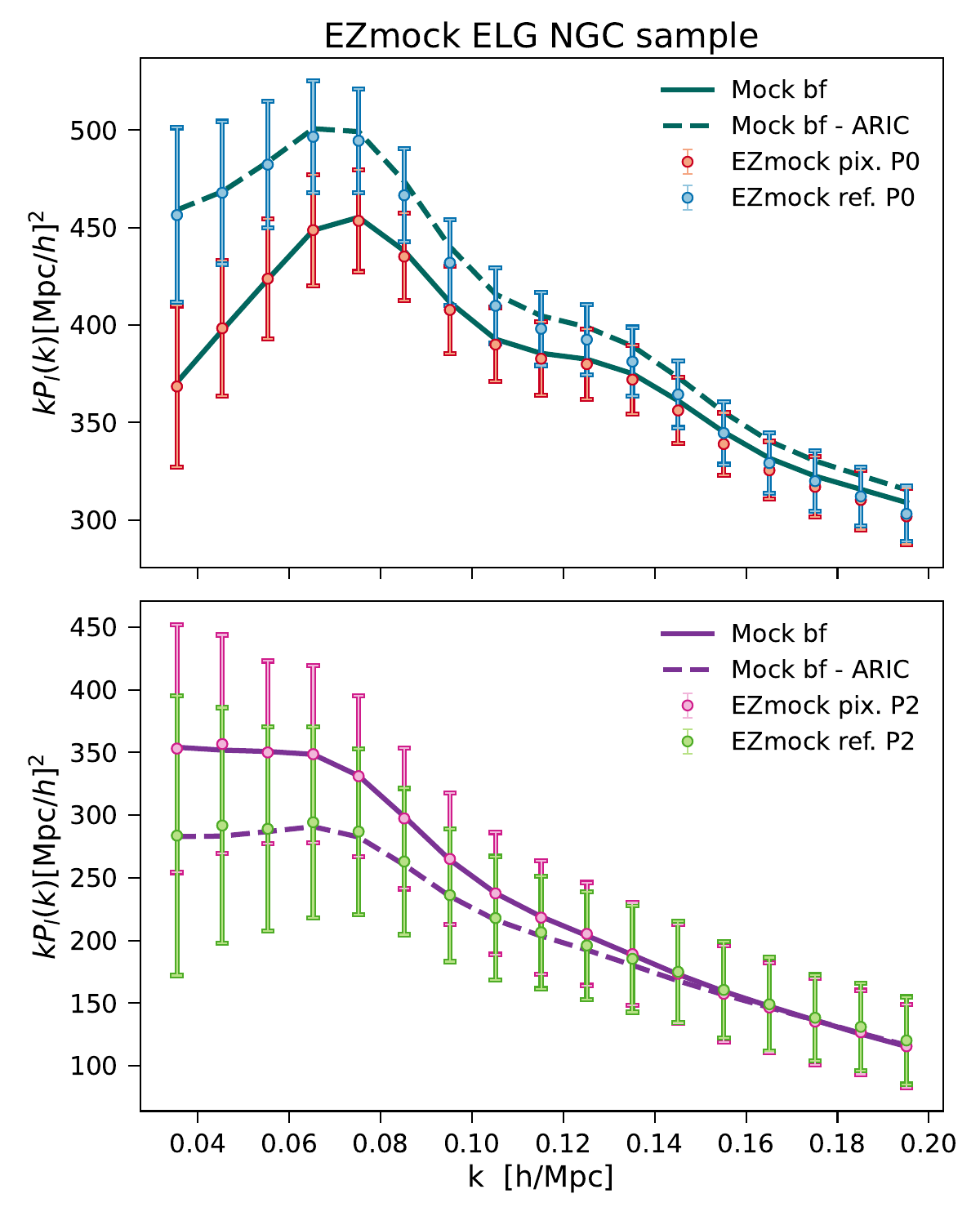}
    \caption{Effect of the radial and angular integral constraint correction on the monopole (upper panel) and quadrupole (lower panel) of the ELG NGC EZmock sample. We show mocks without systematics (EZmock ref.) and with induced radial and angular integral constraint (EZmock pix.). The different mocks are described in Sec.~\ref{sec:LCDMmock}. The integral constraints lead to a suppression of power on large scales for the monopole and enhance power for the quadrupole. The solid lines correspond to the best-fit model fitted to the \textit{pixelated} mock, where the integral constraints are modelled. The dashed lines correspond to the \textit{pixelated} best-fit model, where we did not account for the effect of the angular and radial integral constraint in the theory model. The dashed lines agree with the \textit{reference} mocks on all scales. 
    }
    \label{fig:Systematics}
\end{figure}
    
\subsubsection{Survey window function}
For the eBOSS samples, to address the survey's selection effects on the observed galaxy density, we model the survey window effect on the prediction level. This approach differs from the deconvolution method used for BOSS data (see Sec.~\ref{sec:BOSS}) and follows the procedure presented in~\citet{BOSS:2016psr, Wilson:2015lup, Beutler:2018vpe}.  The approach assumes a local plane-parallel approximation and amounts to a multiplication of the true correlation function with the multipole moments of the window mask $W_n^2$ in configuration space:
\begin{align}
    \xi^\mathrm{W}_l (s) &= \sum_{p,q} A^{p}_{l,q} \frac{2l+1}{2p+1} \mathrm{W}^2_p(s) \xi_q (s)  \\
    &= Q_{l,l'}(s) \xi_{l'} (s) ,
    \label{eq:C_wind}
\end{align}
where $A^{p}_{l,q}$ arises from multiplying two Legendre polynomials together and accounts for the weighted volume average (for its exact form see for example~\citep{Wilson:2015lup}).
As the multipole moments of the configuration space correlation function and the Fourier space power spectrum function form a Hankel transform pair, we can easily evaluate Eq.~\eqref{eq:C_wind} in Fourier space using a 1D FFT algorithm~\citep{Hamilton:1999uv}.
The convolved power spectrum can then be expressed as:
\begin{align}
    P_l^\mathrm{W}(k) &= 4 \pi (-i)^l \int s^2 ds \xi^\mathrm{W}_{l'}(s) j_l(ks) \label{eq:Hankel} \\
    &= \mathrm{W}(k, k')_{l,l'} P_{l'}(k'),
\end{align}
    where the window function for the power spectrum is given by:
    \begin{equation}
        W(k,k')_{l,l'} = \frac{2}{\pi}(-i)^l (i)^{l'} k'^2 \int ds s^2 j_l(ks) Q_{l,l'}(s) j_{l'}(k's).
    \end{equation}
Here, $j_l(ks)$ denotes spherical Bessel functions. Following the approach adopted in previous EFTofLSS full shape  analyses~\citep{DAmico:2019fhj,Ivanov:2021zmi,Chudaykin:2022nru}, we restrict our analysis to multipole moments up to the hexadecapole in $W_n^2$. 

\subsubsection{Integral constraints}
\label{sec:IC}
To define a proper over-density, a mean density of the Universe needs to be assumed. The most common assumption is to set the mean density of the Universe equal to the mean density of the survey. However, this assumption is flawed, particularly for the largest modes (corresponding to the size of the survey), as the sample variance becomes significant. This leads to a phenomenon known as the global integral constraint (IC), where the power spectrum near $k = 0$ is forced to be zero. Since the window function correlates various modes with $k=0$, this effect propagates to other scales, which are important for cosmological measurements. Similarly, the radial survey selection function is estimated from the data, leading to a corresponding radial integral constraint (RIC) which mainly suppresses power on large scales. While the RIC is subdominant for the LRG and QSO data sets~\citep{BOSS:2016psr, deMattia:2019vdg,Gil-Marin:2020bct,Neveux:2020voa}, it represents one of the most significant systematics in the ELG eBOSS data set~\citep{deMattia:2020fkb}. Moreover, the ELG data set exhibits remaining angular photometric systematics. In order to mitigate these angular effects, we remove contaminated modes by averaging the mean density fluctuations in each \textsc{HEALPix}\footnote{\url{http://healpix.jpl.nasa.gov/}}~\citep{Gorski:2004by} pixel to zero, which removes angular modes larger than the pixel scale. This step introduces an additional angular integral constraint. \\

 We correct for the impact of integral constraints on the theoretical power spectrum following the method described in~\citep{deMattia:2019vdg,deMattia:2020fkb}. All three integral constraints are modelled in a similar manner, and the observed correlation function (after window function and integral correction) is described by:
\begin{align}
    \xi^\mathrm{W+IC}_l (s) &= \xi^\mathrm{W}_l (s) \nonumber \\
    &- \sum_{l',s'} s'^2 ds' \frac{4 \pi}{2l'+1} \mathrm{W}_{l,l'}^\mathrm{IC}(s,s') \xi_{l'}(s') - sn \cdot W^\mathrm{IC_{sn}}_l(s),
\end{align}
where $sn$ is the shot noise and $\mathrm{W}_{l,l'}^\mathrm{IC}(s,s')$ is the integral correction window function in configuration space, accounting for the global, radial and angular integral constraints. To provide a complete description, we add the correction to the shot noise contribution already in configuration space (for details, refer to~\citet{deMattia:2019vdg}).
Using the Hankel transformation from Eq.~\eqref{eq:Hankel}, we  derive the window-convolved, integral-constraint-corrected power spectrum multipoles as:
\begin{align}
    P^\mathrm{W+IC}_l (k) &= P^\mathrm{W}_l (k) \nonumber \\
    &-\mathrm{W}^\mathrm{IC}_{l,l'} (k,k') P_{l'}(k') - sn \cdot \mathrm{W}_l^\mathrm{IC_{sn}}(k),
    \label{ARIC_eq}
\end{align}
where  $\mathrm{W}^\mathrm{IC}_{l,l'} (k,k')$ is given by:
\begin{align}
    \mathrm{W}^\mathrm{IC}_{l,l'} (k,k') &= \frac{8}{2l'+1} (-i)^l i^{l'} k'^2 \quad \times \nonumber \\
    &\int s^2 ds \int s'^2 ds' \mathrm{W}_{l,l'}^\mathrm{IC}(s,s') j_l(ks) j_{l'}(k's').
    \label{eq:2dHankel}
\end{align}
To solve Eq.~\eqref{eq:2dHankel}, we utilize a 2D FFTlog algorithm based on~\citet{Fang:2020vhc} and~\citet{Umeh:2020zhp}. Fig.~\ref{fig:Systematics} shows the effect of the combined radial and angular integral constraint correction on the monopole and quadrupole.

\subsection{Data II: complementary data sets}
\label{sec:dataII}
We compare and combine the data sets mentioned in Sec.~\ref{sec:dataI} with additional external data sets. Here we give a brief description of the data sets under consideration:
\begin{itemize}
    \item \textbf{PlanckTTTEEE+lensing:} We use the high $l$ TT,TE, EE and the low $l$ EE and TT power spectra from the Planck PR3-2018 data release~\citep{Planck:2018vyg}. Additionally, we include information from the gravitational lensing potential reconstructed from the temperature and polarisation data of Planck 2018~\citep{Planck:2018lbu}.
    \item \textbf{BAO:} We include complementary BAO data sets that are not correlated with BOSS z1,z3 and eBOSS LRG, ELG and QSO. These include the 6dFGS at $z = 0.106$~\citep{Beutler_2011} and the SDSS DR7 MGS at $z=0.15$~\citep{Ross:2014qpa}. 
    \item \textbf{PantheonPlus:} We include the Pantheon+ SN Ia catalogue, which consists of 1701 light curves of 1550 distinct SNIa spanning a redshift range from $0.001$ to $2.26$~\citep{Brout:2022vxf}.
    \item \textbf{SH0ES:} We also consider SH0ES Cepheid host anchors~\citep{Riess:2021jrx}. When using SH0ES alone, we impose a Gaussian likelihood for $H_0 = 73.04 \pm 1.04 \mathrm{km/s/Mpc}$. When combining it with Pantheon+, we use the PantheonPlusSH0ES likelihood (unless otherwise specified), where the distance calibration for SNIa in Cepheid host galaxies is provided by Cepheids, offering an absolute calibration for the SNIa absolute magnitude $M_B$.
    \item \textbf{BBN:} In the case where we present constraints without Planck, we impose a Big Bang nucleosynthesis (BBN) prior on $\omega_b$ from~\citet{Schoneberg:2019wmt}. This prior incorporates the theoretical prediction from~\citet{Consiglio:2017pot}, the experimental deuterium fraction of~\citet{Cooke:2017cwo}, and the experimental helium fraction of~\citet{Aver:2015iza}. 
\end{itemize}

\subsection{Likelihood}
\label{sec:likelihood}
For BOSS and eBOSS data,
we sample from a Gaussian likelihood of the form:
\begin{equation}
    \ln\left[\mathcal{L} \right] = - \frac{1}{2} \sum_{i,j} (P^\mathrm{M}_l (k_i) - P^\mathrm{D}_l (k_i)) \tilde{C}^{-1}_{ij}(P^\mathrm{M}_{l'} (k_j) - P^\mathrm{D}_{l'} (k_j)),
    \label{eq:G_lik}
\end{equation}
where $P_l^\mathrm{M}$ represents the multipoles of the modelled power spectrum, as described in Sec.~\ref{sec:model}, and $P_l^\mathrm{D}$ are the multipoles obtained by the data.  Due to the limited constraining power of the hexadecapole, our analysis focuses on the monopole and quadrupole ($l=0,2$) only\footnote{Although the hexadecapole may become important in the case of the ELGs, where its power is increased due to the integral constraints (for a more detailed discussion on this, see~\citep{deMattia:2020fkb,Ivanov:2021zmi}), we choose to exclude it. In order to model the hexadecapole, we would need to have an additional free nuisance parameter $c_{r,2}$, which would result in less overall constraining power of the ELGs. The ELGs are already the least constraining data set under consideration.}. The covariance matrices $C$ for the different data sets have been estimated either from 1000 realisations of the EZmocks for each galactic cap (NGC and SGC)~\citep{Zhao:2020bib} in the case of eBOSS or from 2048 realisations for NGC and SGC of the MultiDark-Patchy mocks~\citep{Kitaura:2015uqa,Rodriguez-Torres:2015vqa} in the case of BOSS, respectively. To account for the finite set of mocks, we use the bias-corrected estimator $\tilde{C}$~\citep{Hartlap:2006kj} of the inverse covariance matrix $C$ as:
\begin{equation}
    \tilde{C}^{-1} = \frac{N^\mathrm{real}-N^\mathrm{data}-2}{N^\mathrm{real}-1}C^{-1},
\end{equation}
where $N^\mathrm{real}$ is the total number of realisation of the mocks and $N^\mathrm{data}$ is the number of data points considered in the analysis. \\

Each sample is treated as independent\footnote{ In principle, the eBOSS samples are correlated and have overlapping volume, however the correlation was found to be less than $0.1$~\citep{eBOSS:2020yzd}.}, and the joint likelihood, denoted as $\ln{\left.[\mathcal{L}_\mathrm{joint}(\theta | \phi_\mathrm{joint})\right]}$, is defined as the sum of the individual likelihoods: $\ln{\left.[\mathcal{L}_\mathrm{joint}(\theta | \phi_\mathrm{joint})\right]} = \sum_i \ln{\left.[\mathcal{L}(\theta | \phi_i)\right]}$. Here, $\theta$ represents the shared cosmological parameters, $\phi_\mathrm{joint}$ is the comprehensive set of nuisance parameters ($\phi_\mathrm{joint}=[\phi_1, \phi_2, \ldots, \phi_n]$), and $\ln{\left[\mathcal{L}(\theta | \phi_i)\right]}$ is defined according to Eq.~\eqref{eq:G_lik}. We assume an independent set of EFTofLSS nuisance parameters for each individual sample and galactic cap. Therefore, the parameter space can grow significantly large:
for instance, combining BOSS z1 and z3 for NGC and SGC leads to $4 \times 10 = 40$ nuisance parameters. Combining BOSS z1 with LRGpCMASS, QSO and ELG further increases the number of nuisance parameters to 80. When also considering the Planck likelihoods, the number of nuisance parameters exceeds 100. In order to reduce the dimension of the parameter sampling space, it is worth noting that many of the nuisance parameters in the EFTofLSS theory enter in the power spectrum linearly (and in the likelihood quadratically). This allows for analytical marginalisation over these parameters. Following the procedure in~\citet{DAmico:2019fhj}, we compute the likelihood with free parameters $\{ b_1, c_2, c_4\}$, while marginalising over $\{ b_3,c_{ct}, c_{r,1},c_{r,2}, c_{\epsilon,1},c_\mathrm{mono} , c_\mathrm{quad} \}$, effectively reducing the number of nuisance parameters by more than a factor of two. \\

To obtain theoretical predictions of the power spectrum multipoles, we combine \textsc{CLASS\_EDE}\footnote{\url{https://github.com/mwt5345/class_ede}}~\citep{Hill:2020osr}, an extension of the publicly available Einstein-Boltzmann code \textsc{CLASS}~\citep{Diego_Blas_2011} including early dark energy, with the EFTofLSS code \textsc{PyBird}\footnote{\url{https://github.com/pierrexyz/pybird}}~\citep{DAmico:2020kxu}. To put constraints on cosmological parameters, we perform Markov chain Monte Carlo (MCMC) analyses, where we sample from the posterior distribution via \textit{Metropolis-Hastings}; as implemented in \textsc{MontePython-v3.5}\footnote{\url{https://github.com/brinckmann/montepython_public}}~\citep{Brinckmann:2018cvx}. As a convergence criterion we define a Gelman-Rubin~\citep{Gelman:1992zz} value of $R-1 < 0.01$\footnote{ We examine the $R-1$ values for each individual sampling parameter using \textsc{MontePython}.}. To determine best fit values we use \textsc{iMinuit}\footnote{\url{https://github.com/scikit-hep/iminuit}}~\citep{James:1975dr}. For post-processing chains, we use \textsc{GetDist}\footnote{\url{https://github.com/cmbant/getdist}}~\citep{Lewis:2019xzd}. 

\subsection{Parameters and priors}
\subsubsection{Cosmological parameters}
\label{sec:Cosmological parameters}
Here we discuss the selection of cosmological parameters for our analysis and their corresponding prior.
For runs without Planck data, we impose the following uniform priors on the $\Lambda$CDM parameters,
\begin{align}
    &\omega_{cdm} \in \mathcal{U}[0.05,0.25] , \nonumber \\
    &h \in \mathcal{U}[0.5,1.0] \nonumber , \\
    &\ln (10^{10}A_s) \in \mathcal{U}[1.,4.].
\end{align}
Since the constraining power of BOSS and eBOSS data is not sufficient to tightly constrain the physical baryon density $\omega_{b}$, we impose a BBN prior on $\omega_b$ (as discussed in Sec.~\ref{sec:dataII}). We fix the primordial spectral tilt to its Planck best fit value~\citep{Planck:2018vyg},
\begin{equation}
    n_s = 0.965,
\end{equation}
in the case of $\Lambda$CDM. In the case where we combine BOSS and eBOSS with SH0ES data, we vary $n_s$ according to
\begin{equation}
    n_s \in \mathcal{U}[0.7,1.3].
\end{equation}
for both $\Lambda$CDM and EDE.\\

Additionally, on the level of the linear power spectrum, we include two massless neutrinos and one massive neutrino with $m_\nu = 0.06~\mathrm{eV}$ in our analysis.
For runs including Planck data, we impose wide uniform priors on all the above mentioned parameters. 
For easier comparison with literature, we present cosmological results for the $\Lambda$CDM parameters in terms of the  reduced Hubble constant $h$ and two derived parameters,
the fractional matter abundance $\Omega_m$ and the clustering amplitude $\sigma_8$. \\

Regarding the EDE parameters, we adopt the following uniform priors,
\begin{align}
    &\mathrm{f}_\mathrm{EDE} \in \mathcal{U}[0.001,1.0], \nonumber \\
    &\log_{10} \mathrm{z}_\mathrm{c} \in \mathcal{U}[3.1,4.0], \nonumber \\
    &\theta_i \in \mathcal{U}[0.1,3.1].
\end{align}
Throughout our analysis, we keep the index of the EDE potential fixed to $n=3$, equal to the best fit found in~\citet{Smith:2019ihp}. 

\subsubsection{Priors on nuisance parameters}
\label{sec:priors}
As discussed previously, the EFTofLSS algorithm incorporates 10 nuisance parameters (see Eq.~\eqref{eq:nuis}). 
$c_{r,2}$, $c_\mathrm{mono}$ and $c_4$ are commonly set to zero~\citep{DAmico:2020kxu}, since the signal-to-noise ratios of the data under consideration are too low to properly constrain the two first parameters, while $b_2$ and $b_4$ are almost completely anti-correlated implying that $c_4 \sim 0$. Thus, we restrict our analysis on a submodel of EFTofLSS with the following nuisance parameters:
\begin{equation}
    \{ b_1, c_2, b_3, c_{ct}, c_{r,1},  c_{\epsilon,1},c_\mathrm{quad} \}.
    \label{eq:nuis2}
\end{equation}
The priors for $b_1$ and $c_2$ are respectively set to be $[0,4]$ and $[-4,4]$. 
In order to maintain the perturbative nature of the EFTofLSS, the nuisance parameters are expected to remain of the order of the linear bias $b_1$ ~\citep{DAmico:2019fhj}.
It is therefore common practice to set tight zero-centered Gaussian priors on the nuisance parameters which are marginalised over. Table~\ref{table:bias_priors} shows the prior choices for the marginalised nuisance parameters used in this work, which are in agreement with previous BOSS and eBOSS analyses with \textsc{PyBird}~\citep{DAmico:2019fhj,DAmico:2020kxu,Simon:2022lde,Simon:2022csv}. \\

Recent works have examined different prior choices and their impact on the cosmological parameter constraints~\citep{Carrilho:2022mon, Simon:2022lde, Donald-McCann:2023kpx,Holm:2023laa}. Particularly, \citet{hadzhiyska_cosmology_2023} and \citet{Donald-McCann:2023kpx} demonstrated that the use of a (partial) Jeffreys prior~\citep{jeffreys_theory_1998} can mitigate prior volume effects. Prior volume effects are a commonly seen feature in EFTofLSS where the marginalised posteriors exhibit biases away from their maximum a posteriori (MAP) estimates due to the marginalisation over nuisance parameters. This is especially evident in $\Lambda$CDM when we look at the parameter $\sigma_8$, which is degenerate with $b_1$ (and other nuisance parameters).
In this work, we therefore decide, besides presenting results with a classical Gaussian prior choice, to explore the use of a Jeffreys prior in an extended $\Lambda$CDM context. The Jeffreys prior is notable for its non-informative nature, refraining from favouring any specific parameter region \textit{a priori}. The Jeffreys prior is defined as:
\begin{equation}
    J(\theta) = \sqrt{|F(\theta)|}\ ,
\end{equation}
where $F(\theta)$ is the Fisher information matrix. For a Gaussian likelihood with covariance independent of model parameters $\theta$, this becomes:
\begin{equation}
    F_{ij}(\theta) = \frac{\partial M(\theta)}{\partial \theta_i}C^{-1}\frac{\partial M(\theta)}{\partial \theta_j}^T\ ,
\end{equation}
where $M(\theta)$ is the model and $C$ is the covariance matrix. Due to the involvedness of the partial derivatives, we impose the Jeffreys prior only on the parameters that appear linearly in the model, where the derivatives are trivial. In this case, any volume effect attributed to the linear parameters is mitigated, while volume effects caused by the remaining nuisance parameters ($b_1$ and $c_2$), as well as cosmological parameters, still remain. For this reason, we present cosmological constraints by quoting the $68\%$ credible interval as well as the best-fit, i.e. the MAP. The best-fit value is, by definition, not affected by volume effects~ \citep{DAmico:2022osl,Simon:2022lde} and is therefore a self-diagnostic way to check for remaining volume effects. For further details on the implementation of the partial Jeffreys prior in the marginalised likelihood, we refer the reader to~\citet{ruiy_multi}. 
\begin{table}
 \centering
 \begin{tabular}{p{2cm} p{1.5cm} p{1.5cm} p{1.5cm}} 
 \hline
 \hline
 Parameter & LRG & ELG & QSO\\ 
 \hline
 \hline
 $b_3$ &  $\mathcal{N}(0, 2)$& $\mathcal{N}(0, 2)$& $\mathcal{N}(0, 2)$\\
 $c_{ct}$ & $\mathcal{N}(0, 2)$ & $\mathcal{N}(0, 2)$& $\mathcal{N}(0, 2)$\\
 $c_{r,1}$ & $\mathcal{N}(0, \textbf{8})$ & $\mathcal{N}(0,  \textbf{16})$ &$\mathcal{N}(0, \textbf{16})$\\
 $c_{\epsilon,1}$ &  $\mathcal{N}(0, 2)$& $\mathcal{N}(0, 2)$& $\mathcal{N}(0, 2)$ \\
 $c_\mathrm{quad.}$ &$\mathcal{N}(0, 2)$  & $\mathcal{N}(0, 2)$& $\mathcal{N}(0, 2)$\\
 \hline
 \end{tabular}
\caption{Priors on the marginalised nuisance parameters of the \textsc{PyBird} EFTofLSS model. $\mathcal{N}(\mu, \sigma)$ denotes a normal distribution with mean $\mu$ and standard deviation $\sigma$. Except for the counterterm $c_{r,1}$ the priors are always chosen to be zero centered Gaussians with width 2. The width of the prior on $c_{r,1}$ ensures that the best-fit value lies well within 1$\sigma$. While we find consistent priors with literature~\citep{DAmico:2019fhj,DAmico:2021ymi,Simon:2022adh,Simon:2022csv} for the LRGs and QSOs, the absolute width of the prior on $c_{r,1}$ is different. This is solely due to a redefinition of the perturbative scale $k_R$ (see footnote~\ref{footnote:krkm}). For the ELGs, we find the same width as for QSOs.}
\label{table:bias_priors}
\end{table}

\section{Tests on Mock catalogues}

\label{sec:simulations}
In this section, we present results derived from a set of mock analyses performed to validate the accuracy of our cosmological inference pipeline. Our goal is to ensure unbiased parameter constraints for both $\Lambda$CDM and EDE scenarios. We commence by analysing EZ- and \textsc{PATCHY} mocks within the context of $\Lambda$CDM.  Subsequently, we extend our mock analyses to cosmological models with various different fractional contributions of EDE. In both cases, we investigate the impact of different nuisance parameter priors (refer to Sec.~\ref{sec:priors}) on the cosmological results.

\begin{figure*}
\centering
\includegraphics[width=\textwidth]{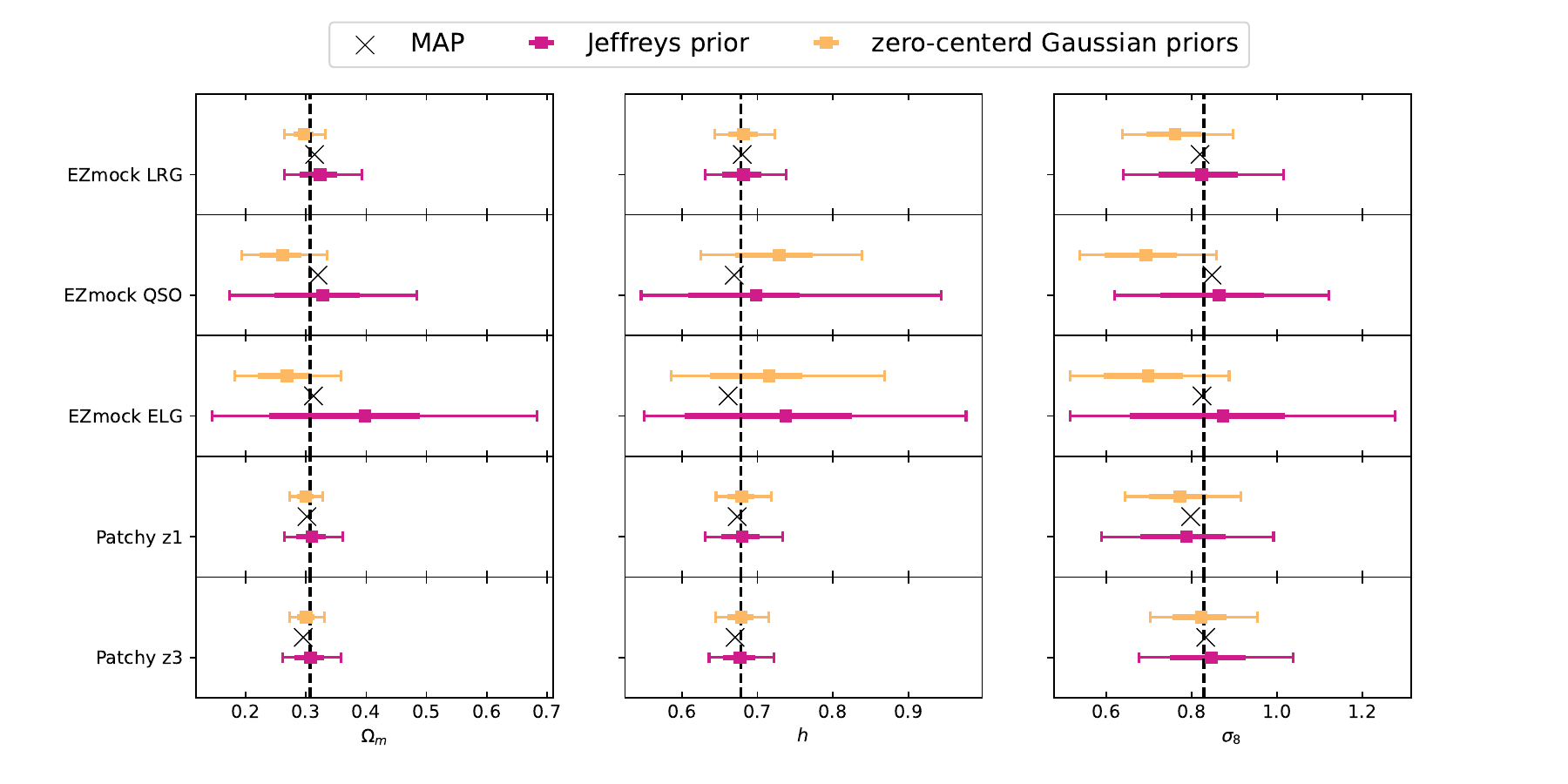}
\caption{Summary of the 1D marginalised posteriors and best-fit values for the cosmological parameters of interest resulting from analyses of the \textit{contaminated} (\textit{pixelated}) EZmocks, and \textsc{PATCHY} mocks as described in Sec.~\ref{sec:LCDMmock}. Coloured squares show the mean,  thick horizontal coloured lines show the width of the 68\% credible interval, thin coloured lines with caps show the 95\% credible interval. The black crosses indicate the MAPs for each corresponding tracer. The black dashed line corresponds to the fiducial cosmologies of the EZmocks and \textsc{PATCHY} mocks.}
\label{fig:Mock_lcdm_marg1D}
\end{figure*}

\subsection{LCDM}
\label{sec:LCDMmock}
We fit the mean of 1000 EZmock realisations~\citep{Zhao:2020bib} and the mean of 2048 MultiDark \textsc{PATCHY} mock realisations~\citep{Kitaura:2015uqa} for each corresponding tracer and redshift bin. The EZmock algorithm is based on the Zel'dovich approximation~\citep{Zeldovich:19170} to model the nonlinear matter density field and populate the field using an effective bias description. The fiducial cosmology in this set of mocks is $\Omega_m=0.307115, \, \Omega_b=0.048206, \, h=0.6777, \, \sigma_8=0.8225, \, n_s=0.9611$. The EZmocks model the systematics and survey geometry in order to match the redshift evolution and clustering properties of eBOSS DR16 LRG, ELG and QSO samples. The 1000 realisations are also used to calculate the covariances matrices used in the official eBOSS analyses~\citep{Gil-Marin:2020bct,Bautista:2020ahg,Tamone:2020qrl,deMattia:2020fkb,Raichoor:2020vio,Neveux:2020voa,Hou:2020rse,eBOSS:2020yzd} and in this work (see section~\ref{sec:likelihood}). Two different sets of mocks are produced: the \textit{reference} mock (without any observational systematics) and the \textit{contaminated} mock (with systematics).
The systematics included are redshift failure, fibre collision, depth dependent radial density and angular systematics. The \textit{contaminated} mock adopts a so-called \textit{shuffled} scheme, where the radial number density distribution $n(z)$ is estimated from the mock galaxy catalogue. This introduces a RIC effect which needs to be accounted for (see section~\ref{sec:IC}, for information on the integral constraints). 
For the ELG sample, the main systematics are of an angular photometric nature~\citep{deMattia:2020fkb}. 
We therefore work with a third set of mocks called \textit{pixelated} mocks. In the \textit{pixelated} mock the measurements are obtained by rescaling weighted randoms in \textsc{HEALPix}~\citep{Gorski:2004by} pixels (nside = 64 ($\simeq 0.84~\mathrm{deg^2}$)) such that the mean density fluctuation in each pixel is 0. 
This leads to an additional angular integral constraint effect, which together with the radial integral constraint (ARIC) biases the clustering measurements on large scales (see Fig.~\ref{fig:Systematics}). The MultiDark \textsc{PATCHY} mock is produced with approximated gravity solvers and analytical-statistical biasing models. The mock is tuned to a reference catalogue from the high resolution N-body simulation BigMultiDark~\citep{Klypin:2014kpa}\footnote{\url{https://www.cosmosim.org/}}, assuming the following fiducial cosmology: $\Omega_m=0.307115, \, \Omega_b=0.048206, \, h=0.6777, \,  \sigma_8=0.8288, \, n_s=0.9611$. It has been calibrated to match the survey geometry, redshift distribution and systematics of the BOSS DR12 galaxy data. As the EZmock for the eBOSS data sets, the 2048 realisations of the MultiDark \textsc{PATCHY} are used to calculate the covariance matrices for the BOSS data sets. For \textsc{PATCHY} as well as EZmocks analyses, we enforce the same k-range as for their corresponding tracers (see Sec.~\ref{sec:dataI}). \\

Fig.~\ref{fig:Mock_lcdm_marg1D} shows the mean (coloured squares) and the 68\% (95\%) credible interval (thick (thin) coloured lines) of the 1D marginalised posteriors on the cosmological parameters $\Omega_m,~h ,~\sigma_8$ from the analysis of the above described mock data for all tracers and redshift bins under consideration. The black dashed line corresponds to the fiducial cosmologies of the mocks and the best-fit values are indicated with black crosses.  We are comparing two different prior choices on the marginalised nuisance parameters as described in Section~\ref{sec:priors}. We observe that depending on the priors applied, the mean of the marginalised posteriors are more or less shifted away from the truth of the mocks. Meanwhile, the best fits are in agreement with the truth. This suggests that these shifts are not due to inaccurate modelling and rather are a form of volume effect.
Depending on the volume in the nuisance parameter space which is integrated over in the process of marginalisation, these effects can be more or less severe (see ~\cite{Simon:2022lde} for more details). This becomes obvious when we compare the shifts in mocks from higher signal-to-noise ratio (SNR) data (BOSS) to lower SNR data (eBOSS), where the impact of the volume effects worsens due to less constraining power of the data.  As a result, the best-fit values may not fall within the 68\% credible interval in the case of the EZmocks when zero-centered Gaussian priors are applied. In order to quantify the shift due to the volume effects, we introduce the following metric commonly used in EFTofLSS analyses (see e.g.~\citet{DAmico:2022osl,Simon:2022lde,Piga:2022mge}: ($\mu - \mu_\mathrm{truth})/\sigma (\mu)$, where we quantify the shift of the marginalised mean $\mu$ away from the truth ($\mu_\mathrm{truth}$) in number of $\sigma$. In the case where zero-centered Gaussian priors are applied, we observe that the mean of the marginalised 1D posteriors shift up to $1.6\sigma$ away from the truth for the EZmocks and up to $0.8\sigma$ for the \textsc{PATCHY} mocks.\\

While we acknowledge that the observed shifts in the considered data sets are moderate, we aim to investigate strategies to alleviate these volume effects. A promising approach, demonstrated in recent analyses~\citep{Donald-McCann:2023kpx,ruiy_multi}, involves the application of a Jeffreys prior on the nuisance parameters which enter linearly in the theory of EFTofLSS. In Fig.~\ref{fig:Mock_lcdm_marg1D}, we illustrate the impact of implementing a Jeffreys prior on the marginalized nuisance parameters, showcasing its influence on the 1D marginalized posterior distributions. We can see that for all samples, the agreement with the fiducial values has increased (where the maximum deviation for the EZmocks is $0.6\sigma$ and $0.4\sigma$ for the \textsc{PATCHY} mocks). Although the Jeffreys prior, as the less informative prior, leads to an inflation of the contours (see Table~\ref{tab:diff_std}) in comparison to the Gaussian prior, the improvement in agreement is not just coming from bigger error bars. Rather, we find that the means of the posteriors shift to their respective truth values and are more consistently located around their best-fit values. Finding unbiased posteriors when imposing a Jeffreys prior on the marginalised nuisance parameters, proves that the observed shifts with the Gaussian priors are due to volume effects and that our analysis pipeline as described in Sec.~\ref{sec:dataandmethodology} is accurate. \\

We note that in the case of very low SNR as for the ELGs and QSOs, also the Jeffreys prior fails to perfectly recover unbiased means for cosmological parameters (especially in $\Omega_m$ and $h$). This suggests the presence of residual volume effects arising from non-trivial degeneracies among nonlinear nuisance parameters. The reasoning behind this is the following: The Jeffreys prior facilitates exploration of an expanded parameter space for linear nuisance parameters compared to Gaussian priors. The size of the expanded parameter space strongly depends on the constraining power of the data. Broadening the range that these parameters can explore inevitably results in a degradation of the constraints on cosmological parameters. In scenarios characterized by low SNR, this exacerbates the impact of volume effects arising from nonlinear parameters.
Nevertheless, imposing a Jeffreys prior on the linear nuisance parameters allows for mean values consistent with the truth within $1\sigma$. 

\begin{table}
    \centering
    \begin{tabular}{p{2.cm} p{1.5cm} p{1.5cm} p{1.5cm}} 
    \hline
    \hline
    \multicolumn{4}{|c|}{$1 - \sigma_{GP}/\sigma_{JP}$}\\
    \hline
    \hline
    Sample & $\Omega_m$ & $h$ & $\sigma_8$\\ 
    \hline
    LRG & 0.49 & 0.32 & 0.31\\ 
    QSO & 0.55 & 0.39  & 0.34\\ 
    ELG & 0.68 & 0.36 & 0.51 \\ 
    z1 & 0.44 & 0.32 & 0.33 \\ 
    z3 & 0.40 & 0.18  & 0.30\\ 
    \hline
    \end{tabular}
\caption{Fractional differences on the constraints of cosmological parameters between using a Jeffreys prior or Gaussian priors on the marginalised nuisance parameters in the inference process for different mock data sets. Utilizing the Jeffreys prior leads to an inflation of the contours for all data sets, while the degree of inflation depends on the constraining power of the data set under consideration.
}
\label{tab:diff_std}
\end{table}

\begin{figure*}
\centering
     \begin{subfigure}[h]{0.49\textwidth}
         \centering
         \includegraphics[width=0.9\textwidth]{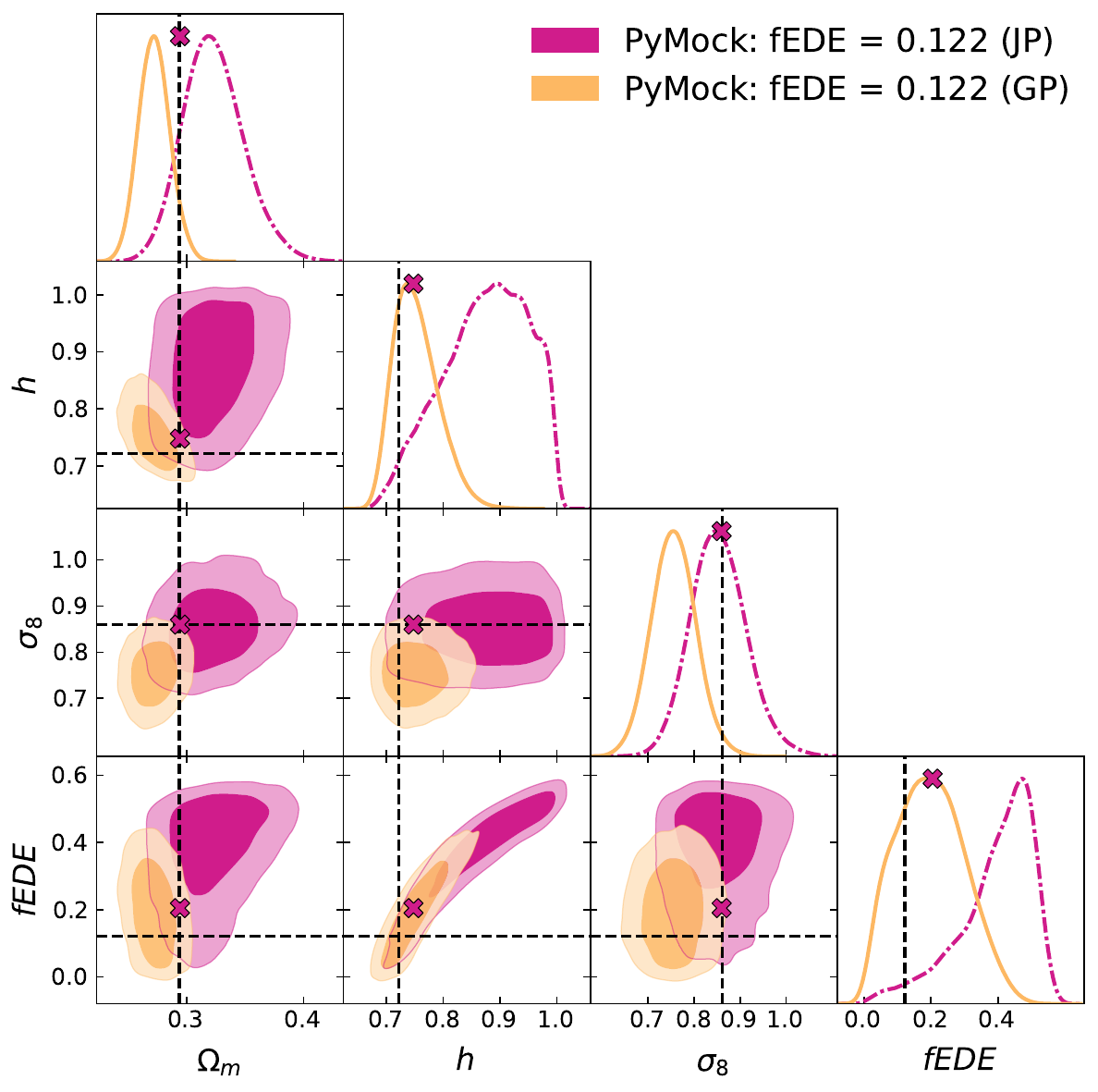}
         \end{subfigure}
     \hfill
     \begin{subfigure}[h]{0.49\textwidth}
         \centering
         \includegraphics[width=0.9\textwidth]{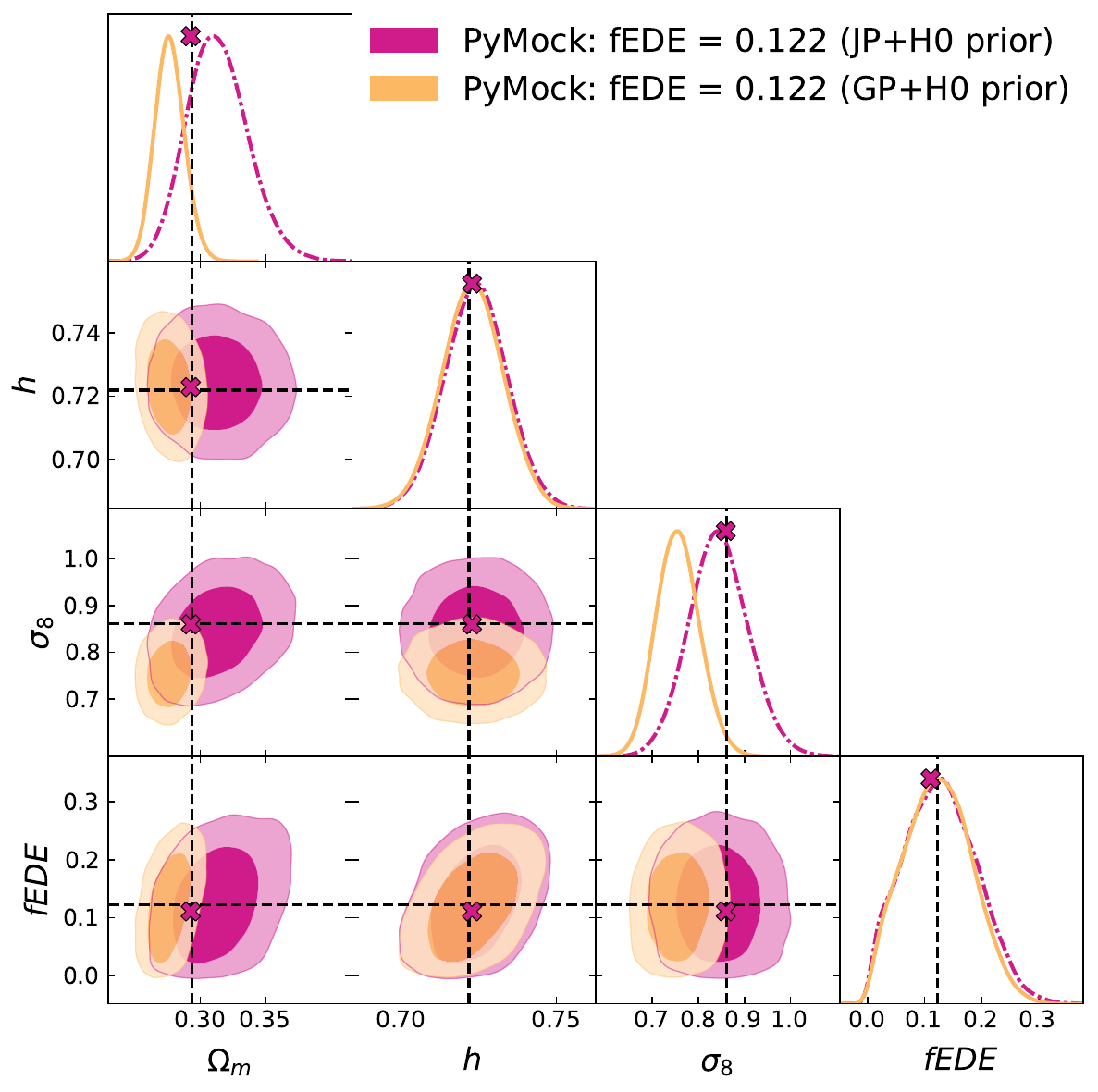}
     \end{subfigure}
     \hfill
     \begin{subfigure}[h]{0.49\textwidth}
         \centering
         \includegraphics[width=0.9\textwidth]{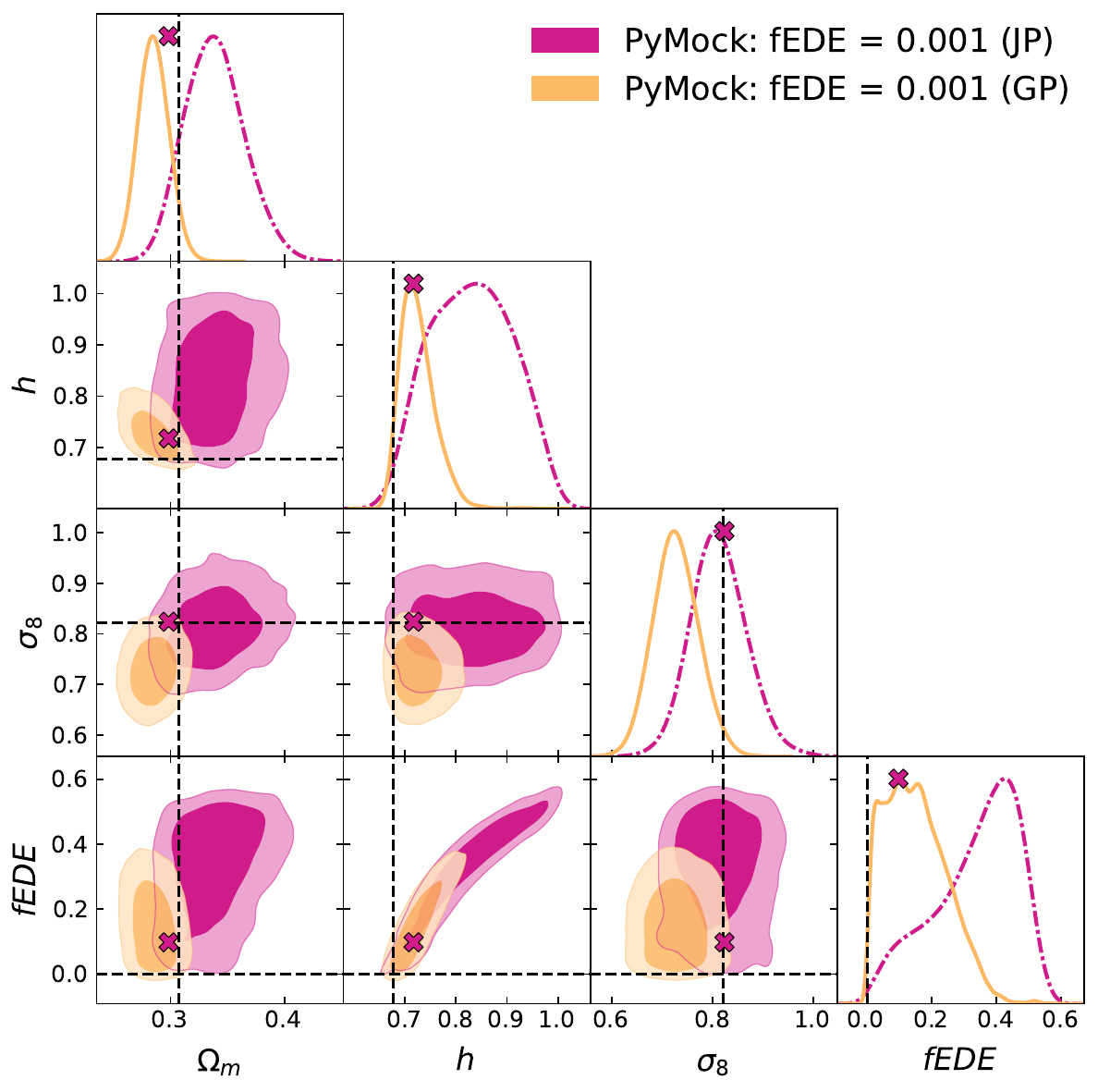}
     \end{subfigure}
     \hfill
     \begin{subfigure}[h]{0.49\textwidth}
         \centering
         \includegraphics[width=0.9\textwidth]{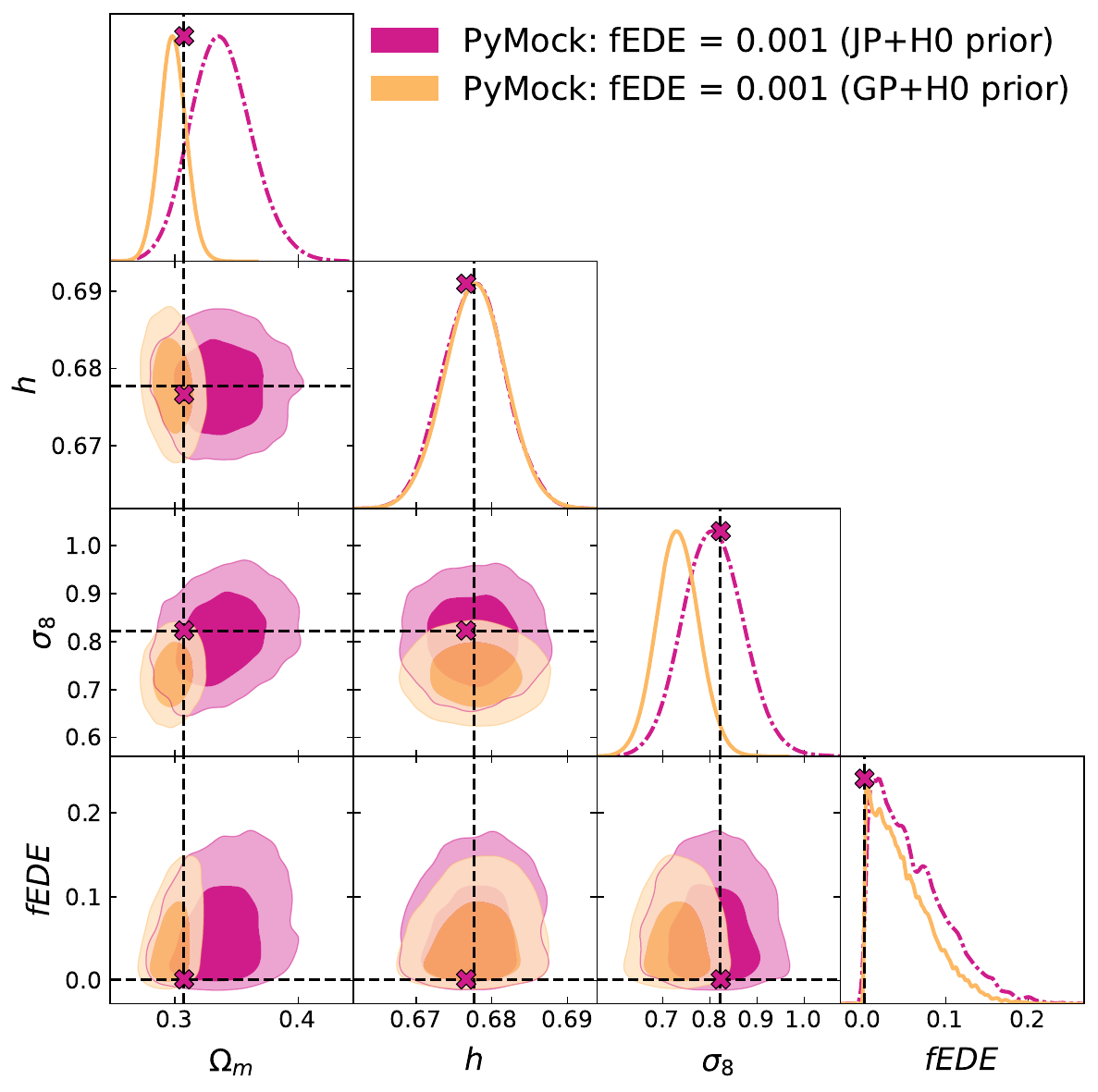}
     \end{subfigure}
     \caption{1D and 2D posterior distributions on the cosmological parameters of EDE resulting from the analyses of two different eBOSS-like PyMocks (defined in Sec.~\ref{sec:EDEmocks}). The upper panel corresponds to mocks with a maximal fractional contribution of EDE of $\sim 12\%$, while the lower panel corresponds to mocks with a minimal EDE contribution of $0.1 \%$. The two contour levels in the off-diagonal elements represent $68\%$ (inner) and $95\%$ (outer) credible intervals and the best-fit values are indicated with crosses. The dashed lines are the fiducial values of the cosmologies used to produce the mocks. We show contours for two different prior choices on the marginalised nuisance parameters in each subplot: Jeffreys prior (violet, dash-dotted line) vs. zero-centered Gaussian priors (orange, solid line). \textit{Left:} The introduction of additional cosmological parameters leads to further projection effects with both prior choices. The Jeffreys prior allows for stronger projection effects than the Gaussian priors since the degeneracy axis $(f_\mathrm{EDE},h$) is less constrained. \textit{Right:} If the eBOSS full shape analysis is combined with another data set, that has more statistical power or is able to break parameter degeneracies (here mimicked by a tight prior on $H_0$), the projection effects are reduced in both cases.}
     \label{fig:ede_pymocks}
\end{figure*}

\begin{figure*}
\centering
     \begin{subfigure}[h]{0.49\textwidth}
         \includegraphics[width=0.9\textwidth]{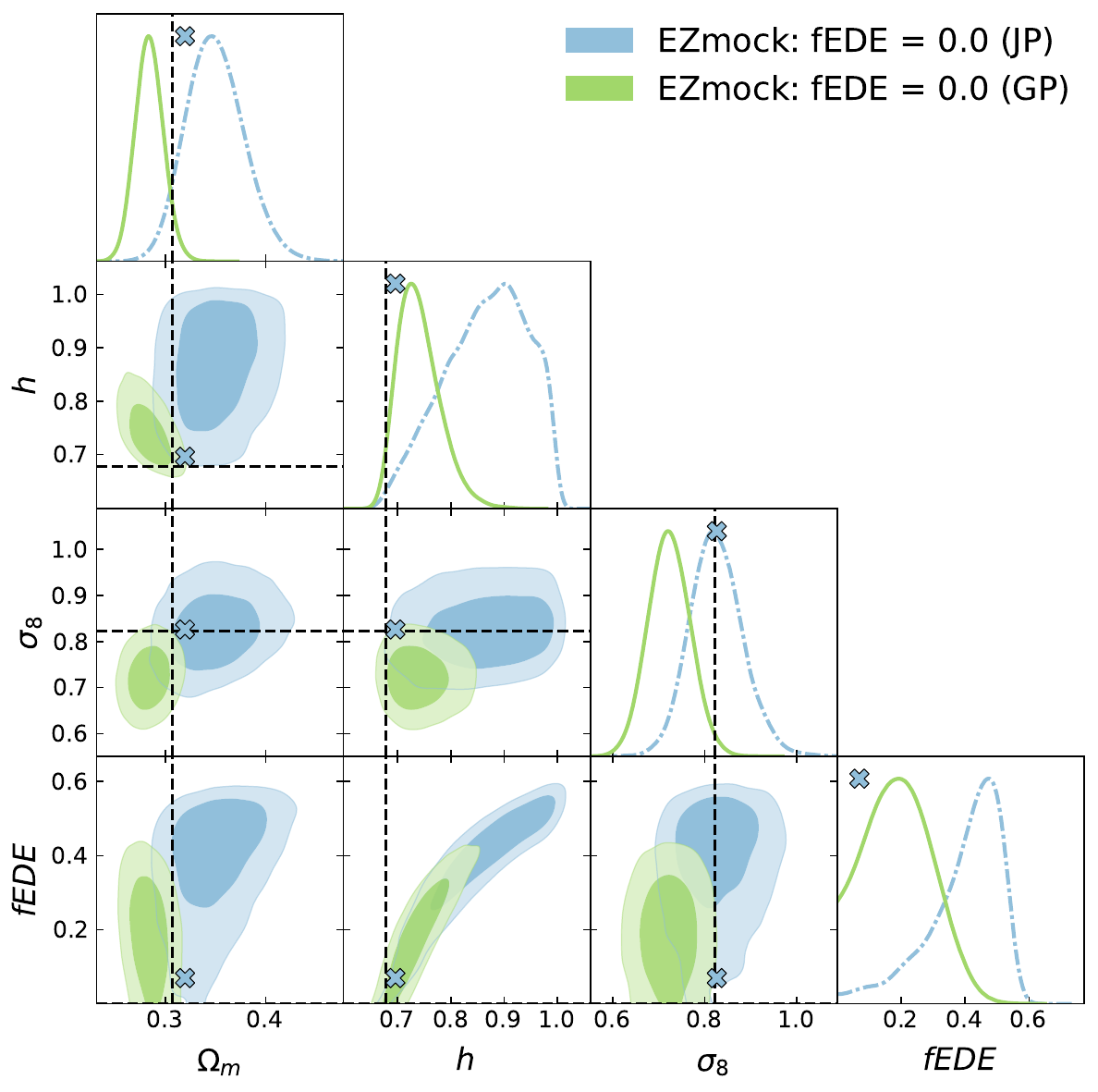}
     \end{subfigure}
     \hfill
     \begin{subfigure}[h]{0.49\textwidth}
         \includegraphics[width=0.9\textwidth]{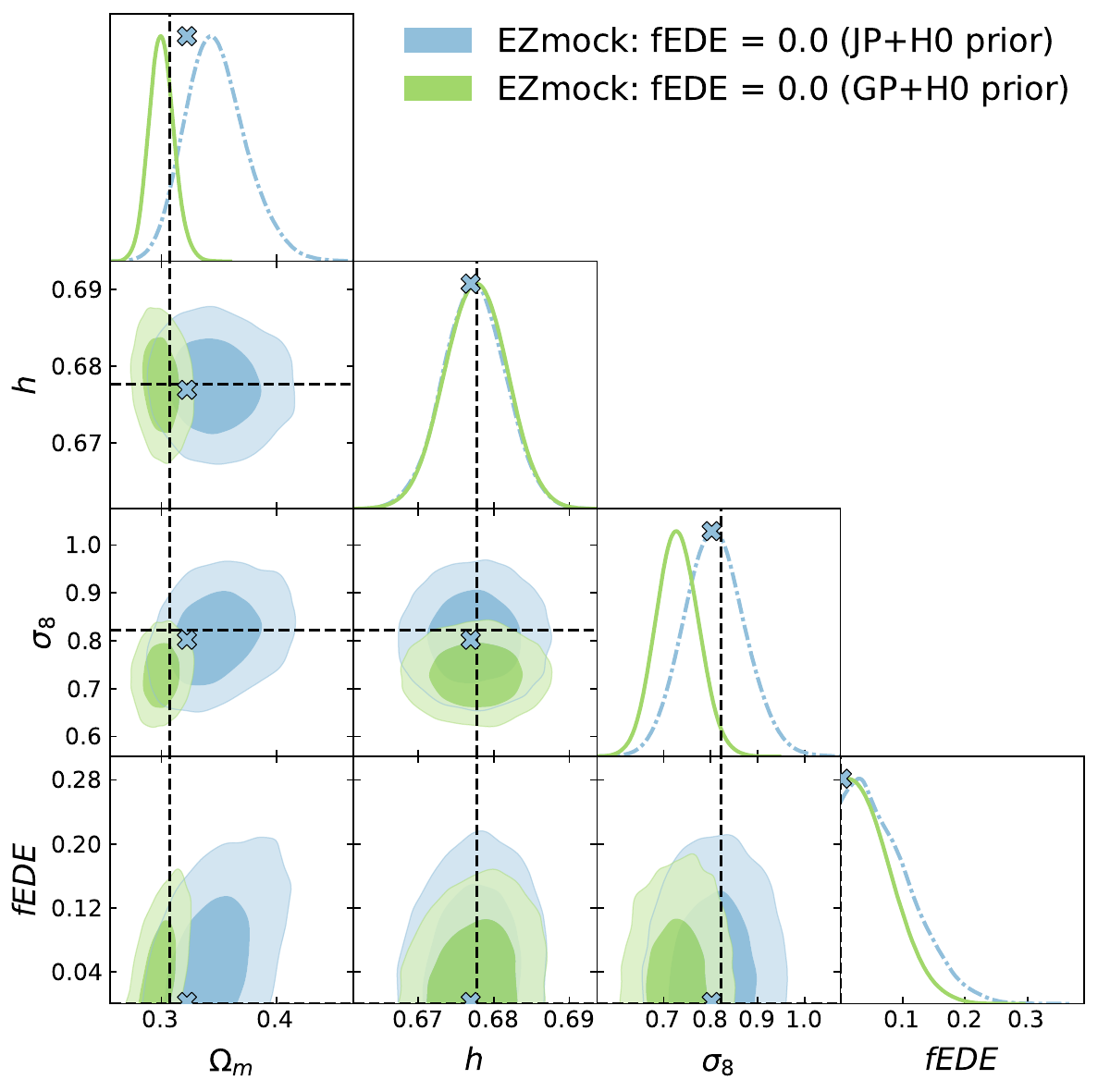}
     \end{subfigure}
     \caption{Same as Fig.~\ref{fig:ede_pymocks} but for \textsc{EZmock} runs, as described in Sec.~\ref{sec:LCDMmock}. The different contours represent the two different prior choices on the linear nuisance parameters of the model: the Jeffreys prior (blue, dash-dotted line) vs. Gaussian priors (green, solid line). Solutions fully consistent with $f_\mathrm{EDE} = 0$ are found, if the mock data is combined with a tight prior on $H_0$ (right side).   }
     \label{fig:ede_ezmocks}
\end{figure*}

\subsection{EDE}
\label{sec:EDEmocks}
As we have seen in the previous section, it is possible to strongly mitigate volume effects by utilizing a partial Jeffreys prior in a $\Lambda$CDM set-up. In this section, we want to test if this is still the case when we extend to beyond $\Lambda$CDM models where possible new degeneracies between parameters can arise. \\

In the case of EDE, we introduce three additional parameters: $f_\mathrm{EDE}$, $\log_{10} z_c$ and $\theta_i$. We want to highlight again that imposing a partial Jeffreys prior as we have done in this analysis, just removes volume effects assigned to parameters which appear linearly in the likelihood. Therefore any volume effect due to non-linear parameters, as for example from EDE parameters, will remain. In order to test the severity of the projection effect coming from the introduction of these new cosmological parameters in our pipeline, it is useful to have mocks where the cosmology, as well as the nuisance parameters are known. For this purpose, we produce two synthetic data vectors with two different $f_\mathrm{EDE}$ values using our pipeline.  We consider two limiting cases: 1) EDE entirely resolving the Hubble tension - an EDE cosmology fixed to the best-fit value of~\citet{Smith:2019ihp} (with $n=3$), corresponding to a maximal EDE contribution of $\sim 12\%$ to the energy density,
\begin{align}
    &f_\mathrm{EDE}=0.122, \log_{10}z_c=3.562, \theta_i=2.83, \nonumber \\
    &\Omega_m=0.293836, \Omega_b=0.011741, h=0.7219, \nonumber \\
    &\sigma_8=0.8606, n_s=0.9889,
\end{align}
and 2) $f_\mathrm{EDE} \rightarrow 0$  - the fiducial EZmock cosmology but with a fractional contribution of EDE of $0.1\%$ (corresponding to the lower bound of the $f_\mathrm{EDE}$ prior):
\begin{align}
    &f_\mathrm{EDE}=0.001, \log_{10}z_c=3.562, \theta_i=2.83, \nonumber \\
    &\Omega_m=0.307115, \Omega_b=0.048206, h=0.6777, \nonumber \\
    &\sigma_8=0.8225, n_s=0.9611,
\end{align}
where we fixed $\log_{10}z_c$ and $\theta_i$ to their best-fit values from~\citet{Smith:2019ihp} as before. We fit the remaining EFT parameters to the mean of 1000 \textit{comtaminated} (or \textit{pixelated}) eBOSS LRG, QSO and ELG EZmocks for both NGC and SGC. This is done by finding the MAP estimate for the 7 EFT parameters we vary, imposing the same uniform priors on $b_1$ and $c_2$ as discussed in Sec.~\ref{sec:priors} and the Gaussian priors mentioned in Table~\ref{table:bias_priors} for the rest of the nuisance parameters. We include the modelling of the systematics according to their EZmock counterparts and use EZmock covariances rescaled by a factor of 10 to find best-fit EFT parameters\footnote{We rescale the covariances such that the synthetic mocks have the same functional form as the simulated mocks on all scales, as in~\citet{Donald-McCann:2023kpx}.}.
We refer to the resulting multipoles as the "\textsc{Py}Mocks" for LRG, QSO and ELG. \\

\begin{table}
    \centering
    \begin{tabular}{p{3.cm} p{1cm} p{1.cm} p{1.cm} p{1.cm}} 
    \hline
    \hline
    \multicolumn{5}{|c|}{$(\mu - \mu_\mathrm{truth})/\sigma (\mu) $}\\
    \hline
    \hline
    Sample & Prior &  \multicolumn{1}{c}{ $\Omega_m$ }& \multicolumn{1}{c}{$h$} & \multicolumn{1}{c}{$\sigma_8$}\\ 
    \hline
    \multirow{2}{3cm}{\textsc{Py}Mocks: $f_\mathrm{EDE} = 0.122$} & JP & \multicolumn{1}{c}{0.83}   & \multicolumn{1}{c}{0.23} & \multicolumn{1}{c}{-0.25} \\ 
    & GP & \multicolumn{1}{c}{-1.51 }&  \multicolumn{1}{c}{0.13}& \multicolumn{1}{c}{ -2.23} \\
    \multirow{2}{3cm}{\textsc{Py}Mocks: $f_\mathrm{EDE} = 0.001$} & JP & \multicolumn{1}{c}{1.23} &\multicolumn{1}{c}{ -0.01} & \multicolumn{1}{c}{-0.22} \\ 
    & GP & \multicolumn{1}{c}{- 0.81} & \multicolumn{1}{c}{0.03 }& \multicolumn{1}{c}{-2.00}\\ 
    \multirow{2}{3cm}{EZmocks} & JP & \multicolumn{1}{c}{1.52} & \multicolumn{1}{c}{- 0.06 } & \multicolumn{1}{c}{- 0.23} \\ 
    & GP & \multicolumn{1}{c}{- 0.62} & \multicolumn{1}{c}{- 0.02} & \multicolumn{1}{c}{-2.07} \\ 
    \hline
    \end{tabular}
 \caption{The difference of the marginalised mean away from the truth value of the mocks in numbers of standard deviations $\sigma$ for the two different prior choices on the linear nuisance parameters: Jeffreys prior (JP) and zero-centered Gaussian priors (GP). The results shown are derived from runs where an additional tight Gaussian prior on $H_0$ centered around or close to the truth of the mocks is imposed. While the Jeffreys prior is able to reduce projection effects in $\sigma_8$, the magnitude of the shift in $\Omega_m$ is on a similar level (while its direction is inversed), highlighting that in the case of EDE additional projection effects appear due to newly introduced degeneracies between cosmological parameters. }
\label{tab:projection_ede}
\end{table}

We analyse the combination of the LRG, QSO and ELG  \textsc{Py}Mocks for the two different contributions of EDE. We fit the monopole and quadrupole of the NGC and SGC synthetic galaxy power spectra for the same $k$-range as for the corresponding data sets, using non-rescaled EZmock covariance matrices. 
We are sampling 3~+~1 cosmological parameters, as well as 12 nuisance parameters, fixing $\{n_s, \omega_b, \log_{10}z_c$, $\theta_i \}$ to their corresponding truth values. The results for the two different configurations of EDE are shown on the left side of Fig.~\ref{fig:ede_pymocks}. As for the $\Lambda$CDM case, we consider two different prior choices on the linear nuisance parameters. Except for $\sigma_8$ where the projection effect is mitigated in case of the Jeffreys prior, the posteriors experience volume effects for both prior choices. These shifts in the posteriors are present even in the case of the partial Jeffreys prior, which indicates that the projection effects have their origin in newly unlocked degeneracies between the cosmological parameters with EDE parameters. 
One very obvious, and expected, degeneracy axis is between $h$ and $f_\mathrm{EDE}$. Indeed, the  main purpose of introducing EDE is to allow for higher $H_0$ values. It is important to notice that the volume effect in case of Gaussian priors is less severe, since the sampled parameter space of the linear nuisance parameters is restricted and very extreme cosmologies (with $f_\mathrm{EDE}$ of $30\%$ and more), which are normally accommodated by large linear nuisance parameters, are excluded\footnote{Due to its uninformative nature, the Jeffreys prior does not make any assumption of the underlying theory model and these large nuisance parameters could, in principle, indicate a breakdown of the model. Recent works~\citep{Braganca:2023pcp,Donald-McCann:2023kpx} investigated additional theoretical priors to ensure that the perturbative nature of the model is preserved as an alternative (or extension) to classical Gaussian priors.}.
Furthermore, the projection effect in $\Omega_m$ which is already slightly present in $\Lambda$CDM for low SNR data is worsened in EDE, due to the new degeneracy between $f_\mathrm{EDE}$ and $h$. In order to showcase the importance that the degeneracy axis between $f_\mathrm{EDE}$ and $h$ holds, we show on the right side of Fig.~\ref{fig:ede_pymocks} mock runs where we additionally impose a prior on $H_0$. For the \textsc{Py}Mocks with $f_\mathrm{EDE} = 0.122$, we impose a prior centered on the truth of the mocks and with a SH0ES-like standard deviation: $H_0 \in \mathcal{N}(72.19, 1.04)$. While for the \textsc{Py}Mocks with $f_\mathrm{EDE} = 0.001$, the prior is  $H_0 \in  \mathcal{N}(67.66, 0.42)$ and is motivated by  TT,TE,EE+lowE+lensing+BAO constraints from Planck2018. 
Imposing this additional prior clearly reduces the volume effects within $h$ and $f_\mathrm{EDE}$ for both cases and improves on the projection effect visible in $\Omega_m$. The strong degeneracy of ($f_\mathrm{EDE}$, $h$) makes it hard to determine the correct best-fit values. Without an additional prior the best-fit values, especially of $f_\mathrm{EDE}$, are shifted away from the truth up to $0.7\sigma$. Introducing the $H_0$ prior and constraining this degeneracy axis, allows us to recover best-fit values consistent with the truth (maximal deviation of $0.2\sigma$). 
As a last point, we want to mention that only in the cases where this additional $H_0$ prior is applied, the pipeline is able to clearly distinguish between a high ($\sim 12 \%$) and a vanishing ($0.1 \%$) contribution of EDE. This is true even in the case of the Gaussian priors, where arguably the volume effects in $f_\mathrm{EDE}$ are less severe. \\

In order to test the robustness of our EDE pipeline, we show results from mock data which were produced independently of EFTofLSS. We fit the monopole and quadrupole of the combined \textit{contaminated} (\textit{pixelated}) eBOSS LRG,QSO and ELG EZmocks for both NGC and SGC, as already described in Sec.~\ref{sec:LCDMmock}. The EZmocks assume a flat  $\Lambda$CDM cosmology, corresponding to $f_\mathrm{EDE} = 0$. The results are shown in Fig.~\ref{fig:ede_ezmocks} again for two different prior choices and with (right) and without (left) imposing a tight $H_0$ prior. Consistent with the synthetic mocks, the degeneracy axis between $h$ and $f_\mathrm{EDE}$ is visible on the left side, leading to volume effects and the presence of a peak in $f_\mathrm{EDE}$. Imposing an $H_0$ prior, as for the synthetic data, resolves the volume effects in $h$ and $f_\mathrm{EDE}$ and  leaves a reduced projection effect in $\Omega_m$. Table~\ref{tab:projection_ede} quantifies the shift of the mean in regards to the truth for all EDE mock runs where an additional $H_0$ prior is applied. The deviation in terms of $\sigma$ in $\Omega_m$ for all mock sets (and all prior choices) is between $0.62 \sigma$ and $1.51 \sigma$, highlighting the remaining projection effect in $\Omega_m$, while the projection effect in $\sigma_8$ is clearly reduced in the Jeffreys prior case (from $\leq 2.23 \sigma$ (GP) to $\leq 0.23 \sigma$ (JP)). The best-fit values are shifted in regards to the truth by a maximum of $0.6\sigma$. While this is slightly higher than the shift for the synthetic mocks ($0.2\sigma$), it is unclear if this is due to an inaccuracy in the non-linear modelling of EFTofLSS or the EZmocks.\\

We conclude that although the partial Jeffreys prior is able to resolve all volume effects in $\Lambda$CDM, this is not necessarily true for extended $\Lambda$CDM scenarios, where additional degeneracies between cosmological parameters can arise. It is beyond the scope of this work to implement a Jeffreys prior on the nuisance parameters, as well as on cosmological parameters and we leave this for future work. We therefore express caution about the use of a partial Jeffreys prior in an extended $\Lambda$CDM analysis utilizing EFTofLSS, as long as the data set under consideration has no significant statistical power or the data set is not combined with external data sets (mimicked here by an additional tight $H_0$ prior), so that the degeneracies between the parameters affected by volume effects can be broken. Moving on to EDE data runs, we henceforth will not show EFTofLSS only constraints and will always show combinations with either Planck or SH0ES data\footnote{ Since the constraining power on $z_c$ and $\theta_i$ from current data sets is weak, we are aware that our analysis will have some residual volume effects coming from degeneracies between the EDE parameters which might not be entirely resolved by combining with external data sets~\citep{Smith:2020rxx,Hill:2020osr,Herold:2021ksg}.
}.

\section{Constraints on $\Lambda$CDM}
\label{sec:LCDM}

\begin{figure*}
\centering
\includegraphics[width=\textwidth]{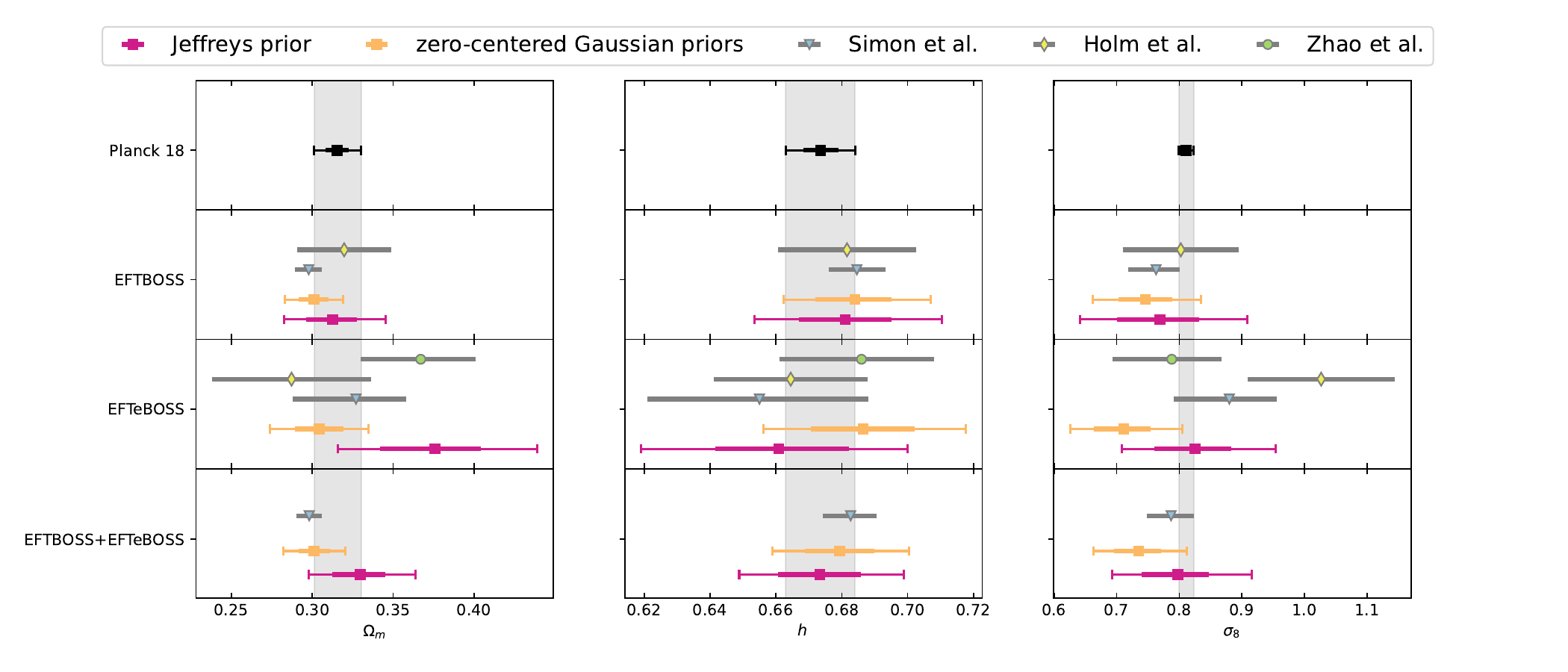}
\caption{Summary of the 1D marginalised posteriors on cosmological parameters resulting from the EFTBOSS and EFTeBOSS analyses, as well as their combined analysis. Coloured squares show the mean,  thick horizontal coloured lines show the width of the 68\% credible interval, thin coloured lines with caps show the 95\% credible interval.  In addition to the results from this work, we show results from the PlanckTTTEEE+lowl+lowE+lens analysis and results from~\citet{Simon:2022csv,ruiy_multi,Holm:2023laa}. The grey band, reflecting the $95\%$ credible interval of the Planck results, has been plotted to aid comparison. For~\citet{Simon:2022csv} (triangle,blue) and~\citet{ruiy_multi} (circle, green), we only indicate the 68\% credible intervals.~\citet{Holm:2023laa} (diamond, yellow) refers to the profile likelihood results with reconstructed 68\% confidence intervals using the Neyman construction. }
\label{fig:LCDM_eftonly}
\end{figure*}

\begin{table*}
    \centering
    \begin{tabular}{ p{1.5cm} p{2.cm} p{2.cm} p{2.cm}p{2.cm} p{2.cm} p{2.cm} p{2.cm} } 
    \hline
    \hline
    &\multicolumn{2}{c}{BOSS} &\multicolumn{2}{|c|}{eBOSS} &\multicolumn{2}{|c|}{BOSSz1+eBOSS} & \multicolumn{1}{c}{Planck}  \\
    &\multicolumn{1}{c}{GP}&\multicolumn{1}{c}{JP}&\multicolumn{1}{c}{GP}&\multicolumn{1}{c}{JP} &\multicolumn{1}{c}{GP}&\multicolumn{1}{c}{JP}&\\
    \hline
    \hline
    \multirow{2}{2cm}{ $\Omega_{m}$}  &  \multicolumn{2}{c}{$0.3069$}& \multicolumn{2}{c}{$0.3548$}& \multicolumn{2}{c}{$0.3240$ }& \multicolumn{1}{c}{$0.3158$}  \\
     & \multicolumn{1}{c}{$0.301_{-0.0095}^{+0.0095}$} &  \multicolumn{1}{c}{$0.3125_{-0.016}^{+0.016}$}& \multicolumn{1}{c}{$0.3043_{-0.015}^{+0.015}$} &  \multicolumn{1}{c}{$0.3758_{-0.035}^{+0.029}$}&\multicolumn{1}{c}{$0.3012_{-0.01}^{+0.0097}$ }& \multicolumn{1}{c}{$0.3302_{-0.018}^{+0.016}$}&\multicolumn{1}{c}{$0.3153\pm0.0073$ } \\
    \hline
    \multirow{2}{2cm}{ $h$}  &  \multicolumn{2}{c}{$0.6799$}& \multicolumn{2}{c}{$0.6625$}& \multicolumn{2}{c}{$0.6707$}& \multicolumn{1}{c}{$0.6732$}   \\
     & \multicolumn{1}{c}{$0.684_{-0.012}^{+0.011}$} & \multicolumn{1}{c}{$0.681_{-0.015}^{+0.014}$} & \multicolumn{1}{c}{$0.6865_{-0.016}^{+0.016}$} &  \multicolumn{1}{c}{$0.6609_{-0.02}^{+0.022}$}& \multicolumn{1}{c}{$0.6793_{-0.011}^{+0.01}$ }& \multicolumn{1}{c}{$0.6733_{-0.013}^{+0.013}$}&\multicolumn{1}{c}{$0.6736\pm0.0054$}  \\
    \hline
    \multirow{2}{2cm}{ $\sigma_8$} &\multicolumn{2}{c}{$0.8062$  }& \multicolumn{2}{c}{$0.8746$}& \multicolumn{2}{c}{$0.8544$}& \multicolumn{1}{c}{ $0.8120$}    \\
     & \multicolumn{1}{c}{$0.746_{-0.047}^{+0.042}$ } &\multicolumn{1}{c}{$0.7685_{-0.071}^{+0.065}$ } & \multicolumn{1}{c}{$0.7112_{-0.049}^{+0.044}$} &  \multicolumn{1}{c}{$0.8254_{-0.067}^{+0.06}$}& \multicolumn{1}{c}{$0.735_{-0.041}^{+0.036}$}& \multicolumn{1}{c}{$0.7938_{-0.053}^{+0.055}$}&\multicolumn{1}{c}{$0.8111\pm0.0060$}\\
    \hline
    \hline
    \multirow{2}{2cm}{ $\omega_{cdm}$}  &\multicolumn{2}{c}{$0.1185$}& \multicolumn{2}{c}{$0.1324$}& \multicolumn{2}{c}{$0.1224$ }& \multicolumn{1}{c}{$0.12011$}   \\
     & \multicolumn{1}{c}{$0.1175_{-0.0053}^{+0.005}$} &\multicolumn{1}{c}{$0.1217_{-0.011}^{+0.0092}$} & \multicolumn{1}{c}{$0.12_{-0.006}^{+0.0063}$} &  \multicolumn{1}{c}{$0.1405_{-0.011}^{+0.011}$}& \multicolumn{1}{c}{$0.1156_{-0.0048}^{+0.0047}$}& \multicolumn{1}{c}{$0.1264_{-0.0089}^{+0.0087}$}&\multicolumn{1}{c}{$0.1200 \pm 0.0012$}  \\
    \hline
    \multirow{2}{2cm}{ $10^2\omega_{b}$}  & \multicolumn{2}{c}{$2.271$}& \multicolumn{2}{c}{$2.269$} & \multicolumn{2}{c}{$2.272$}& \multicolumn{1}{c}{$2.2383$}    \\
     & \multicolumn{1}{c}{$2.27_{-0.04}^{+0.036}$} & \multicolumn{1}{c}{$2.269_{-0.04}^{+0.035}$}  & \multicolumn{1}{c}{$2.269_{-0.04}^{+0.036}$} &  \multicolumn{1}{c}{$2.268_{-0.039}^{+0.037}$}&\multicolumn{1}{c}{$2.27_{-0.039}^{+0.037}$}& \multicolumn{1}{c}{$2.268_{-0.041}^{+0.036}$}&\multicolumn{1}{c}{$2.237 \pm 0.015$} \\
    \hline
    \multirow{2}{2cm}{ $ln(10^{10}A_s)$}  &  \multicolumn{2}{c}{$3.051$}& \multicolumn{2}{c}{$3.088$}& \multicolumn{2}{c}{$3.133$ }& \multicolumn{1}{c}{$3.0448$}    \\
     & \multicolumn{1}{c}{$2.903_{-0.13}^{+0.13}$ } & \multicolumn{1}{c}{$2.919_{-0.18}^{+0.17}$ }& \multicolumn{1}{c}{$2.779_{-0.15}^{+0.15}$} &  \multicolumn{1}{c}{$2.903_{-0.17}^{+0.16}$}&\multicolumn{1}{c}{ $2.898_{-0.12}^{+0.11}$ }& \multicolumn{1}{c}{$2.943_{-0.14}^{+0.14}$}&\multicolumn{1}{c}{$3.044\pm0.014$}  \\
    \hline
    \hline
    $N_\mathrm{data}$ & \multicolumn{2}{c}{153} & \multicolumn{2}{c}{213}&\multicolumn{2}{c}{289 }&\multicolumn{1}{c}{2352} \\
    \hline
    $N_\mathrm{param}$ & \multicolumn{2}{c}{4 + 8} & \multicolumn{2}{c}{4 + 12}&\multicolumn{2}{c}{4 + 16}&\multicolumn{1}{c}{6 + 21} \\
    \hline
    $\chi^2$ & \multicolumn{2}{c}{$110$} & \multicolumn{2}{c}{$158$}&\multicolumn{2}{c}{$219$}&\multicolumn{1}{c}{2775} \\
    \hline
    \hline
    \end{tabular}
\caption{ Best-fit and mean $\pm$ $1\sigma$ bounds for the cosmological parameters of $\Lambda$CDM measured from the full shape analysis of BOSS, eBOSS and the combined BOSSz1+eBOSS data (always in combination with a BBN prior). We also present parameter constraints from the~PlanckTTTEEE+lowl+lowE+lens analysis. We present the mean and $1\sigma$ bounds of the full shape analyses for the two different prior choices on the linear nuisance parameters: Jeffreys prior (JP) and zero-centered Gaussian priors (GP). 
$N_\mathrm{data}$ indicates the number of fitted data points for each data set combination. We consider  $N_\mathrm{BOSSz1} = N_\mathrm{BOSSz3} = 76$~\citep{BOSS:2016wmc},  $N_\mathrm{LRG} = N_\mathrm{QSO} = 72$ and $N_\mathrm{ELG} = 68$~\citep{eBOSS:2020yzd}, $N_\mathrm{BBN} = 1$~\citep{Schoneberg:2019wmt} and $N_\mathrm{Planck} = 2352$~\citep{Planck:2018vyg,Planck:2018lbu}.  $N_\mathrm{param}$ is the combination of the number of varied cosmological and nuisance parameters.
The $\chi^2$ values are calculated
at the best-fit points.}
\label{tab:results_lcdm_eftonly}
\end{table*}

\begin{table}
    \centering
    \begin{tabular}{p{2cm} p{1cm} p{1.1cm} p{1.1cm} p{1.1cm}} 
    \hline
    \hline
    \multicolumn{5}{|c|}{$N_\sigma = |\mu_{Planck} - \mu_{EFT}|/\sqrt{\sigma_{Planck}^2 + \sigma_{EFT}^2}$}\\
    \hline
    \hline
    Sample & Prior & $\Omega_m$ & $h$ & $\sigma_8$\\ 
    \hline
    \multirow{2}{3cm}{BOSS} & JP & 0.16 & 0.46 & 0.65\\ 
    &GP & 1.19 & 0.79  & 1.53\\
    \hline
    \multirow{2}{3cm}{eBOSS} & JP & 1.69 & 0.56 & 0.21\\ 
    &GP & 0.66 & 0.76  & 2.25\\
    \hline
    \multirow{2}{3cm}{BOSSz1+eBOSS} & JP & 0.77 & 0.02 & 0.31\\ 
    &GP & 1.16 & 0.47  & 2.09\\
    \hline
    \end{tabular}
\caption{The average level of agreement between the 1D marginalised posteriors resulting from the full shape analyses of BOSS, eBOSS, and the combined BOSSz1+eBOSS and the Planck 2018 results. For each data set, we show results obtained with two prior choices on the linear nuisance parameters of the full shape analyses: Jeffreys prior (JP) and zero-centered Gaussian priors (GP).}
\label{tab:diff_Nplanck}
\end{table}

\begin{table*}
    \centering
    \begin{tabular}{ p{1.5cm} p{2.cm} p{2.cm} p{2.cm}p{2.cm} p{2.cm} p{2.cm} p{2.cm} } 
    \hline
    \hline
     & \multicolumn{2}{|c|}{BOSS + Baseline} & \multicolumn{2}{|c|}{BOSSz1 + eBOSS + Baseline} & \multicolumn{2}{|c|}{BOSSz1 + eBOSS +  SH0ES + Baseline} & \multicolumn{1}{|c|}{Baseline} \\
    &\multicolumn{1}{c}{GP}&\multicolumn{1}{c}{JP}&\multicolumn{1}{c}{GP}&\multicolumn{1}{c}{JP} &\multicolumn{1}{c}{GP}&\multicolumn{1}{c}{JP}&\\
    \hline
    \hline
    
    \multirow{2}{2cm}{ $\Omega_{m}$}  &  \multicolumn{2}{c}{$0.3144$}& \multicolumn{2}{c}{$0.3137$}& \multicolumn{2}{c}{$0.3061$ }& \multicolumn{1}{c}{$0.3219$}  \\
     & \multicolumn{1}{c}{$0.3137_{-0.0058}^{+0.006}$} &  \multicolumn{1}{c}{$0.3169_{-0.0062}^{+0.0061}$}& \multicolumn{1}{c}{$0.3139_{-0.0058}^{+0.0058}$} &  \multicolumn{1}{c}{$0.3202_{-0.0061}^{+0.006}$}&\multicolumn{1}{c}{$0.3_{-0.0052}^{+0.0051}$ }& \multicolumn{1}{r}{$0.3039_{-0.0055}^{+0.0052}$}&\multicolumn{1}{c}{$0.3184_{-0.0067}^{+0.0067}$ } \\
    \hline
    \multirow{2}{2cm}{ $h$}   &  \multicolumn{2}{c}{$0.6745$}& \multicolumn{2}{c}{$0.6741$}& \multicolumn{2}{c}{$0.6810$ }& \multicolumn{1}{c}{$0.6689$}  \\
     & \multicolumn{1}{c}{$0.6749_{-0.0046}^{+0.0041}$} &  \multicolumn{1}{c}{$0.6726_{-0.0045}^{+0.0044}$}& \multicolumn{1}{c}{$0.6747_{-0.0044}^{+0.0041}$} &  \multicolumn{1}{c}{$0.6702_{-0.0044}^{+0.0043}$}&\multicolumn{1}{c}{$0.6855_{-0.004}^{+0.004}$ }& \multicolumn{1}{r}{$0.6825_{-0.0041}^{+0.0041}$}&\multicolumn{1}{c}{$0.6715_{-0.0049}^{+0.0048}$ }  \\
    \hline
    \multirow{2}{2cm}{ $\sigma_8$} &  \multicolumn{2}{c}{$0.8095$}& \multicolumn{2}{c}{$0.8064$}& \multicolumn{2}{c}{$0.8099$ }& \multicolumn{1}{c}{$0.8094$}    \\
     & \multicolumn{1}{c}{$0.8104_{-0.006}^{+0.0062}$} &  \multicolumn{1}{c}{$0.8116_{-0.0059}^{+0.0059}$}& \multicolumn{1}{c}{$0.8097_{-0.0058}^{+0.0058}$} &  \multicolumn{1}{c}{$0.8125_{-0.0058}^{+0.006}$}&\multicolumn{1}{c}{$0.8074_{-0.0064}^{+0.006}$ }& \multicolumn{1}{r}{$0.8095_{-0.0063}^{+0.0061}$}&\multicolumn{1}{c}{$0.8124_{-0.0059}^{+0.0061}$ }\\
    \hline
    \hline
    \multirow{2}{2cm}{ $\omega_{cdm}$}   &  \multicolumn{2}{c}{$0.12$}& \multicolumn{2}{c}{$0.1196$}& \multicolumn{2}{c}{$0.1187$ }& \multicolumn{1}{c}{$0.1211$}  \\
     & \multicolumn{1}{c}{$0.1198_{-0.00095}^{+0.00099}$} &  \multicolumn{1}{c}{$0.1203_{-0.00098}^{+0.001}$}& \multicolumn{1}{c}{$0.1198_{-0.00092}^{+0.00099}$} &  \multicolumn{1}{c}{$0.1208_{-0.001}^{+0.00094}$}&\multicolumn{1}{c}{$0.1177_{-0.00089}^{+0.00087}$ }& \multicolumn{1}{r}{$0.1183_{-0.0009}^{+0.0009}$}&\multicolumn{1}{c}{$0.1206_{-0.0011}^{+0.0011}$ } \\
    \hline
    \multirow{2}{2cm}{ $10^2\omega_{b}$}   &  \multicolumn{2}{c}{$2.241$}& \multicolumn{2}{c}{$2.231$}& \multicolumn{2}{c}{$2.257$ }& \multicolumn{1}{c}{$2.229$}   \\
     & \multicolumn{1}{c}{$2.24_{-0.013}^{+0.013}$} &  \multicolumn{1}{c}{$2.236_{-0.014}^{+0.013}$}& \multicolumn{1}{c}{$2.24_{-0.014}^{+0.013}$} &  \multicolumn{1}{c}{$2.232_{-0.014}^{+0.014}$}&\multicolumn{1}{c}{$2.262_{-0.014}^{+0.013}$ }& \multicolumn{1}{r}{$2.257_{-0.013}^{+0.014}$}&\multicolumn{1}{c}{$2.234_{-0.014}^{+0.014}$ } \\
    \hline
    \multirow{2}{2cm}{ $ln(10^{10}A_s)$}   &  \multicolumn{2}{c}{$3.040$}& \multicolumn{2}{c}{$3.033$}& \multicolumn{2}{c}{$3.050$ }& \multicolumn{1}{c}{$3.032$}    \\
     & \multicolumn{1}{c}{$3.044_{-0.015}^{+0.014}$} &  \multicolumn{1}{c}{$3.043_{-0.014}^{+0.014}$}& \multicolumn{1}{c}{$3.042_{-0.014}^{+0.014}$} &  \multicolumn{1}{c}{$3.042_{-0.014}^{+0.014}$}&\multicolumn{1}{c}{$3.052_{-0.016}^{+0.014}$ }& \multicolumn{1}{r}{$3.053_{-0.016}^{+0.014}$}&\multicolumn{1}{c}{$3.043_{-0.015}^{+0.014}$ } \\
    \hline
    \multirow{2}{2cm}{ $n_s$}   &  \multicolumn{2}{c}{$0.9663$}& \multicolumn{2}{c}{$0.9684$}& \multicolumn{2}{c}{$0.9697$ }& \multicolumn{1}{c}{$0.9635$}    \\
     & \multicolumn{1}{c}{$0.9659_{-0.0039}^{+0.0037}$} &  \multicolumn{1}{c}{$0.9648_{-0.0039}^{+0.0038}$}& \multicolumn{1}{c}{$0.9657_{-0.0036}^{+0.0038}$} &  \multicolumn{1}{c}{$0.9637_{-0.0039}^{+0.0038}$}&\multicolumn{1}{c}{$0.9714_{-0.0038}^{+0.0037}$ }& \multicolumn{1}{r}{$0.9699_{-0.0037}^{+0.0036}$}&\multicolumn{1}{c}{$0.9644_{-0.004}^{+0.004}$ } \\
    \hline
    \multirow{2}{2cm}{ $\tau_\mathrm{reio}$}   &  \multicolumn{2}{c}{$0.0516$}& \multicolumn{2}{c}{$0.0487$}& \multicolumn{2}{c}{$0.0525$ }& \multicolumn{1}{c}{$0.04675$}    \\
     & \multicolumn{1}{c}{$0.05405_{-0.0072}^{+0.0073}$} &  \multicolumn{1}{c}{$0.05321_{-0.0073}^{+0.0071}$}& \multicolumn{1}{c}{$0.05315_{-0.0068}^{+0.0071}$} &  \multicolumn{1}{c}{$0.05192_{-0.0069}^{+0.007}$}&\multicolumn{1}{c}{$0.06005_{-0.008}^{+0.007}$ }& \multicolumn{1}{r}{$0.05931_{-0.0081}^{+0.0069}$}&\multicolumn{1}{c}{$0.05289_{-0.0074}^{+0.0072}$ } \\
    \hline
    \hline
    $N_\mathrm{data}$ & \multicolumn{2}{c}{4086} & \multicolumn{2}{c}{4222}&\multicolumn{2}{c}{4090}&\multicolumn{1}{c}{3934} \\
    \hline
    $N_\mathrm{param}$ & \multicolumn{2}{c}{6 + 30 } & \multicolumn{2}{c}{6 + 38 }&\multicolumn{2}{c}{6 + 38 }&\multicolumn{1}{c}{6 + 22 } \\
    \hline
    $\chi^2$ & \multicolumn{2}{c}{4298} & \multicolumn{2}{c}{4409}&\multicolumn{2}{c}{4326}&\multicolumn{1}{c}{4187} \\
    \hline
     AIC& \multicolumn{2}{c}{4370} & \multicolumn{2}{c}{4497}&\multicolumn{2}{c}{4414}&\multicolumn{1}{c}{4243} \\
    \hline
    \hline
    \end{tabular}
\caption{ Best-fit and mean $\pm$ $1\sigma$ bounds for the cosmological parameters of $\Lambda$CDM measured from the combined full shape analysis of BOSS and BOSSz1+eBOSS with the baseline, as well as with the full PantheonPlusSH0ES likelihood. The cosmological results from the baseline come from the PlanckTTTEEE+lowl+lowE+lens+ext.BAO+PantheonPlus analysis. We present the mean and $1\sigma$ bounds of the full shape analyses for the two different prior choices on the linear nuisance parameters: Jeffreys prior (JP) and zero-centered Gaussian priors (GP). 
$N_\mathrm{data}$ indicates the number of fitted data points for each data set combination. We consider  $N_\mathrm{BOSSz1} = N_\mathrm{BOSSz3} = 76$~\citep{BOSS:2016wmc}, $N_\mathrm{LRG} = N_\mathrm{QSO} = 72$ and $N_\mathrm{ELG} = 68$~\citep{eBOSS:2020yzd}, $N_\mathrm{BAO} = 2$~\citep{Beutler_2011,Ross:2014qpa}, $N_\mathrm{PantheonPlus} = 1580$~\citep{Brout:2022vxf}, $N_\mathrm{PantheonPlusSH0ES} = 1448$~\citep{Brout:2022vxf,Riess:2021jrx} and $N_\mathrm{Planck} = 2352$~\citep{Planck:2018vyg,Planck:2018lbu}.  $N_\mathrm{param}$ is the combination of the number of varied cosmological and nuisance parameters. The corresponding Akaike
Information Criterion (AIC) and $\chi^2$ values are calculated
at the best-fit points. 
}
\label{tab:results_lcdm_combined}
\end{table*}

\begin{figure*}
\centering
\includegraphics[width=0.5\textwidth]{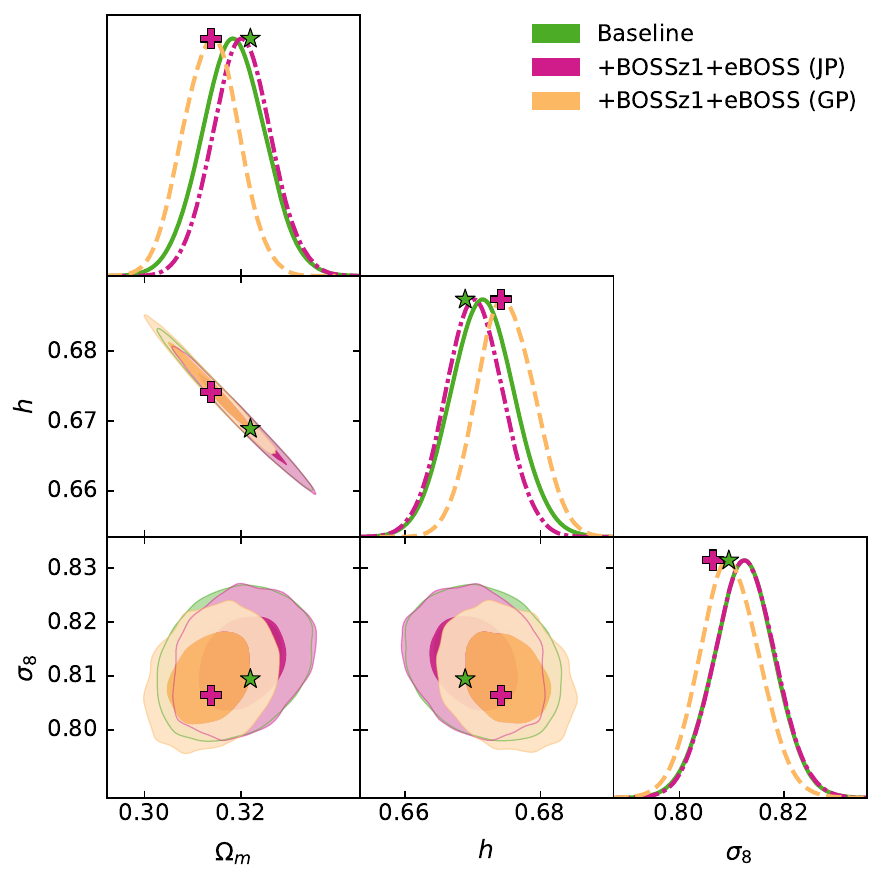}
\caption{1D and 2D posterior distributions on the cosmological parameters of the combined full shape analysis of BOSSz1+eBOSS in combination with the baseline. The baseline corresponds to PlanckTTTEEE+lowl+lowE+ext.BAO+PantheonPlus (green, solid line). The two contour levels in the off-diagonal elements represent $68\%$ (inner) and $95\%$ (outer) credible intervals and the best-fit values are indicated with corresponding markers. We compare two different prior choices of the BOSSz1+eBOSS analysis: Jeffreys prior (JP) (violet, dash-dotted line) and zero-centered Gaussian priors (GP) (orange, dashed line).}
\label{fig:LCDM_comp}
\end{figure*}

In this section, we present cosmological constraints for the $\Lambda$CDM model coming from the full shape analysis of the public eBOSS data. Fig.~\ref{fig:LCDM_eftonly} and Table~\ref{tab:results_lcdm_eftonly} summarizes our main findings.  For the EFT only analysis of eBOSS and BOSS, we fix $n_s$ to its Planck best-fit value and impose a BBN prior on $\omega_b$, as described in Sec.~\ref{sec:Cosmological parameters}. Furthermore, we perform combined analyses of full shape with other LSS surveys such as the full shape of BOSS, external BAO data, PantheonPlus, as well as with Planck data (see Sec.~\ref{sec:dataI} $\&$~\ref{sec:dataII} for details). The results of the combined analysis are shown in Fig.~\ref{fig:LCDM_comp} and Table~\ref{tab:results_lcdm_combined}. \\

\subsection{Full Shape analysis of eBOSS and BOSS}
We start by presenting full shape data constraints only. We show 1D marginalized constraints on cosmological parameters for the full shape analysis of BOSS and eBOSS individually, as well as the combined BOSSz1+eBOSS analysis in Fig~\ref{fig:LCDM_eftonly}. The coloured squares correspond  to the mean of the posteriors and the thick (thin) coloured lines indicate the 68\% (95\%) credible intervals. For comparison, we also show results from the CMB PlanckTTTEE+lowl+lowE+lens analysis in black. As for the mock runs, we show full shape results with two different priors on the linear nuisance parameters: the Jeffreys prior in violet and Gaussian priors in orange. Table~\ref{tab:results_lcdm_eftonly} summarises our findings. \\

We quantify the level of agreement between different configurations and different data sets by the number of $\sigma$ between the peak values of two marginalised distributions:
\begin{equation}
    N_\sigma = \frac{|\mu_1 - \mu_2|}{\sqrt{\sigma_1^2 + \sigma_2^2}},
\end{equation}
where $\mu_i$ and $\sigma_i$ are the mean and $1\sigma$ errors calculated from the 1D marginalised posteriors. When the distributions are asymmetric, we use the $1\sigma$ errors between the corresponding peaks. Comparing the level of agreement between Planck and the full shape analyses, as presented in Table~\ref{tab:diff_Nplanck}, we find that the analyses assuming a Jeffreys prior show an overall better agreement with Planck than the analyses with Gaussian priors. The agreement for $\sigma_8$ with Planck is especially improved for all three data set combinations if a Jeffreys prior is applied, resulting in no notable disagreement between Planck and large-scale structure data. This suggests that certain minor discrepancies that have been seen between previous EFTofLSS analysis and Planck could be due to different prior choices of EFTofLSS nuisance parameters. For a more thorough  study of different analysis choices of EFTofLSS and their influence on parameter constraints, we refer the reader to \citet{Donald-McCann:2023kpx}. The improved agreement with Planck in the case where a Jeffreys prior is assumed is in general two fold. On one hand, the peak of the posterior shifts towards the Planck values. On the other hand, the widths of the 68\% confidence intervals are enlarged. For BOSS, this broadening can be as large as $ \sim 70\%$, while the agreement for all parameters is improved by at least a factor of $\gtrsim 1.7$. The only case where the Jeffreys prior decreases the agreement with Planck compared to the Gaussian priors is in the parameter $\Omega_m$ for the eBOSS full shape analysis, where the posterior of $\Omega_m$ is shifted to high values. As already discussed in Sec.~\ref{sec:LCDMmock}, this is most likely due to residual volume effects coming from degeneracies between EFT parameters that enter non-linearly in the one-loop power spectrum, which can become important in the case of low SNR data. From eBOSS alone, we constrain $\Omega_m$ to a precision of $5\%$ (9\%), $h$ to 2\% (3\%) and $\sigma_8$ to 7\% (8\%) at $1\sigma$ level for the runs with Gaussian priors (Jeffreys prior). For the constraints from the combined BOSSz1+eBOSS data, we reconstruct $\Omega_m$, $h$, and $\sigma_8$ to 3\% (5\%), 2\% (2\%), and 6\% (7\%) precision at 68\% credible interval. \\

We compare our constraints on $\Lambda$CDM from the full shape analyses of BOSS, eBOSS and their combination with constraints from previous EFTofLSS analyses, e.g.~\citet{Simon:2022csv,ruiy_multi,Holm:2023laa}. The results of these comparisons are shown in Fig~\ref{fig:LCDM_eftonly}, together with the comparison to Planck, and should be understood with the caveat that there are some differences in the data sets included, modelling choices and priors on the cosmological parameters $n_s$ and $\omega_b$.~\citet{Simon:2022csv} derives full shape constraints using BOSS multipoles, eBOSS QSO multipoles and the combination of the two. The assumption of the priors on the linear nuisance parameters corresponds to the Gaussian priors used in this work.~\citet{Holm:2023laa} performs a profile likelihood analysis of EFTofLSS applied to the BOSS and eBOSS QSO data set. Since the Jeffreys prior has the effect of at least partially cancelling the contribution of the Laplace term in the full shape likelihoods, we expect comparable results between these two analyses. The analysis of~\citet{Holm:2023laa} further differs in that the spectral index $n_s$ is a free parameter, while it is fixed to its Planck best-fit value in our analysis and  in~\citet{Simon:2022csv}.~\citet{ruiy_multi} performs a full shape analysis of the eBOSS LRG data assuming a Jeffreys prior. The analysis of \citet{ruiy_multi} also differs in that the hexadecapole $P_4(k)$ is included in addition to the monopole and quadrupole. In order to fit $P_4(k)$ an additional nuisance parameter $c_{r,2}$ is varied. Each of the three full shape analyses presented in Fig.~\ref{fig:LCDM_eftonly} can be approximated by one of the above described EFTofLSS literature works in the sense that similar data combinations and analysis assumptions were made. It is important to notice that this is the first work where all three different tracers (LRG, QSO and ELG) of the eBOSS survey are analysed simultaneously within EFTofLSS, while~\citet{Simon:2022csv,ruiy_multi,Holm:2023laa} use sub-sets of these. 

We note the overall good agreement between the full shape analyses presented in this work and their respective literature results. The full shape analyses of BOSS and the combination of BOSS and eBOSS agree with the literature within $\lesssim 1\sigma$ under the assumption of zero-centered Gaussian priors; similar agreement is seen comparing the profile likelihood analysis with the Jeffreys prior analysis. The results of the eBOSS analysis are slightly less in agreement, mainly because the most constraining eBOSS data set (LRG) was not taken into account in~\citet{Simon:2022csv,Holm:2023laa}. For a more thorough comparison with~\citet{Holm:2023laa} it would be interesting to see how the consistency level behaves if $n_s$ is freed in an eBOSS QSO analysis where the Jeffreys prior is imposed. For the results of the eBOSS analysis, assuming a Jeffreys prior, we find good agreement with the eBOSS LRG analysis of~\citet{ruiy_multi}, with results within $\lesssim 1\sigma$.\\

\subsection{Combination with external Data Sets }
In a last step, we present EFTofLSS analysis of eBOSS and BOSS data in combination with Planck, external BAO data and PantheonPlus in Fig.~\ref{fig:LCDM_comp} and table~\ref{tab:results_lcdm_combined}. We show results for the analysis assuming a Jeffreys prior, as well as with classical zero-centered Gaussian priors on the linear nuisance parameters and indicate the best fits. Including BOSS and eBOSS full shape analysis improves constraints coming from Planck+BAO+PantheonPlus on $\Omega_m$, $h$, $\sigma_8$ by 14\% (10\%), 16\% (9\%) and 5\% (2\%) for the analysis imposing Gaussian priors (Jeffreys prior).

\section{Constraints on EDE}
\label{sec:EDEresults}
\begin{table*}
    \centering
    \begin{tabular}{ p{1.5cm} p{2.cm} p{2.cm} p{2.cm} p{2.cm} p{2.cm} p{2.cm} p{2.cm} } 
    \hline
    \hline
      & \multicolumn{2}{|c|}{BOSS  + Baseline}& \multicolumn{2}{|c|}{BOSSz1 + eBOSS + Baseline} & \multicolumn{2}{|c|}{BOSSz1 + eBOSS +  SH0ES + Baseline} & \multicolumn{1}{|c|}{Baseline}\\
    &\multicolumn{1}{c}{GP}&\multicolumn{1}{c}{JP}&\multicolumn{1}{c}{GP}&\multicolumn{1}{c}{JP} &\multicolumn{1}{c}{GP}&\multicolumn{1}{c}{JP}&\\
    \hline
    \hline
    
    \multirow{2}{2cm}{ $\Omega_{m}$}  & \multicolumn{2}{c}{$0.3071$}&  \multicolumn{2}{c}{$0.3156$}& \multicolumn{2}{c}{$0.3038$ }& \multicolumn{1}{c}{$0.3193$}  \\
     & \multicolumn{1}{c}{$0.3126_{-0.0059}^{+0.0059}$} &  \multicolumn{1}{c}{$0.3156_{-0.006}^{+0.0066}$} & \multicolumn{1}{c}{$0.3129_{-0.0059}^{+0.006}$} &  \multicolumn{1}{c}{$0.3185_{-0.0062}^{+0.0061}$} &\multicolumn{1}{c}{$0.3005_{-0.0053}^{+0.0051}$ }& \multicolumn{1}{r}{$0.3061_{-0.0058}^{+0.0054}$} &\multicolumn{1}{c}{$0.3172_{-0.0069}^{+0.0068}$ } \\
    \hline
    \multirow{2}{2cm}{ $h$}  & \multicolumn{2}{c}{$0.6857$}&  \multicolumn{2}{c}{$0.6759$}& \multicolumn{2}{c}{$0.7319$ }& \multicolumn{1}{c}{$0.6817$} \\
     & \multicolumn{1}{c}{$0.6812_{-0.008}^{+0.0053}$} &  \multicolumn{1}{c}{$0.6794_{-0.0085}^{+0.0055}$} & \multicolumn{1}{c}{$0.6803_{-0.0078}^{+0.0052}$} &  \multicolumn{1}{c}{$0.6796_{-0.011}^{+0.0061}$} &\multicolumn{1}{c}{$0.7173_{-0.0086}^{+0.0082}$ }& \multicolumn{1}{r}{$0.7203_{-0.0087}^{+0.0082}$} &\multicolumn{1}{c}{$0.6789_{-0.0095}^{+0.0062}$ }  \\
    \hline
    \multirow{2}{2cm}{ $\sigma_8$}  & \multicolumn{2}{c}{$0.8161$}&  \multicolumn{2}{c}{$0.8207$}& \multicolumn{2}{c}{$0.8528$ }& \multicolumn{1}{c}{$0.8220$}   \\
     & \multicolumn{1}{c}{$0.8146_{-0.0082}^{+0.007}$} &  \multicolumn{1}{c}{$0.8165_{-0.0089}^{+0.0072}$} & \multicolumn{1}{c}{$0.8136_{-0.0082}^{+0.0066}$} &  \multicolumn{1}{c}{$0.8199_{-0.011}^{+0.0075}$} &\multicolumn{1}{c}{$0.8379_{-0.0098}^{+0.0094}$ }& \multicolumn{1}{r}{$0.8475_{-0.0094}^{+0.009}$} &\multicolumn{1}{c}{$0.8182_{-0.0097}^{+0.007}$ } \\
    \hline
    \multirow{2}{2cm}{ $f_\mathrm{EDE}$}  & \multicolumn{2}{c}{$0.0256$}&  \multicolumn{2}{c}{$0.0087$}& \multicolumn{2}{c}{$0.1702$ }& \multicolumn{1}{c}{$0.0415$}    \\
     & \multicolumn{1}{c}{$<0.0575$} &  \multicolumn{1}{c}{$<0.0614$} & \multicolumn{1}{c}{$<0.0584$} &  \multicolumn{1}{c}{$<0.0752$} &\multicolumn{1}{c}{$0.1179_{-0.022}^{+0.025}$ }& \multicolumn{1}{r}{$0.1399_{-0.022}^{+0.023}$} &\multicolumn{1}{c}{$<0.0635$ } \\
    \hline
    \multirow{2}{2cm}{ $\log_{10}z_c$}  & \multicolumn{2}{c}{$3.6232$}&  \multicolumn{2}{c}{$3.4968$}& \multicolumn{2}{c}{$3.5550$ }& \multicolumn{1}{c}{$3.6027$}    \\
     & \multicolumn{1}{c}{unconstr.} &  \multicolumn{1}{c}{$>3.27$} & \multicolumn{1}{c}{unconstr.} &  \multicolumn{1}{c}{$>3.30$} &\multicolumn{1}{c}{$3.621_{-0.11}^{+0.017}$ }& \multicolumn{1}{r}{$3.575_{-0.038}^{+0.031}$} &\multicolumn{1}{c}{unconstr. } \\
    \hline
    \multirow{2}{2cm}{ $\theta_i$}  & \multicolumn{2}{c}{$2.5916$}&  \multicolumn{2}{c}{$2.9075$}& \multicolumn{2}{c}{$2.7653$ }& \multicolumn{1}{c}{$2.5876$}    \\
     & \multicolumn{1}{c}{unconstr.} &  \multicolumn{1}{c}{unconstr.} & \multicolumn{1}{c}{unconstr.} &  \multicolumn{1}{c}{unconstr.} &\multicolumn{1}{c}{$2.747_{-0.11}^{+0.12}$ }& \multicolumn{1}{r}{$2.729_{-0.062}^{+0.089}$} &\multicolumn{1}{c}{unconstr. } \\
    \hline
    \hline
    \multirow{2}{2cm}{ $\omega_{cdm}$}   & \multicolumn{2}{c}{$0.1212$}&  \multicolumn{2}{c}{$0.1211$}& \multicolumn{2}{c}{$0.1394$ }& \multicolumn{1}{c}{$0.1252$} \\
     & \multicolumn{1}{c}{$0.1218_{-0.0024}^{+0.0012}$} &  \multicolumn{1}{c}{$0.1225_{-0.0027}^{+0.0014}$} & \multicolumn{1}{c}{$0.1217_{-0.0023}^{+0.0011}$} &  \multicolumn{1}{c}{$0.124_{-0.0036}^{+0.0016}$} &\multicolumn{1}{c}{$0.1311_{-0.0035}^{+0.0031}$ }& \multicolumn{1}{r}{$0.1354_{-0.0035}^{+0.0033}$} &\multicolumn{1}{c}{$0.123_{-0.0029}^{+0.0016}$ } \\
    \hline
    \multirow{2}{2cm}{ $10^2\omega_{b}$}    & \multicolumn{2}{c}{$2.2522$}&  \multicolumn{2}{c}{$2.2420$}& \multicolumn{2}{c}{$2.2669$ }& \multicolumn{1}{c}{$2.2529$}   \\
     & \multicolumn{1}{c}{$2.25_{-0.019}^{+0.017}$} &  \multicolumn{1}{c}{$2.246_{-0.019}^{+0.017}$} & \multicolumn{1}{c}{$2.249_{-0.018}^{+0.016}$} &  \multicolumn{1}{c}{$2.246_{-0.023}^{+0.016}$} &\multicolumn{1}{c}{$2.286_{-0.022}^{+0.022}$ }& \multicolumn{1}{r}{$2.276_{-0.02}^{+0.021}$} &\multicolumn{1}{c}{$2.246_{-0.021}^{+0.016}$ }  \\
    \hline
    \multirow{2}{2cm}{ $ln(10^{10}A_s)$}    & \multicolumn{2}{c}{$3.0535$}&  \multicolumn{2}{c}{$3.0615$}& \multicolumn{2}{c}{$3.0679$ }& \multicolumn{1}{c}{$3.0463$}    \\
     & \multicolumn{1}{c}{$3.047_{-0.015}^{+0.014}$} &  \multicolumn{1}{c}{$3.046_{-0.015}^{+0.014}$} & \multicolumn{1}{c}{$3.045_{-0.015}^{+0.014}$} &  \multicolumn{1}{c}{$3.047_{-0.015}^{+0.015}$} &\multicolumn{1}{c}{$3.066_{-0.015}^{+0.015}$ }& \multicolumn{1}{r}{$3.069_{-0.014}^{+0.014}$} &\multicolumn{1}{c}{$3.047_{-0.016}^{+0.014}$ } \\
    \hline
    \multirow{2}{2cm}{ $n_s$}    & \multicolumn{2}{c}{$0.9735$}&  \multicolumn{2}{c}{$0.9676$}& \multicolumn{2}{c}{$0.9926$ }& \multicolumn{1}{c}{$0.9708$}  \\
     & \multicolumn{1}{c}{$0.9689_{-0.006}^{+0.0047}$} &  \multicolumn{1}{c}{$0.9685_{-0.0063}^{+0.0046}$} & \multicolumn{1}{c}{$0.9684_{-0.0059}^{+0.0045}$} &  \multicolumn{1}{c}{$0.9686_{-0.0074}^{+0.0049}$} &\multicolumn{1}{c}{$0.9894_{-0.0064}^{+0.0059}$ }& \multicolumn{1}{r}{$0.9903_{-0.0061}^{+0.006}$} &\multicolumn{1}{c}{$0.9685_{-0.0067}^{+0.0049}$ }  \\
    \hline
    \multirow{2}{2cm}{ $\tau_\mathrm{reio}$}    & \multicolumn{2}{c}{$0.0580$}&  \multicolumn{2}{c}{$0.0588$}& \multicolumn{2}{c}{$0.0497$ }& \multicolumn{1}{c}{$0.0504$}    \\
    & \multicolumn{1}{c}{$0.05414_{-0.0075}^{+0.007}$} &  \multicolumn{1}{c}{$0.05308_{-0.0073}^{+0.0071}$} & \multicolumn{1}{c}{$0.05351_{-0.007}^{+0.0071}$} &  \multicolumn{1}{c}{$0.05175_{-0.007}^{+0.0073}$} &\multicolumn{1}{c}{$0.05649_{-0.0075}^{+0.0074}$ }& \multicolumn{1}{r}{$0.05405_{-0.0072}^{+0.0069}$} &\multicolumn{1}{c}{$0.05305_{-0.0076}^{+0.0071}$ }  \\
    \hline
    \hline
    $N_\mathrm{data}$ & \multicolumn{2}{c}{4086} & \multicolumn{2}{c}{4222}&\multicolumn{2}{c}{4090}&\multicolumn{1}{c}{3934} \\
    \hline
    $N_\mathrm{param}$ & \multicolumn{2}{c}{9 + 30 } & \multicolumn{2}{c}{9 + 38 }&\multicolumn{2}{c}{9 + 38 }&\multicolumn{1}{c}{9 + 22 } \\
    \hline
    $\chi^2$ & \multicolumn{2}{c}{4297} & \multicolumn{2}{c}{4407}&\multicolumn{2}{c}{4291}&\multicolumn{1}{c}{4185} \\
     \hline
     AIC& \multicolumn{2}{c}{4375 } & \multicolumn{2}{c}{4501}&\multicolumn{2}{c}{4385}&\multicolumn{1}{c}{4247 } \\
    \hline
    \hline
    \end{tabular}
\caption{Best-fit and mean $\pm$ $1\sigma$ bounds for the cosmological parameters of EDE measured from the combined full shape analysis of BOSS and BOSSz1+eBOSS with the baseline, as well as with the full PantheonPlusSH0ES likelihood. The cosmological results from the baseline come from the PlanckTTTEEE+lowl+lowE+lens+ext.BAO+PantheonPlus analysis. We present the mean and $1\sigma$ bounds of the full shape analyses for the two different prior choices on the linear nuisance parameters: Jeffreys prior (JP) and zero-centered Gaussian priors (GP).
$N_\mathrm{data}$ indicates the number of fitted data points for each data set combination. We consider  $N_\mathrm{BOSSz1} = N_\mathrm{BOSSz3} = 76$~\citep{BOSS:2016wmc}, $N_\mathrm{LRG} = N_\mathrm{QSO} = 72$ and $N_\mathrm{ELG} = 68$~\citep{eBOSS:2020yzd}, $N_\mathrm{BAO} = 2$~\citep{Beutler_2011,Ross:2014qpa}, $N_\mathrm{PantheonPlus} = 1580$~\citep{Brout:2022vxf}, $N_\mathrm{PantheonPlusSH0ES} = 1448$~\citep{Brout:2022vxf,Riess:2021jrx} and $N_\mathrm{Planck} = 2352$~\citep{Planck:2018vyg,Planck:2018lbu}.  $N_\mathrm{param}$ is the combination of the number of varied cosmological and nuisance parameters. 
The corresponding Akaike
Information Criterion (AIC) and $\chi^2$ values are calculated
at the best-fit points.}
\label{tab:results_ede_combined}
\end{table*}

We present constraints on cosmological parameters within EDE through a full shape analysis of eBOSS and BOSS data, incorporating various external data sets. We begin with constraints from LSS data combined with Planck, probing whether EDE can yield high values of $H_0$ without relying on information from the local distance ladder. Subsequently, we integrate SH0ES data and assess their impact on EDE constraints. The primary outcomes of this study are summarized in Fig.\ref{fig:EDE_base} and detailed in Table~\ref{tab:results_ede_combined}. To gauge the potential of EDE in addressing the $H_0$ tension, we evaluate its effectiveness after the inclusion of eBOSS data and examine its preference over $\Lambda$CDM in Table~\ref{tab:bic_comparison}. Looking into the near future, we present the potential of Planck-free constraints on EDE from eBOSS and BOSS in Fig.~\ref{fig:eft_only_ede} and Table~\ref{tab:results_ede_eft}. While  we acknowledge the presence of projection effects at the current stage, analyses of Stage-4 LSS survey have the potential to strongly mitigate these effects.
 
\subsection{Combination with external Data Sets}
First, we investigate whether large-scale structure data in combination with Planck supports an EDE energy fraction capable of entirely resolving the Hubble tension. In Table~\ref{tab:results_ede_combined} we summarise our results of a joint analysis of the full shape likelihood of BOSS and eBOSS data with Planck, external BAO and PantheonPlus data. We refer to the combination of PlanckTTTEEE+lens+ext.BAO+PantheonPlus data as our baseline data set. Fig.~\ref{fig:EDE_base} presents the marginalised posteriors on $\{h, \Omega_m, \sigma_8, f_\mathrm{EDE}\}$ as well the corresponding best-fit values and $\chi^2$ statistics. The contours of all parameters can be found in Fig.~\ref{fig:EDE_comb_gp} in Appendix~\ref{app:appendix}. As in the previous section, we consider both a Gaussian prior and a Jeffreys prior in our analysis (Section~\ref{sec:priors} for details on the different prior choices). Assuming Gaussian priors we find that while EDE predicts a higher $H_0$ value than $\Lambda$CDM, the increase is small.
The combination of baseline+BOSSz1+eBOSS gives an upper bound on $f_\mathrm{EDE} < 0.0584$ (95\%CL), while the baseline+BOSS leads to $f_\mathrm{EDE} < 0.0575$. Both these analyses lead to tighter upper bounds on $f_\mathrm{EDE}$ compared to the baseline ($f_\mathrm{EDE} < 0.0635$), indicating the importance of including the full shape analysis of large-scale structure data in the EDE discussion.
Our results indicate that the analysis of large-scale structure data in combination with Planck does not favour an EDE energy fraction able to entirely resolve the Hubble tension. These findings agree with recent literature~\citep{Hill:2020osr,Ivanov:2019pdj,DAmico:2020ods,Simon:2022adh,Simon:2023hlp}.\\

\begin{figure*}
\centering
     \begin{subfigure}[h]{0.49\textwidth}
         \includegraphics[width=0.9\textwidth]{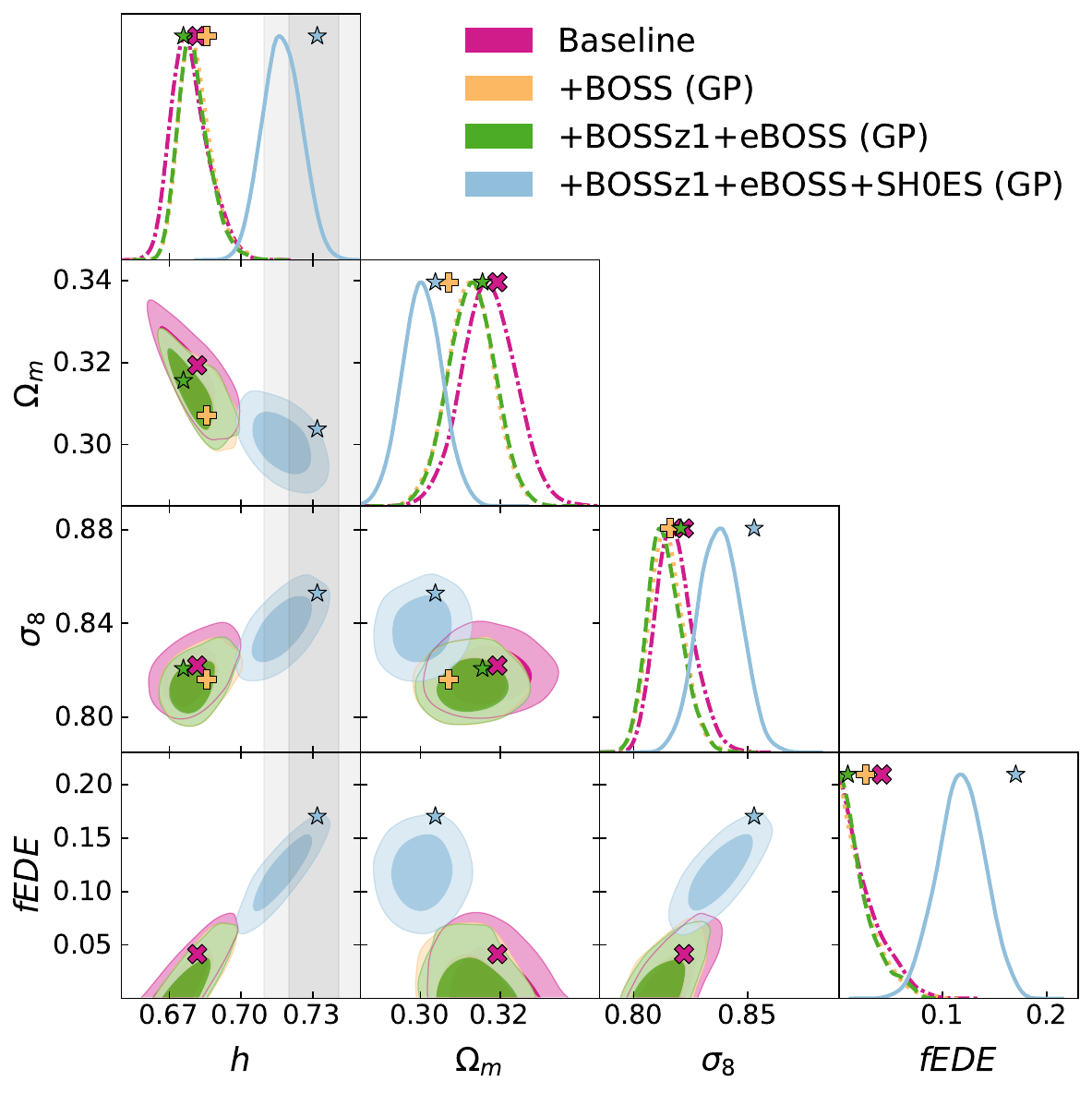}
         
     \end{subfigure}
     \hfill
     \begin{subfigure}[h]{0.49\textwidth}
         \centering
         \includegraphics[width=0.9\textwidth]{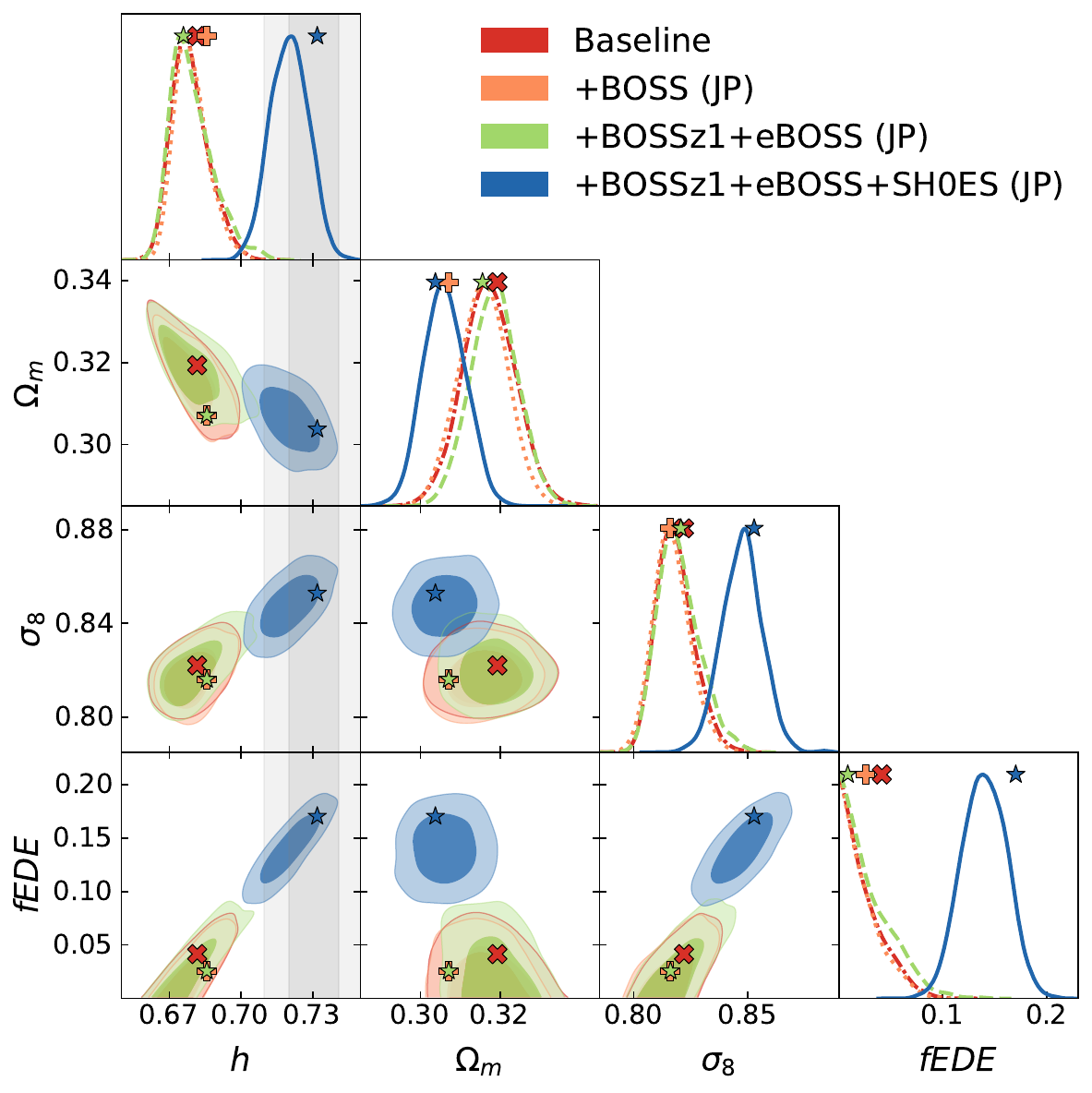}
    \end{subfigure}
    \hfill

\caption{1D and 2D posterior distributions on the cosmological parameters of the baseline (PlanckTTTEEE+lens+ext.BAO+PantheonPlus) (dot-dashed), baseline+EFTBOSS (dotted), baseline+EFTBOSS+EFTeBOSS (dashed) and baseline+EFTBOSS+EFTeBOSS+SH0ES (solid), where we included the full likelihood of PantheonPlusSH0ES. The two contour levels in the off-diagonal elements represent $68\%$ (inner) and $95\%$ (outer) credible intervals and the best-fit values are indicated with corresponding markers. The grey bands correspond to the $1\sigma$ and $2\sigma$ bounds coming from the latest SH0ES analysis~\citep{Riess:2021jrx}. \textit{Left:} zero-centerd Gaussian priors on the linear nuisance parameters. \textit{Right:} Jeffreys prior imposed on the linear nuisance parameters of the EFTofLSS analyses. 
}
\label{fig:EDE_base}
\end{figure*} 
Next, we examine the impact of adopting a Jeffreys prior instead of Gaussian priors on our results. While Jeffreys priors have been effective in mitigating projection effects that pose challenges to the robustness of the EFTofLSS analysis within the framework of LCDM, the scenario is more nuanced when considering the EDE model. As was pointed out in various previous works~\citep{Schoneberg:2021qvd,Smith:2020rxx,Murgia:2020ryi,Herold:2021ksg}, posteriors in EDE analyses are highly non-Gaussian and volume effects can appear upon marginalisation. In the limit of $f_\mathrm{EDE} \rightarrow 0$, the model becomes equivalent to $\Lambda$CDM but with two additional redundant parameters ($z_c$ and $\theta_i$).
This can lead to an artificial preference for $f_\mathrm{EDE} = 0$ in the marginalised posteriors. The previous results should therefore be interpreted with care.
Recent literature~\citep{Herold:2021ksg,Herold:2022iib,Reeves:2022aoi,Cruz:2023cxy} complements Bayesian analyses of EDE with a profile likelihood approach, which are free of volume projection effects. However, profile likelihoods introduce increased numerical complexity, particularly with a large number of parameters.

When employing the Jeffreys prior in the context of EDE, we can effectively mitigate projection effects arising from the linear nuisance parameters of the EFTofLSS. However, it is important to note that projection effects stemming from newly introduced EDE parameters persist. 
Nevertheless, as detailed in Section~\ref{sec:EDEmocks}, the joint application of full shape analyses on both BOSS and eBOSS data sets, assuming a Jeffreys prior, is anticipated to result in fewer volume projection effects compared to analyses employing Gaussian priors, as long as we include external data sets possessing sufficient constraining power to address degeneracies among EDE parameters.
The 2D posteriors coming from the full shape analysis of BOSS and eBOSS with a Jeffreys prior in combination with the baseline are shown on the right of  Fig.~\ref{fig:EDE_base}. As for the analysis with the Gaussian priors, the full contours can be found in in Fig.~\ref{fig:EDE_comb_jp} in Appendix~\ref{app:appendix}. The corresponding reconstructed posteriors and best fit values are given in Table~\ref{tab:results_ede_combined}.
 Employing a Jeffreys prior, we find an upper bound on $f_\mathrm{EDE} < 0.0614$ (95\% CL) when baseline+BOSS is analysed and $f_\mathrm{EDE} < 0.0752$ (95\% CL) when baseline+BOSSz1+eBOSS is analysed. In both cases, we infer consistently higher upper bounds on $f_\mathrm{EDE}$ as for analyses with Gaussian priors. \\

\begin{table*}
    \centering
    \begin{tabular}{p{2.7cm} p{1cm} p{1cm}p{1cm} p{1cm} p{1cm} p{1cm} p{1cm} p{1cm} p{1cm} p{1cm} } 
    \hline
    \hline
    &\multicolumn{5}{|c|}{$\Lambda$CDM }&\multicolumn{5}{|c|}{EDE }\\
    \hline
    \hline
    Planck high-l TTTEEE & - & 2347.6 & 2346.8 & 2349.4 & 2348.0 & - & 2345.6& 2346.1& 2346.1 & 2352.3 \\ 
    \hline
    Planck low-l EE & - & 395.8 & 395.8 & 395.8 & 397.2& - & 395.7&396.8 & 397.2 & 395.9\\ 
    \hline
    Planck low-l TT & - & 23.5 & 23.1 & 22.5 & 22.7 & - & 22.8 &22.2 & 23.3 & 21.3\\ 
    \hline
    Planck lensing & - & 9.0 & 8.9 & 9.2 & 8.9 & -  & 9.6 &  9.0& 9.5& 10.6\\ 
    \hline
    BAO small-z  & -&  0.8 & 1.0 & 1.1 & 1.5& - & 0.8& 1.4 & 1.0& 1.6\\ 
    \hline
    BOSS z1 & 60.4 &  - &  59.0  & 59.2  & 59.3 & 59.4  & - & 59.4& 59.1& 60.4 \\ 
    \hline
    BOSS z3 &  -&  - & 52.8 & - & - & - &  -& 50.3& -& -\\ 
    \hline
    eBOSS LRG & 54.8 &  - & - & 54.8 & 55.3 & 53.9 &  -& - & 54.7& 54.3 \\ 
    \hline
    eBOSS QSO & 46.4 &  -& - & 41.0 & 41.2 & 40.5 &  -& - & 40.6& 41.1\\ 
    \hline
    eBOSS ELG & 64.6 &  - & - & 64.8& 65.9 & 64. 0 &  -& - &  64.8& 64.4\\ 
    \hline
    Pantheon+ & -  &  1410.1 & 1410.9 & 1410.9 & - &  - &1410.4 & 1411.9 & 1410.7 & - \\ 
    \hline
    PantheonPlusSH0ES & - &  - & - & -  & 1326.1 & & -& - & - & 1289.2\\ 
    \hline
    SH0ES & 3.3 & - & - & -  &  - &  0.1 & -& - & - & -\\ 
    \hline
    $\omega_b$ & 0.5 & - & - & -  &  - &  0.2 & -& - & - & -\\ 
    \hline
    \hline 
    $\chi^2_{\mathrm{EDE}} - \chi^2_{\Lambda\mathrm{CDM}} $ & &  &  &  & & -12 & -2 & -1& -2 & -35\\ 
    \hline
    $\mathrm{AIC}_{\mathrm{EDE}} -\mathrm{AIC}_{\Lambda\mathrm{CDM}} $ & &  &  &  & & -6  & +4& +5 & +4& -29 \\ 
    \hline
    \hline 
    
    \end{tabular}
\caption{Table of best-fit $\chi^2$ contributions from individual experiments for $\Lambda$CDM and EDE when fitted to different data set combinations: BOSSz1+eBOSS+SH0ES, Baseline+BOSS, Baseline+BOSSz1+eBOSS, Baseline+BOSSz1+eBOSS+SH0ES. The individual total $\chi^2$ can be found in Table~\ref{tab:results_lcdm_combined} for $\Lambda$CDM constrained by baseline+EFT, Table~\ref{tab:results_ede_combined} for EDE constrained by baseline+EFT and Table~\ref{tab:results_ede_eft} for EFT+SH0ES constraints on $\Lambda$CDM and EDE. We report the difference in $\chi^2$ between EDE and $\Lambda$CDM, as well as the difference in AIC.}
\label{tab:bic_comparison}
\end{table*}

With the inclusion of SH0ES data, we reconstruct $f_\mathrm{EDE} = 0.1179_{-0.022}^{+0.025}$ with $h = 0.7173_{-0.0086}^{+0.0082}$~km/s/Mpc and $f_\mathrm{EDE} =0.1399_{-0.022}^{+0.023}$ with $h =0.7203_{-0.0087}^{+0.0082}$~km/s/Mpc for analyses with the assumption of Gaussian prior and Jeffreys prior, respectively.
Both correspond to a more than $5\sigma$ detection of a non-zero $f_\mathrm{EDE}$. The consistency of $h$ with the SH0ES constraint $h= 0.7304 \pm 0.0104$~km/s/Mpc~\citep{Riess:2021jrx} is at the  $1\sigma$ level for the analysis with Gaussian priors and the $0.8\sigma$ level for the analysis with the Jeffreys prior. In comparison, for $\Lambda$CDM, the consistency is much poorer ($4.1\sigma$ and $4.3\sigma$, respectively).
In line with previous work, the inclusion of SH0ES data not only results in a higher value of $h$ and, consequently, $f_\mathrm{EDE}$, but also induces a shift in the spectral tilt $n_s$ within EDE relative to $\Lambda$CDM. For both choices of priors, we observe a value within $2\sigma$ of scale-invariant value of $n_s = 1$. The resolution of the Hubble tension through EDE would thus carry important implications for models seeking to describe the primordial Universe~\citep{DAmico:2021fhz, Kallosh:2022ggf, Ye:2021nej, Smith:2022hwi, Jiang:2022qlj}.

\subsection{The Hubble tension and EDE}

Finally, we quantify how well EDE is able to resolve the tension  and discuss improvements in the overall fit to the data by examining the change in $\chi^2$ assuming EDE and $\Lambda$CDM, respectively (see Table~\ref{tab:bic_comparison}). Naively, one would expect an improvement in $\chi^2$ equal to the number of additional parameters. A common way to present data comparison is therefore with the reduced $\chi^2$. But while the number of degrees of freedom can be estimated for models where the fitting parameters enter in a linear way, the effective number of degrees of freedom is unknown for models with non-linear fitting parameters~\citep{andrae2010and}. 
As a way to look beyond the regular $\chi^2$ statistic and attempt a fairer model comparison of EDE with $\Lambda$CDM, we, therefore, turn to the Akaike Information Criterion (AIC)~\citep{Trotta:2008qt}, which penalises models with additional degrees of freedom: 
\begin{equation}
    \mathrm{AIC} \equiv -2 \ln \mathcal{L}_\mathrm{max} + 2 N_\mathrm{param},
\end{equation}
where $N_\mathrm{param}$ is the number of fitted parameters (cosmological and nuisance) and $\mathcal{L}_\mathrm{max}$ is the maximum likelihood value. We present $\Delta$AIC relative to $\Lambda$CDM for the different data set combinations in Table~\ref{tab:bic_comparison}, where the model with minimum AIC is preferred. While the improvement in $\chi^2$ is close to what is expected $\Delta \chi^2 \simeq 3$, the AIC shows that $\Lambda$CDM is preferred over EDE for all data combinations where no additional information from  SH0ES is added. Only when considering SH0ES data, the AIC favors EDE over $\Lambda$CDM, with a substantial $\Delta \mathrm{AIC} = -29$. The improvement of $\Delta \chi^2 = -35$ is mainly coming from the fit to the PantheonPlusSH0ES likelihood, where there is an improvement of $\Delta \chi^2_\mathrm{PantheonPlusSH0ES} = -36.9$, while the overall fit to Planck is just slightly worsened by $\Delta \chi^2_\mathrm{Planck} = +3.3$. This demonstrates the overall attraction of EDE:  the fit to Planck is maintained, while $H_0$ is compatible with SH0ES. \\

In order to address the question of how the inclusion of SH0ES data is impacting the fit of a certain model $\mathcal{M}$ to  a given data set $\mathcal{D}$, we compute the tension metric $Q_\mathrm{DMAP}$ as discussed in~\citet{Raveri:2018wln,Schoneberg:2021qvd}:
\begin{equation}
    Q_\mathrm{DMAP} \equiv \sqrt{\chi^2_\mathcal{M} (\mathcal{D}+\mathrm{SH0ES}) - \chi^2_\mathcal{M}(\mathcal{D})} ,
\end{equation}
where we compare the difference of the maximum a posterior (DMAP) values upon the addition of a $H_0$ prior~\citep{Riess:2021jrx}\footnote{It was demonstrated in~\citet{Simon:2023hlp} that incorporating the SH0ES prior onto the Pantheon+ likelihood yields constraints on EDE models equivalent to those from the full PantheonPlusSH0ES likelihood~\citep{Riess:2021jrx}. For simplicity, we calculate $Q_\mathrm{DMAP}$ by substituting the PantheonPlusSH0ES likelihood with PantheonPlus along with the Gaussian prior on $H_0$ from~\citet{Riess:2021jrx} in the corresponding minimization process of the chains.}.
In terms of the $Q_\mathrm{DMAP}$ metric, the tension between SH0ES and the baseline+BOSSz1+eBOSS is reduced from $5.2 \sigma$ in $\Lambda$CDM to $3.0 \sigma$ in EDE. For baseline+BOSS alone, the tension is lowered to $2.0 \sigma$ in the case of EDE. The difference in $Q_\mathrm{DMAP}$ for EDE between baseline+BOSSz1+eBOSS and baseline+BOSS is primarily attributed to a poorer fit to the PlanckTTTEEE likelihood.
\citet{Schoneberg:2021qvd} established criteria for classifying the success of a given model in addressing the $H_0$ tension. Models were categorized based on three individual tests: Achieving  a good fit to all data with a minimal $Q_\mathrm{DMAP}$, significantly improving the fit over $\Lambda$CDM according to the AIC and the ability of allowing for high values of $H_0$ without incorporating a local distance ladder prior.
As per~\citet{Schoneberg:2021qvd}, passing the $Q_\mathrm{DMAP}$ criterion required reducing the tension to $\leq 3\sigma$, and the AIC criterion must suggest more than a weak preference over $\Lambda$CDM following the Jeffreys scale~\citep{jeffreys_theory_1998,Nesseris:2012cq}, specifically with $\Delta\mathrm{AIC} \leq -6.91$. Although each criterion has its advantages and drawbacks, they are computationally efficient and thought to complement one another. A more detailed Bayes factor study~\citep{Kass:1995loi} is, therefore, deferred to future work, and the question of whether EDE is a suitable model to explain the tension is addressed here using the aforementioned criteria. While EDE is incapable of accommodating sufficiently high $H_0$ values to resolve the tension without SH0ES data, as discussed earlier, it passed the $Q_\mathrm{DMAP}$ and AIC criterion in previous works~\citep{Schoneberg:2021qvd,Simon:2023hlp}. Our analysis, incorporating the full shape analysis of BOSSz1+eBOSS, suggests that EDE clearly passes the AIC test (with $\Delta\mathrm{AIC} = -29$), while the $Q_\mathrm{DMAP}$ test with $Q_\mathrm{DMAP} = 3\sigma$ represents the upper limit for passing the test. Although eBOSS data currently does not decisively exclude EDE as a model to address the $H_0$ tension, it reveals a tendency that the inclusion of large-scale structure data exerts increasing pressure on EDE  as a solution to the tension. Consequently, it will be crucial to assess the implications of Stage-4 large-scale structure data, such as from DESI~\citep{DESI:2016fyo} or Euclid~\citet{Amendola:2016saw}, on our understanding of EDE.\\

\subsection{Full Shape analysis in combination with SH0ES}
In a final step, we discuss the constraining power of a Planck-free analysis of EDE with a full shape analysis of eBOSS and BOSS, including a BBN prior as defined in Sec.~\ref{sec:dataII}. For a discussion about the impact of different BBN priors on the Hubble tension and EDE constraints, we refer the reader to~\citet{Takahashi:2023twt}. As we have discussed in Sec.~\ref{sec:EDEmocks}, we will not show full shape only constraints due to possible strong volume projection effects. In order to minimize potential volume effects, we are combining the full shape analysis of BOSS and eBOSS with SH0ES. The results are presented in Table~\ref{tab:results_ede_eft} and the relevant $\chi^2$ statistics can be found in Table~\ref{tab:bic_comparison}. We show the associated reconstructed 2D posteriors in Fig.~\ref{fig:eft_only_ede} and compare our results with the posteriors reconstructed from PlanckTTTEE+lowl+lowE+lens+SH0ES. Best-fit values are indicated by cross markers for both analyses. 
Upon incorporating SH0ES data, we derive the following EDE parameter constraints: $f_\mathrm{EDE} = 0.2539_{-0.13}^{+0.089}$ with $h = 0.7289_{-0.011}^{+0.011}$ km/s/Mpc and $f_\mathrm{EDE} = 0.1794_{-0.06}^{+0.085}$ with $h = 0.7328_{-0.01}^{+0.011}$ km/s/Mpc for analyses assuming Gaussian and Jeffreys priors, respectively. The consistency of $h$ with the SH0ES constraint $h= 0.7304 \pm 0.0104$~km/s/Mpc~\citep{Riess:2021jrx} is within $<0.1\sigma$ for the analysis with Gaussian priors and $<0.2\sigma$ for the analysis with the Jeffreys prior. In comparison, for $\Lambda$CDM, the consistency is within $<1.5\sigma$ and $<0.8\sigma$, respectively. In the case of Planck combined with SH0ES, we obtain $f_\mathrm{EDE} = 0.1084_{-0.024}^{+0.031}$ with $h = 0.7157_{-0.01}^{+0.0091}$ km/s/Mpc. When contrasting the difference in $\chi^2$ between $\Lambda$CDM and EDE, we observe an improvement of $\Delta \chi^2 = -12$ for the full shape only constraints within EDE. This improvement primarily stems from the fit to the SH0ES prior, where there is a specific enhancement of $\Delta \chi^2_\mathrm{SH0ES} = -3.2$. Furthermore, we note an enhancement in the eBOSS QSO data set of $\Delta \chi^2_\mathrm{QSO} = -5.9$, as the QSOs favor a higher $\Omega_m$ value, which is more likely accommodated within the EDE model. The full shape analysis of BOSSz1+eBOSS+SH0ES hints at a mild preference for EDE over $\Lambda$CDM with $\Delta$AIC = -6. \\

However, it is important to exercise caution in interpreting these results at this stage. As highlighted in the earlier mock analysis (Sec.~\ref{sec:EDEmocks}), eBOSS faces substantial challenges related to projection effects when considering EDE, both with Gaussian priors and a Jeffreys prior on the linear nuisance parameters. These effects arise due to the increasing significance of nonlinear degeneracies. Comparing the best-fit values with the mean of the posteriors in Fig.~\ref{fig:eft_only_ede}, it becomes evident that imposing an $H_0$ prior is insufficient to fully account for all projection effects when varying all EDE parameters. 
Consequently, we conclude that while, in principle, we can extend constraints beyond $\Lambda$CDM models with BOSS and eBOSS full shape analyses alone, the observed projection effects need to be treated cautiously. To achieve full shape constraints comparable to Planck, we must await forthcoming large-scale structure surveys such as DESI or Euclid, where these projection effects are expected to play a subdominant role due to the enhanced constraining power of the data sets.

\begin{figure*}
\centering
\includegraphics[width=0.9\textwidth]{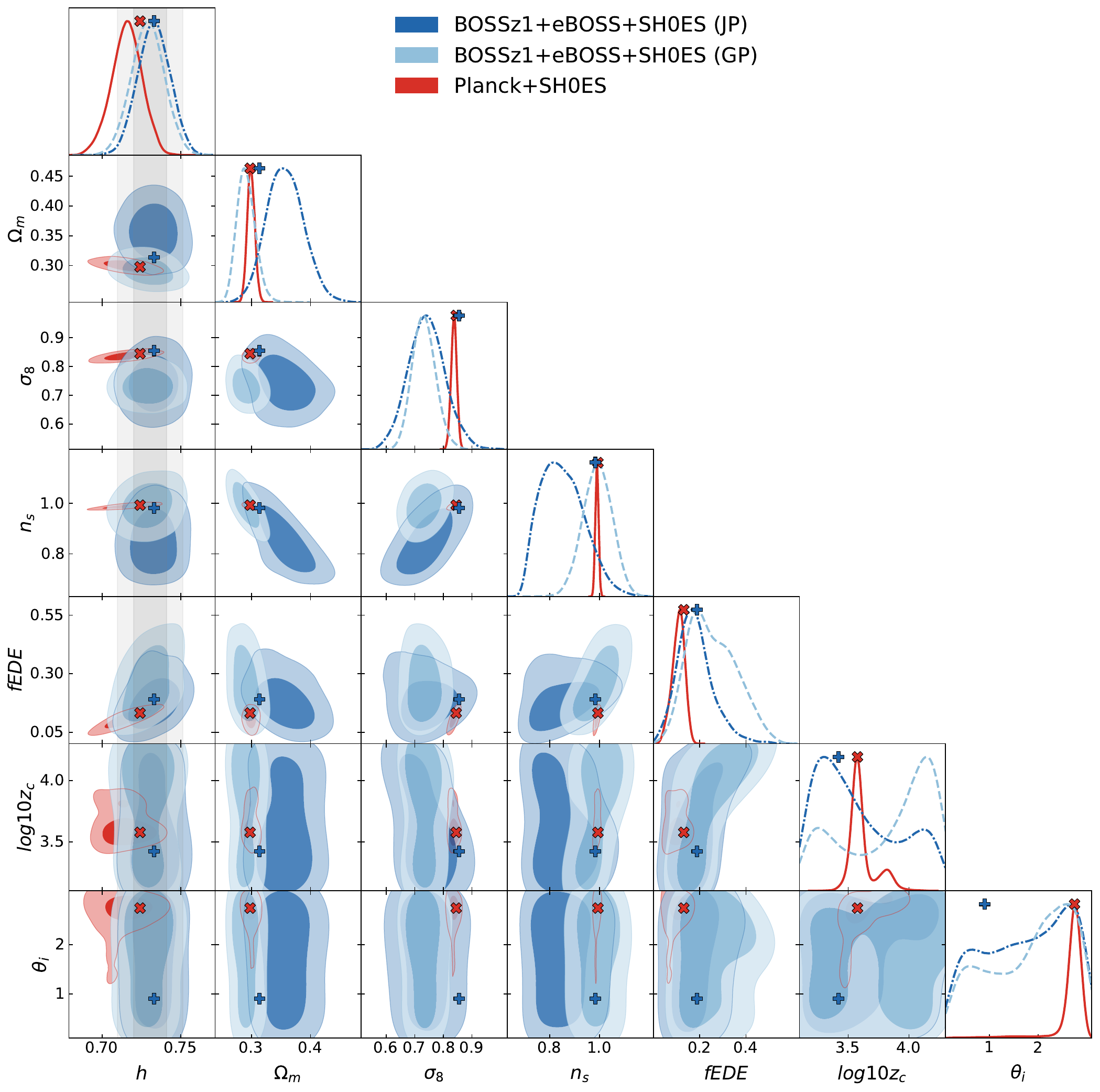}
\caption{1D and 2D posterior distributions on the cosmological parameters of the combined full shape analysis of BOSSz1 and eBOSS in combination with SH0ES (imposing a BBN prior~\citep{Schoneberg:2019wmt}) for the two different prior choices: Jeffreys prior (dark blue, dot-dashed line) and zero-centered Gaussian priors (light blue, dashed line). The two contour levels in the off-diagonal elements represent $68\%$ (inner) and $95\%$ (outer) credible intervals and the best-fit values are indicated with corresponding markers. We also show results from PlanckTTTEEE+lowl+lowE+lens in combination with SH0ES (red, solid line) for comparison. The grey bands correspond to $1\sigma$ and $2\sigma$ bounds of the latest SH0ES analysis~\citep{Riess:2021jrx}.}
\label{fig:eft_only_ede}
\end{figure*}

\begin{table*}
    \centering
    \begin{tabular}{ p{1.5cm} p{2.cm} p{2.cm} p{2.cm} p{2.cm} p{2.cm}} 
    \hline
    \hline
     &\multicolumn{2}{|c|}{BOSSz1+eBOSS+SH0ES}&\multicolumn{2}{|c|}{BOSSz1+eBOSS+SH0ES} & \multicolumn{1}{c}{PlanckTTTEEE+lens+SH0ES}  \\
     &\multicolumn{2}{|c|}{$\Lambda$CDM}&\multicolumn{2}{|c|}{EDE}& \multicolumn{1}{|c|}{EDE} \\
    &\multicolumn{1}{c}{GP}&\multicolumn{1}{c}{JP}&\multicolumn{1}{c}{GP}&\multicolumn{1}{c}{JP}&\\
    \hline
    \hline
    \multirow{2}{2cm}{ $\Omega_{m}$}  &  \multicolumn{2}{c}{0.3417}  &  \multicolumn{2}{c}{$0.3138$}& \multicolumn{1}{c}{$0.2976$}  \\
     & \multicolumn{1}{c}{$0.3092_{-0.018}^{+0.014}$}& \multicolumn{1}{c}{$0.3845_{-0.027}^{+0.026}$ }& \multicolumn{1}{c}{$0.2909_{-0.017}^{+0.014}$} &  \multicolumn{1}{c}{$0.3574_{-0.033}^{+0.03}$}&\multicolumn{1}{c}{$0.2992_{-0.0067}^{+0.0063}$ } \\
    \hline
    \multirow{2}{2cm}{ $h$} &  \multicolumn{2}{c}{$0.7116$} &  \multicolumn{2}{c}{$0.7330$}& \multicolumn{1}{c}{$0.7240$}   \\
    & \multicolumn{1}{c}{$0.7107_{-0.009}^{+0.0088}$}& \multicolumn{1}{c}{$0.7197_{-0.0095}^{+0.0093}$} & \multicolumn{1}{c}{$0.7289_{-0.011}^{+0.011}$} & \multicolumn{1}{c}{$0.7328_{-0.01}^{+0.011}$} &\multicolumn{1}{c}{$0.7157_{-0.0091}^{+0.01}$}  \\
    \hline
    \multirow{2}{2cm}{ $\sigma_8$} &\multicolumn{2}{c}{0.7874} &\multicolumn{2}{c}{$0.8558$  }& \multicolumn{1}{c}{ $0.8450$}    \\
     & \multicolumn{1}{c}{$0.7162_{-0.041}^{+0.039}$ }& \multicolumn{1}{c}{$0.7208_{-0.053}^{+0.045}$ }& \multicolumn{1}{c}{$0.7331_{-0.045}^{+0.039}$ } &\multicolumn{1}{c}{$0.742_{-0.069}^{+0.059}$ } &\multicolumn{1}{c}{$0.8372_{-0.01}^{+0.01}$}\\
    \hline
    \multirow{2}{2cm}{ $f_\mathrm{EDE}$} & \multicolumn{2}{c}{$-$} & \multicolumn{2}{c}{$0.1897$}& \multicolumn{1}{c}{$0.1319$}    \\
     & \multicolumn{1}{c}{$-$}& \multicolumn{1}{c}{$-$}& \multicolumn{1}{c}{$0.2539_{-0.13}^{+0.089}$}& \multicolumn{1}{c}{$0.1794_{-0.085}^{+0.06}$} &\multicolumn{1}{c}{$0.1084_{-0.024}^{+0.031}$ } \\
    \hline
    \multirow{2}{2cm}{ $\log_{10}z_c$}  & \multicolumn{2}{c}{$-$} & \multicolumn{2}{c}{$3.4216$}& \multicolumn{1}{c}{$3.5765$}    \\
     & \multicolumn{1}{c}{$-$}& \multicolumn{1}{c}{$-$}& \multicolumn{1}{c}{$3.83^{+0.48}_{-0.65}$} &  \multicolumn{1}{c}{$3.59^{+0.57}_{-0.46}$} & \multicolumn{1}{c}{$3.614_{-0.11}^{+0.032}$}  \\
    \hline
    \multirow{2}{2cm}{ $\theta_i$} & \multicolumn{2}{c}{$-$} & \multicolumn{2}{c}{$0.9013$}&  \multicolumn{1}{c}{$2.748$}    \\
     & \multicolumn{1}{c}{$-$}& \multicolumn{1}{c}{$-$}& \multicolumn{1}{c}{$1.8^{+1.2}_{-1.4}$} &  \multicolumn{1}{c}{$1.75^{+0.89}_{-0.89}$}&\multicolumn{1}{c}{$2.684_{-0.068}^{+0.22}$ } \\
    \hline
    \hline
    \multirow{2}{2cm}{ $\omega_{cdm}$}  &\multicolumn{2}{c}{$0.1495$}&\multicolumn{2}{c}{$0.1453$}& \multicolumn{1}{c}{$0.1326$}   \\
     & \multicolumn{1}{c}{$0.1326_{-0.01}^{+0.0084}$}& \multicolumn{1}{c}{$0.1758_{-0.016}^{+0.016}$}& \multicolumn{1}{c}{$0.1311_{-0.0093}^{+0.0074}$} &\multicolumn{1}{c}{$0.1686_{-0.019}^{+0.016}$}&\multicolumn{1}{c}{$0.1298_{-0.0036}^{+0.0038}$}  \\
    \hline
    \multirow{2}{2cm}{ $10^2\omega_{b}$}  & \multicolumn{2}{c}{$2.2899$}& \multicolumn{2}{c}{$2.2667$}&  \multicolumn{1}{c}{$2.2825$}    \\
     & \multicolumn{1}{c}{$2.296_{-0.039}^{+0.038}$}& \multicolumn{1}{c}{$2.283_{-0.04}^{+0.036}$}& \multicolumn{1}{c}{$2.274_{-0.039}^{+0.037}$} & \multicolumn{1}{c}{$2.267_{-0.04}^{+0.036}$} &\multicolumn{1}{c}{$2.284_{-0.021}^{+0.022}$} \\
    \hline
    \multirow{2}{2cm}{ $ln(10^{10}A_s)$}  &  \multicolumn{2}{c}{$2.8076$}&  \multicolumn{2}{c}{$3.0780$}& \multicolumn{1}{c}{$3.0762$}    \\
     & \multicolumn{1}{c}{$2.714_{-0.14}^{+0.13}$ }& \multicolumn{1}{c}{$2.506_{-0.16}^{+0.13}$  }& \multicolumn{1}{c}{$2.906_{-0.15}^{+0.14}$ } & \multicolumn{1}{c}{$2.702_{-0.23}^{+0.19}$ }&\multicolumn{1}{c}{$3.07_{-0.017}^{+0.015}$}  \\
     \hline
    \multirow{2}{2cm}{ $n_s$}  & \multicolumn{2}{c}{$0.8310$}& \multicolumn{2}{c}{$0.9818$}&  \multicolumn{1}{c}{$0.9931$}    \\
     & \multicolumn{1}{c}{$0.8945_{-0.05}^{+0.05}$}& \multicolumn{1}{c}{$0.7606_{-0.061}^{+0.015}$}& \multicolumn{1}{c}{$0.9908_{-0.059}^{+0.062}$} & \multicolumn{1}{c}{$0.8519_{-0.11}^{+0.071}$} &\multicolumn{1}{c}{$0.9891_{-0.0065}^{+0.007}$} \\
    \hline
    \hline
    $N_\mathrm{data}$ & \multicolumn{2}{c}{288}& \multicolumn{2}{c}{288} &\multicolumn{1}{c}{2352} \\
    \hline
    $N_\mathrm{param}$ & \multicolumn{2}{c}{5+16}& \multicolumn{2}{c}{8 + 16} &\multicolumn{1}{c}{8 + 21 } \\
    \hline
    $\chi^2$ & \multicolumn{2}{c}{$230$}& \multicolumn{2}{c}{$218$} &\multicolumn{1}{c}{2775} \\
    \hline
     AIC& \multicolumn{2}{c}{272 }& \multicolumn{2}{c}{266 } & \multicolumn{1}{c}{2833} \\
    \hline
    \hline
    \end{tabular}
\caption{Best-fit and mean $\pm$ $1\sigma$ bounds for the cosmological parameters of $\Lambda$CDM and EDE measured from the combined full shape analysis of BOSSz1+eBOSS+SH0ES, with a BBN prior on $\omega_b$, confronted with the EDE analysis of Planck+SH0ES.  We present the mean and $1\sigma$ bounds of the full shape analyses for the two different prior choices on the linear nuisance parameters: Jeffreys prior (JP) and zero-centered Gaussian priors (GP). We consider  $N_\mathrm{BOSSz1} = N_\mathrm{BOSSz3} = 76$~\citep{BOSS:2016wmc},  $N_\mathrm{LRG} = N_\mathrm{QSO} = 72$ and $N_\mathrm{ELG} = 68$~\citep{eBOSS:2020yzd}, $N_\mathrm{H_0} = 1$~\citep{Riess:2021jrx}, $N_\mathrm{BBN} = 1$~\citep{Schoneberg:2019wmt} and $N_\mathrm{Planck} = 2352$~\citep{Planck:2018vyg,Planck:2018lbu}. The corresponding Akaike
Information Criterion (AIC) and $\chi^2$ values are calculated
at the best-fit points.
}
\label{tab:results_ede_eft}
\end{table*}

\section{Conclusions}
\label{sec:conclusion}

In this paper, we present constraints on EDE derived from a full shape analysis applied to the complete eBOSS DR16 data sets, encompassing LRG, QSO, and ELG data. For the full shape analysis of eBOSS and BOSS data in combination with Planck, external BAO measurements, PantheonPlus (baseline), and SH0ES data, we derive values of $H_0 = 71.73_{-0.86}^{+0.82}$ km/s/Mpc and $H_0 = 72.03_{-0.87}^{+0.82}$ km/s/Mpc for analyses assuming Gaussian priors and a Jeffreys prior on the linear nuisance parameters, respectively. Irrespective of the priors applied, consistency with SH0ES is found at a level of less than $1\sigma$ for EDE, contrasting with the $>4\sigma$ consistency for $\Lambda$CDM. Utilizing the Akaike Information Criterion (AIC) to assess the preference for EDE over $\Lambda$CDM, accounting for the additional parameters introduced by EDE, reveals a mild preference for $\Lambda$CDM in scenarios involving the baseline dataset, baseline+BOSS, and baseline+BOSSz1+eBOSS. Only with the inclusion of additional information from SH0ES does the AIC distinctly favor EDE over $\Lambda$CDM, indicating a significant preference with $\Delta$AIC = -29. For an evaluation of the residual Hubble tension using the $Q_\mathrm{DMAP}$ metric, we observe a reduction of the tension to $3\sigma$ in EDE, in contrast to $5.2\sigma$ in $\Lambda$CDM, considering the baseline+BOSSz1+eBOSS dataset. The introduction of eBOSS data, however, elevates the tension in EDE from $2\sigma$ to $3\sigma$. Although eBOSS does not currently exclude EDE, its inclusion indicates a trend that additional large-scale structure data increases the pressure of EDE of simultaneously fitting various data sets.\\

We have also demonstrated that using additional large-scale structure data in EDE analyses could significantly limit the model as a potential resolution to the Hubble tension. In particular, full shape analyses exhibit the ability to impose stringent constraints on EDE parameters. Using the full shape analysis of eBOSS and BOSS data, while incorporating a BBN prior and SH0ES data, yields $H_0$ values of $72.9_{-1.1}^{+1.1}$ km/s/Mpc and $73.3_{-1.0}^{+1.1}$ km/s/Mpc, depending on the application of Gaussian or Jeffreys priors. The analysis suggests a mild preference for EDE over $\Lambda$CDM with $\Delta$AIC = -6. While we are capable of placing constraints on EDE parameters through full shape-only analyses, we showed that it is crucial to understand the impact of possible remaining projection effects on the cosmological constraints.\\

While the projection effects coming from the marginalisation of linear nuisance parameters are expected to be reduced in Stage-IV surveys, we have proposed methods to address them within EDE for eBOSS data by exploring different prior choices. Building on recent successes in addressing projection effects in $\Lambda$CDM through the application of a Jeffreys prior to linear nuisance parameters, we explored its effectiveness in mitigating projection effects in EDE and compared it to zero-centered Gaussian priors which are widely used in literature. Our comprehensive mock study, covering $\Lambda$CDM and EDE, suggests that while the Jeffreys prior is able to mitigate projection effect within $\Lambda$CDM, in the case of EDE - where the introduction of additional parameters leads to degeneracies between non-linear cosmological parameters - significant projection effects persist with both prior choices. Full shape analysis in the context of beyond $\Lambda$CDM achieves unbiased posteriors only when combined with data sets capable of breaking these newly introduced parameter degeneracies.\\

We defer to future work the examination of the impact of a Jeffreys prior on non-linear parameters within models beyond $\Lambda$CDM in EFTofLSS, as well as the direct comparison with a profile likelihood analysis.

\section*{Acknowledgements}
We would like to thank Seshadri Nadathur and Arnaud de Mattia for useful discussions regarding the eBOSS data sets. We also would like to thank Harry Desmond for insights about the Jeffreys prior and Pierre Zhang regarding questions about \textsc{Pybird}. RG and JD-M were supported by STFC studentships. RZ is supported by NSFC grants 11925303 and 11890691, and is also supported by the Chinese Scholarship Council (CSC) and the University of Portsmouth. KK, DB and RC are supported by STFC grant ST/W001225/1.
TS is supported by the European Research Council (ERC) under the
European Union’s HORIZON-ERC-2022 (Grant agreement No. 101076865). For the purpose of open access, the author(s) has applied a Creative Commons Attribution (CC BY) licence to any Author Accepted Manuscript version arising. 

\section*{Data Availability}
Supporting research data are available on request from the corresponding author.

\bibliographystyle{mnras}
\bibliography{ref}

\appendix

\section{Contours}
\label{app:appendix}
In this appendix, we show the full contours discussed in Sec.~\ref{sec:EDEresults}. We show full shape constraints of BOSS and eBOSS on all cosmological parameters. Fig.~\ref{fig:EDE_comb_gp} corresponds to the analysis assuming Gaussian priors on the linear nuisance parameters, while Fig.~\ref{fig:EDE_comb_jp} corresponds to the analysis imposing a Jeffreys prior on the linear nuisance parameters. Different coloured markers indicate the respective best-fit values.
\begin{figure*}
\centering
\includegraphics[width=0.9\textwidth]{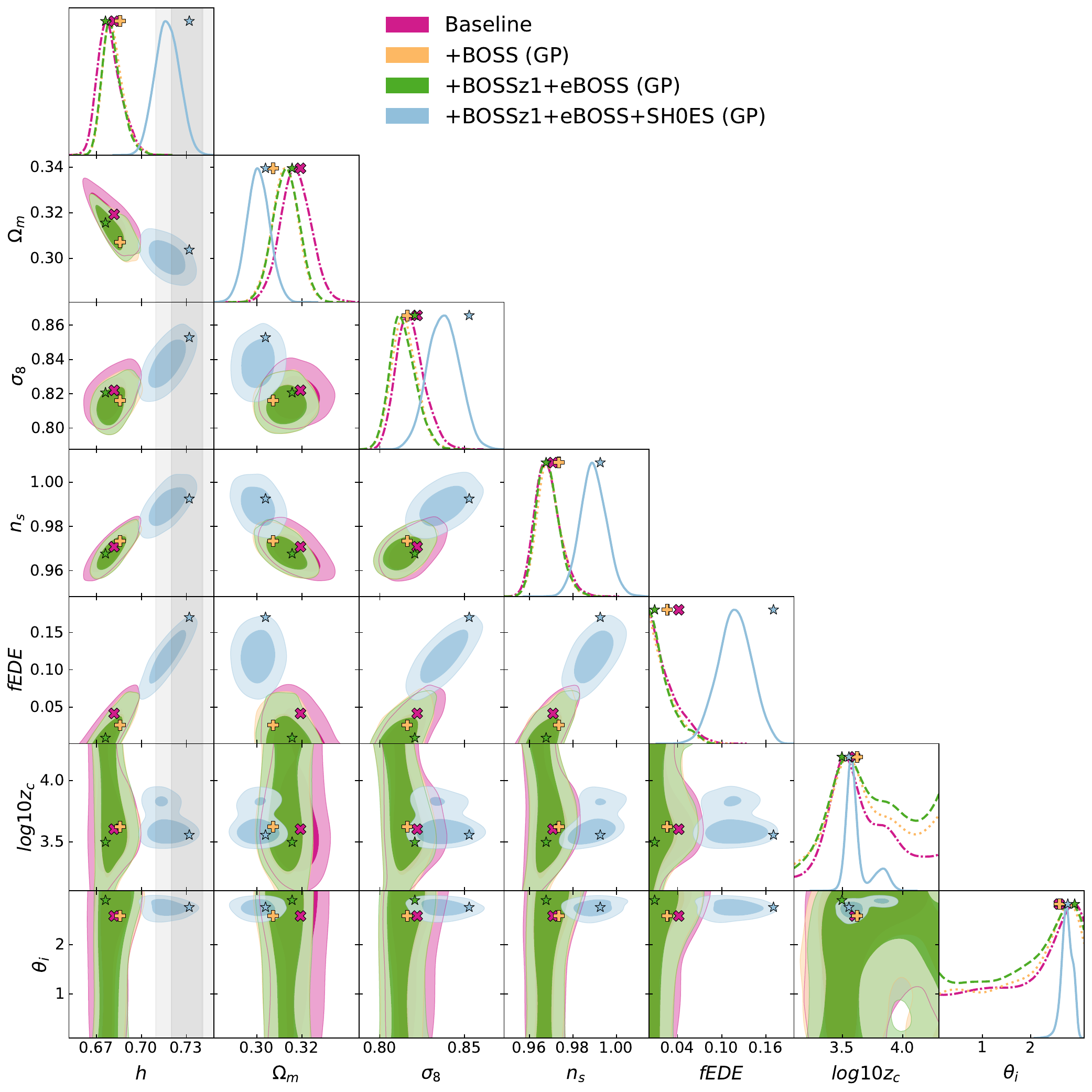}
\caption{The full 1D and 2D posterior distributions on the cosmological parameters of the baseline (PlanckTTTEEE+lens+ext.BAO+PantheonPlus) (dot-dashed), baseline+BOSS (dotted), baseline+BOSSz1+eBOSS (dashed) and baseline+BOSSz1+eBOSS+SH0ES (solid), where we included the full likelihood of PantheonPlusSH0ES.  We show results with zero-centered Gaussian priors imposed on the linear nuisance parameters. }
\label{fig:EDE_comb_gp}
\end{figure*}

\begin{figure*}
\centering
\includegraphics[width=0.9\textwidth]{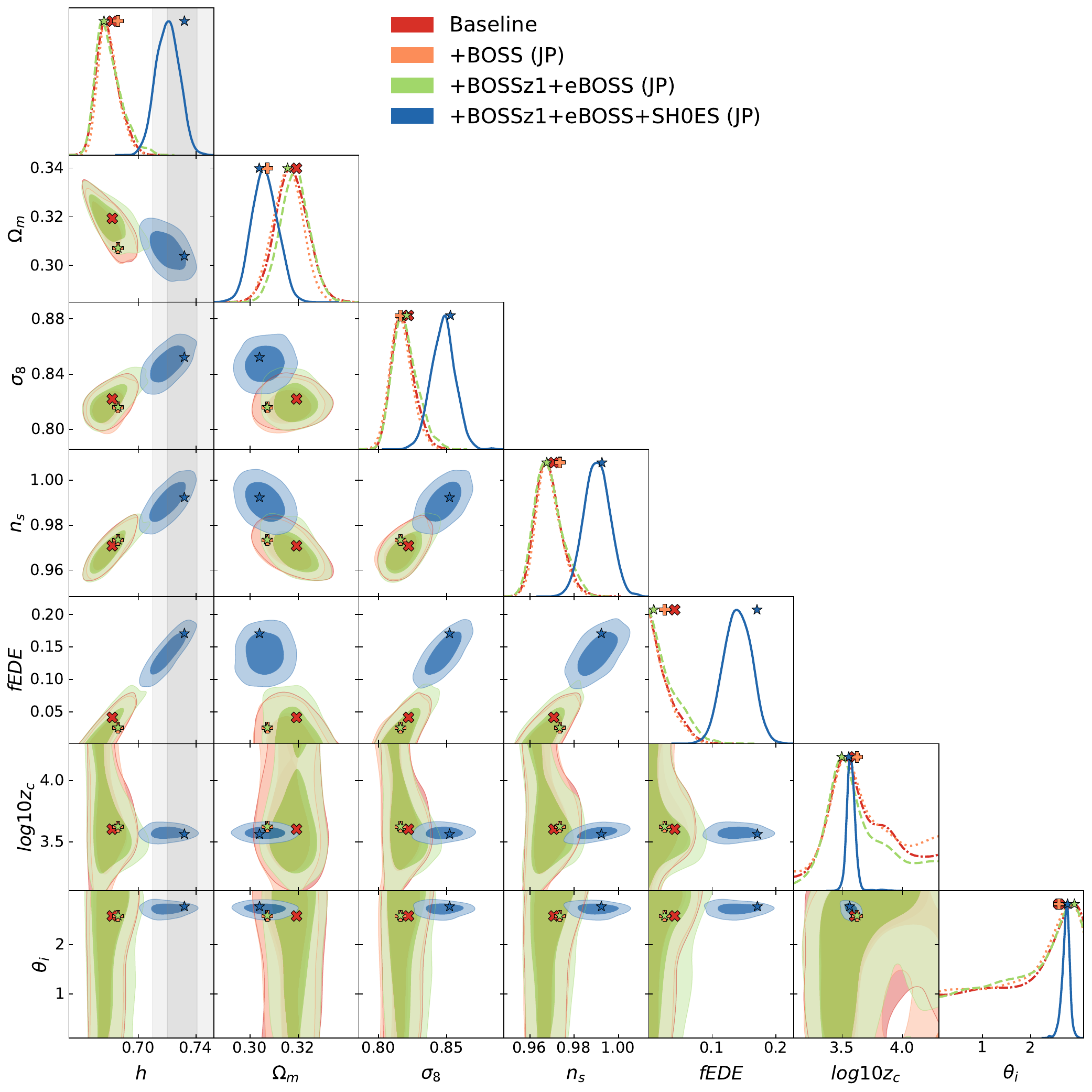}
\caption{The full 1D and 2D posterior distributions on the cosmological parameters of the baseline (PlanckTTTEEE+lens+ext.BAO+PantheonPlus) (dot-dashed), baseline+BOSS (dotted), baseline+BOSSz1+eBOSS (dashed) and baseline+BOSSz1+eBOSS+SH0ES (solid), where we included the full likelihood of PantheonPlusSH0ES.  We show results with a Jeffreys prior imposed on the linear nuisance parameters. }
\label{fig:EDE_comb_jp}
\end{figure*}



\bsp	
\label{lastpage}
\end{document}